\pdfoutput=1 
\documentclass[12pt]{report}
\usepackage[utf8]{inputenc}
\usepackage{geometry}
\usepackage{float}
\usepackage{units}
\usepackage{mathrsfs}
\usepackage{mathtools}
\usepackage{pdfpages}
\usepackage{amsmath}
\usepackage{amssymb}
\usepackage{stackrel}
\usepackage{graphicx}
\usepackage{setspace}
\PassOptionsToPackage{normalem}{ulem}
\usepackage{ulem}
\makeatletter

\InputIfFileExists{t2aenc.def}{}{%
  \errmessage{File `t2aenc.def' not found: Cyrillic script not supported}}
\DeclareRobustCommand{\cyrtext}{%
  \fontencoding{T2A}\selectfont\def\encodingdefault{T2A}}
\DeclareRobustCommand{\textcyr}[1]{\leavevmode{\cyrtext #1}}

\newcommand{\lyxmathsym}[1]{\ifmmode\begingroup\def\b@ld{bold}
  \text{\ifx\math@version\b@ld\bfseries\fi#1}\endgroup\else#1\fi}

\@ifundefined{date}{}{\date{}}

\usepackage{graphics}
\usepackage{setspace}

\usepackage{geometry}
\usepackage{tikz}
\usepackage{tikz-cd}
\usepackage{amssymb}
\usepackage{quiver}
\usepackage{pdfpages}
\usepackage{comment}
\usepackage[T2A]{fontenc}
\date{}

\usepackage{cite}

\usepackage{color}

\usepackage{tocloft}
\cftpagenumbersoff{subsection}
\cftpagenumbersoff{subsubsection}

\RequirePackage{tikz-cd}
\RequirePackage{amssymb}
\usetikzlibrary{calc}
\usetikzlibrary{decorations.pathmorphing}

\tikzset{curve/.style={settings={#1},to path={(\tikztostart)
    .. controls ($(\tikztostart)!\pv{pos}!(\tikztotarget)!\pv{height}!270:(\tikztotarget)$)
    and ($(\tikztostart)!1-\pv{pos}!(\tikztotarget)!\pv{height}!270:(\tikztotarget)$)
    .. (\tikztotarget)\tikztonodes}},
    settings/.code={\tikzset{quiver/.cd,#1}
        \def\pv##1{\pgfkeysvalueof{/tikz/quiver/##1}}},
    quiver/.cd,pos/.initial=0.35,height/.initial=0}

\tikzset{tail reversed/.code={\pgfsetarrowsstart{tikzcd to}}}
\tikzset{2tail/.code={\pgfsetarrowsstart{Implies[reversed]}}}
\tikzset{2tail reversed/.code={\pgfsetarrowsstart{Implies}}}
\tikzset{no body/.style={/tikz/dash pattern=on 0 off 1mm}}

\makeatother

\begin{document}



\title{Compositional nonlinear audio signal processing with Volterra series}

\author{Jake Araujo-Simon}

\date{}

\maketitle

\section*{\centering Abstract}

We present a compositional theory of nonlinear audio signal processing
based on a categorification of the Volterra series. We begin by augmenting the
classical definition of the Volterra series so that it is functorial with
respect to a base category whose objects are temperate distributions
and whose morphisms are certain linear transformations. This motivates the derivation
of formulae describing how the outcomes of nonlinear transformations
are affected if their input signals are linearly processed --
e.g., translated, modulated, sampled, or periodized. We then consider
how nonlinear audio systems, themselves, change, and introduce
as a model thereof the notion of morphism of Volterra series, which we exhibit as both
a type of lens map and natural transformation. We show how
morphisms can be parameterized and used to generate indexed families
(e.g., sequences) of Volterra series, which are well-suited to model
nonstationary or time-varying nonlinear phenomena. We then describe how
Volterra series and their morphisms organize into a category, which we call
\emph{Volt}. We exhibit the operations of sum, product, and series composition
of Volterra series as monoidal products on \emph{Volt}, and identify,
for each in turn, its corresponding universal property. In particular,
we show that the series composition of Volterra series is associative.
We then bridge between our framework and the subject at the heart
of audio signal processing: time-frequency analysis. Specifically, we show that
a known equivalence, between a class of second-order Volterra series and the bilinear
time-frequency distributions (TFDs), can be extended to one between certain
higher-order Volterra series and the so-called polynomial TFDs. We end by outlining
potential avenues for future work, including the incorporation of nonlinear
system identification techniques and the potential extension of our
theory to the settings of graph and topological audio
signal processing.\pagebreak{}

\tableofcontents{}

\cleardoublepage{}

\section*{\centering Notation\protect \\
}

$\omega=2\pi f$---angular frequency\\
$\boldsymbol{\omega}\in\mathbb{R}^{n}$---an $n-$dimensional vector\\
$f:X\rightarrow Y$---a map from a space $X$ to a space $Y$\\
$D^{C}$---the space of mappings from $C$ to $D$\\
$S(\mathbb{R})$---the semicategory of signals\\
$S^{'}(\mathbb{R})$---the category of temperate distributions \\
$F:S^{'}(\mathbb{R})\rightarrow S^{'}(\mathbb{R})$---the Fourier
transform\\
$s\in S(\mathbb{R}^{n})$---an $n-$signal\\
$\hat{s}=F(s)\in S^{'}(\mathbb{R}^{n})$---an $n-$spectrum\\
$T^{k}(W)$---the $k^{\text{th}}-$tensor power of a vector space
$W$\\
$T(W)=\bigoplus_{k=0}^{\infty}T^{k}(W)$---the tensor algebra of
$W$\\
$s^{\otimes j}$---the $j-$fold iterated tensor product of $s$
with itself\\
$(B\triangleleft A)(-)=B(A(-))$---the series composition of $A$
and $B$

\cleardoublepage\pagenumbering{arabic}

\typeout{} 
\chapter*{Introduction}

\markboth{}{}

\addcontentsline{toc}{chapter}{\numberline{0}Introduction}A nonlinear
system $V$ is one whose response to a set of inputs may be more than
the sum of its responses to each of the inputs independently; that
is, its behavior may fail to satisfy the properties of superposition
\begin{equation}
V[s+s^{'}]\ne V[s]+V[s^{'}]\label{eq:failure of superposition}
\end{equation}
and scaling
\[
V[\alpha s]\ne\alpha V[s]
\]
for all signals $s,s'$ and scalar $\alpha$.\footnote{The class of nonlinear systems includes, by this definition, the linear
ones. The term nonlinear might, therefore, be better thought of as
an acronym for \emph{no}t\emph{ n}ecessarily\emph{ linear}.}

But nonlinearity is only a negativistic notion. The departure of a
nonlinear system from linearity could, instead, be written positively,
in terms of one or more \emph{nonlinear effects}: for example
\begin{equation}
V[s+s^{'}]=V[s]+V[s^{'}]+\gamma(s,s'),\label{eq:nonlinear effect}
\end{equation}
where $\gamma$ is a function that now encodes the nonlinear part
and depends on both the inputs $s$ and $s^{'}$. But how do nonlinear
effects arise, and how can the function $\gamma$ be described?

The \emph{Volterra series} (VS) is a universal\footnote{Strictly speaking, the Volterra series is a universal model only for
nonlinear systems that have \emph{fading memory} - i.e., for which
the influence of inputs from the distant past on the present output
trends asymptotically to zero \cite{boyd_fading_1985}. However, this
is an immanently physical condition.} and combinatorially complete model of nonlinear dynamics that provides
a concrete answer to these questions. Roughly, it says that nonlinear
effects arise as \emph{sums of weighted interactions between the samples
of} (various combinations of) \emph{the input signals}. More precisely,
a Volterra series consists of a collection of weights that measure
the strength of a nonlinear system's response to each possible (multiplicative)
interaction of the samples amongst each possible combination of input
signals, taken with repetition (allowing for self-interactions) and counting
permutations.

As the focus of this work is on the nonlinear processing of conventional
audio, and on the structure of nonlinear audio systems and their transformations,
it deals near-to-exclusively with nonlinear systems that process one
or many streams of audio data. However, the abstract form of the Volterra
series is ammenable to generalization, and capable of describing
nonlinear systems that process inputs from other types of domains, such as
signals over graphs or topological spaces. Thus, the relevance is to the broader
context of nonlinear systems theory and topological signal processing.

The domain of audio moreover provides a fruitful testbed for the development
of a general theory of nonlinear signal processing, since nonlinear
effects in audio abound, and their representations in the frequency
domain are relatively well understood both mathematically and intuitively.
For example, intermodulation, harmonic and inharmonic distortion,
and nonlinear resonance are all classes of nonlinear phenomena that
are routinely \emph{experienced } by audio engineers
and musicians. The frequency domain form of the VS affords, in this
light, a mathematical syntax that maps well onto an audio-based semantics; and conversely,
the audio-based semantics provides a natural lens by which to interpret the mathematics.

\section*{Overview}

In Chapter 1, we motivate the VS and briefly survey the core variants
that have been introduced in the literature, observing the various
facets of the same mathematical object that are illuminated by the
different definitions. We begin by reviewing the univariate Volterra
series in both its time and frequency domain forms. We then prepare
the lift to the multivariate case by rewriting the univariate series
using the tensor product, and recall the universal property of the
tensor product, as well as the tensor power and tensor algebra constructions.
We then review the multivariate VS and show how, in a certain precise
sense, it is combinatorially complete. We then briefly review the
VS with parametrized kernels, and catalogue the Volterra series representations
of some elementary linear and nonlinear operators.

In Chapter 2, we begin the process of categorifying the Volterra series, by extending
the domain and codomain of its action beyond a space of mere isolated signals to one that includes the linear transformations
between them. That is, we redefine the Volterra series as an endofunctor
on the category, $S^{'}(\mathbb{R})$, of tempered distributions and convolutors between them, which we introduce
in Appendix A. In particular, we ask what the change in the result would be of processing an input signal by
a nonlinear system, if the input were linearly transformed; can we derive an associated linear
transformation between the corresponding nonlinear outputs? We give concrete answers to this
question in the cases of four elementary types of linear transformations: translation, modulation,
periodization, and sampling.

In Chapter 3, we introduce a mechanism for how nonlinear systems modeled
by Volterra series, themselves, change; that is, we define a notion of \emph{morphism
of Volterra series}, which we exhibit as a type of lens map and show is a natural
transformation between them. Volterra series morphisms enable us to speak
about the networks of ways that nonlinear systems are
related to one another functionally, and to define a category in which Volterra series,
themselves, are the objects, which we call \emph{Volt}.

In Chapter 4, we study three basic ways of combining Volterra series: sum, product,
and series composition - and for each, we write down the universal property satisfied by the
mode of combination. In Appendix B we prove the associativity of the most complex of these
three: the series composition product, $\triangleleft$. In so doing,
we effectively exhibit \emph{Volt }as a symmetric monoidal category,
with series composition as the monoidal (tensor) product. Thus, we establish a categorical
framework in which nonlinear systems and their interconnections can
be modeled in a rigorous diagrammatic way by Volterra series, in a
manner analogous to how circuit diagrams or signal flow graphs are
used to model the interconnections of linear ones.

Having built up the theoretical framework and categorical structure
of Volterra series in Chapters 2, 3, and 4, in Chapter 5 we turn our focus
to a subject at the heart of audio signal processing: time-frequency distributions (TFDs).
We first recapitulate a result of Powers and Nam \cite{nam_volterra_2003}, in which they
showed that the time-frequency shift-covariant TFDs of Cohen's class
are equivalent to a certain class of double, second-order Volterra series.
We then extend the equivalence to the realm of time-\emph{multi}frequency,
or higher-order time-frequency distributions (HO-TFD), and specifically
to the higher-order Wigner-Ville distribution (HO-WVD) and polynomial
Wigner-Ville distribution (PWVD). Thus we enable the incorporation of higher
time-frequency analysis into the categorical framework developed in the previous chapters.

We end in Chapter 6 with some concluding remarks and a prospectus
for future work, including possible extensions of our theory to the contexts of
graph and topological signal processing, as well as the development
of analysis and identification methods for the decomposition of complex nonlinear
systems into simpler ones.

\section*{Prior Work}

Nonlinear systems in audio are ubiquitous. Practically all analog
musical instruments are examples: the string on a violin, which exhibits
some stiffness; a metal drum; or the reed of a clarinet mouthpiece.
So, too, in analog electronics, diodes, transistors, inductors and
transformers are nonlinear--as are the audio circuits containing
them, such as mixers, modulators, and rectifiers. In reality, it is
only our idealizations of physical systems that are truly linear. Yet, nonlinear
systems have rarely been treated in full generality and systematically,
with bespoke approaches instead being taken to tasks of modeling particular
nonlinear systems; or else machine learning techniques have been employed,
which potentially offer little insight for a human operator into the
physical mechanisms of action underlying the system behavior.

In the 1950's, Norbert Wiener and others revived and modernized the
theory of analytic functionals due to Volterra\footnote{Norbert Wiener had become interested in Volterra series decades earlier,
and in the late 1940's the theory was privately developed for military and defense purposes by
researchers at MIT. It was not until the late
1950's that the research was disclosed, and a modern
theory of Volterra series made its way into the public domain.} \cite{volterra_theory_1959}. It has since been used in a vast array
of engineering contexts: see, e.g., \cite{cheng_volterra-series-based_2017}
for a recent overview and \cite{rugh_nonlinear_1981,mathews_polynomial_2000}
for textbook introductions to the subject. However, despite its success,
two primary factors have limited the application of the Volterra series
as a model. The first is that the Volterra series is computationally
complex, with complexity increasing rapidly in both the series order of nonlinearity and memory.
The second is that Volterra series-based methods have typically
been applied to problems in an \emph{ad hoc}, rather than a general,
manner. For example, a common approach to nonlinear system identification
is to model the unknown system under test as a single, monolithic
Volterra series, and then solve for its parameters using harmonic
probing methods; however, these methods become computationally intractable
for highly nonlinear systems\cite{peyton_jones_new_2018,novak_nonlinear_2010,novak_synchronized_2015}.
As a result, simplified Volterra models, such as memoryless polynomial
or Hammerstein series, have often been used instead \cite{rebillat_prediction_2010,novak_nonlinear_2010,novak_nonparametric_2014,rebillat_identification_2011}.

More recently, neural networks (NN) have dominated the field of nonlinear
approximation. Work has been done on the relationship between Volterra
series and NNs; see, \cite{hakim_volterra_1991,stegmayer_towards_2004}
for Volterra series-based representations of NNs; \cite{marmarelis_relation_1994}
for a technique to estimate VS kernels from a feedforward network;
\cite{marmarelis_volterra_1997} on the relationship between VS and
3-layer perceptrons; and, more recently, \cite{roheda_volterra_2020}
on Volterra Neural Networks.

While the use of Volterra series in audio contexts has been more limited,
applications have included the modeling of complex effects generators
such as guitar pedals \cite{novak_chebyshev_2010}, a Moog ladder
filter \cite{helie_volterra_2010}, speech waveforms \cite{despotovic_nonlinear_2012,patil_nonlinear_2013},
ubiquitous elements such as acoustic transducers, microphones, and
loudspeakers -- which operate in linear regimes only up to limiting
points, past which they exhibit (harmonic or inharmonic) distortion
\cite{hoerr_using_2019,loriga_nonlinear_2017,mezghani-marrakchi_nonlinear_2014,orcioni_identification_2018,rebillat_identification_2011,rebillat_prediction_2010,reed_practical_1996}
-- and musical instruments \cite{tronchin_modelling_2017}. Volterra
series have also been of theoretical interest; in particular, a subfamily
of Volterra series was shown equivalent to the Time-Frequency Distributions,
of fundamental importance to the theory of signal processing
\cite{nam_volterra_2003}.

The interconnection of nonlinear systems is more complex and has been
less studied than in the linear case; however, there is precedence
for a systematic approach. In \cite{rugh_nonlinear_1981}, Rugh described
so-called \emph{interconnection structured systems} modeled by Volterra
series, and gave formulas for how to compute the kernels of such systems.
In \cite{chen_modeling_1995}, Chen catalogued the Volterra series
models of a variety of different types of nonlinear systems, and gave
wiring, or signal-flow, diagrams for each. In \cite{helie_volterra_2010},
Hélie described various products of Volterra series, including the
composition product, and used an interconnection of Volterra series
to model a complex audio effect. In \cite{carassale_modeling_2010},
Carassale and Kareem catalogued the rules for sum, product, and series
composition of Volterra series as well as the Volterra series representations
of some simple systems, emphasizing how a modular approach to nonlinear
system modeling enables the application of heterogeneous methods to
various parts of the modeling process; and in \cite{carassale_synthesis_2014},
they gave a similar account for multi-variate Volterra series. In
\cite{rebillat_identification_2011,novak_nonparametric_2014}, similar
treatments were given of the interconnections for Diagonal Volterra,
or Hammerstein Series, focusing on the analysis of such systems using
the exponential swept sine method, with applications to audio.

Graph-based methods have also arisen, in the context of graph signal
processing (GSP), providing another approach to the interconnection
of audio systems. GSP is an emerging theory that extends linear signal
processing to the case of signals indexed over graphs; see \cite{shuman_emerging_2013}
for an overview. Vertex-frequency graph signal processing (VFGSP)
has also arisen as a set of tools for the joint localization in the
graph frequency and vertex domains, playing, in relation to GSP, a
role analogous to that played by time-frequency methods in relation
to classical signal processing \cite{stankovic_vertex-frequency_2020}.
A continuous version of GSP, graphon signal processing, is also a
topic of active research \cite{ruiz_graphon_2021,morency_graphon_2021,ghandehari_noncommutative_2022,beck_signal_2023}.
The more general theory of topological signal processing has also
recently emerged, which uses, in place of graphs, topological models
such as simplicial complexes and hypergraphs to model the signal domain,
and assigns data over these spaces using sheaves\cite{barbarossa_topological_2020,puschel_discrete_2021,sardellitti_topological_2021,robinson_topological_2014}.
Recently, Volterra series were introduced in a GSP context as graph
Volterra models, which were then upgraded in \cite{leus_topological_2021}
to topological Volterra filters.

Although often not explicitly stated as such, sheaf theoretic and
topological accounts are very close in spirit to category theoretic
ones\footnote{Indeed, category theory sprang out of work done \cite{eilenberg_general_1945}
by Eilenberg and MacLane in algebraic topology, to which sheaves are
central.}. Although still uncommon in the practice of signal processing, category theory
is already ubiqitous as a formalizing language within the fields
that the theory of signal processing draws principally from, such as harmonic analysis, the
theory of Hilbert spaces, quantum mechanics, measure theory, and functional
analysis. Outside of audio, there have been some fruitful interactions
between category theory and signal processing: for example, in topological
signal processing \cite{robinson_nyquist_nodate,robinson_topological_2014};
statistical signal processing \cite{pastor_mathematical_2020}; and
the formalization of a diagrammatic calculus for signal flow graphs
\cite{di_lavore_monoidal_2022}; as well as a unified treatment of
various signal processing concepts via the notion of functoriality
\cite{samant_functorial_2018}. There is also at least one application
of topos theory (a rich subfield of category theory) to the treatment
of music theory \cite{mazzola_topos_2017}. The theory of `algebraic
signal processing', developed in \cite{puschel_algebraic_2012} is
also resemblant of a categorical approach. There has also been work
at the intersection of signal processing and temporal logic, which
made use of Volterra series \cite{deshmukh_logical_2020}. Very recently,
there has been a line of work developed at the intersection of audio,
topological signal processing, and sheaf theory \cite{essl_causal_2022,essl_deforming_2022,essl_topological_2020,essl_topologizing_2021,essl_topology_2022},
which has touched upon and implicated many category theoretic concepts.

Category theory is, itself, sometimes regarded as a theory of interconnection,
or composition. In particular, the recent book \cite{poly_book}
develops a theory of interaction that takes place within the category,
Poly, of polynomial functors and natural transformations between them.
This work is particularly salient to our own and one which we drew inspiration
from, since Volterra series appear to be, in a way, to signals what
polynomials are to sets.

\subsubsection*{A note on style}

We attempt to strike in this work a compromise, in both style and
level of rigour, between what would be more characteristic of the
signal processing and engineering literature - from which have learned
most of what we know about Volterra series - and what would be more
idiomatic to a mathematical field such as category theory, whose notation
and typical manner of discourse leaves little to no room for ambiguity.
In particular, we follow, but loosely, the convention of always specifying,
for any mapping (such as a signal or system) $F$, its \emph{type} alongside the definition of
its \emph{action}; i.e., we write $F:A\rightarrow B$, where $A$
is \emph{$F$}'s \emph{domain }and $B$ is its \emph{codomain, }alongside
any formula for how to compute the \emph{image} $F(a)$$\in B$ of
each element $a\in A$.

\typeout{} 
\chapter{The Volterra series}

In this chapter, we introduce the Volterra series classically and
at increasing levels of generality. See \cite{rugh_nonlinear_1981}
or \cite{mathews_polynomial_2000} for a standard exposition of the
subject. We begin, in section 1.1, with the univariate Volterra series,
which models a nonlinear system that processes a single input signal.
In section 1.2 we review multilinear maps and the tensor product,
and in sections 1.3 and 1.4 we review the multivariate Volterra series
and its parameterized version. We then survey the representation of
a few elementary systems by Volterra series. We refer liberally to
the spaces $S(\mathbb{R})$ and $S^{'}(\mathbb{R})$, of (Schwartz)
signals and tempered distributions, respectively, but relegate their
constructions to Appendix A.

\section{Volterra series basics}

The (univariate) Volterra series is an operator, $V:S(\mathbb{R})\rightarrow S(\mathbb{R})$
acting on the space of signals, $S(\mathbb{R})$. Its output is defined
as a sum of the diagonal outputs of a series of multilinear operators,
which in the time domain is written
\begin{equation}
y(t)=V[s](t)\coloneqq\sum_{j=0}^{\infty}V_{j}[s](t)\label{eq:}
\end{equation}
\begin{equation}
=V_{0}+\sum_{j=1}^{\infty}\int_{\boldsymbol{\tau}_{j}\in\mathbb{R}^{j}}v_{j}(\boldsymbol{\tau}_{j})\prod_{r=1}^{j}s(t-\tau_{r})d\tau_{r}\label{eq:univariate Volterra series; time domain}
\end{equation}
\\
where each \emph{homogeneous Volterra operator} (VO),\emph{ }$V_{j}$,
for $j\ge1$, is defined by its action
\[
y_{j}(t)=V_{j}[s](t)=\int_{\boldsymbol{\tau}_{j}\in\mathbb{R}^{j}}v_{j}(\boldsymbol{\tau}_{j})\prod_{r=1}^{j}s(t-\tau_{r})d\tau_{r},
\]
and where the zeroth-order operator is given by 
\[
y_{0}(t)=V_{0}[s](t)=v_{0},
\]
where $v_{0}$ is a constant that does not depend on the input\footnote{We will often drop the zeroth-order term from the notation and write
$\sum_{j=0}^{\infty}\int_{\boldsymbol{\tau}_{j}\in\mathbb{R}^{j}}v_{j}(\boldsymbol{\tau}_{j})\prod_{r=1}^{j}s(t-\tau_{r})d\tau_{j}$
instead of $v_{0}+\sum_{j=1}^{\infty}\int_{\boldsymbol{\tau}_{j}\in\mathbb{R}^{j}}v_{j}(\boldsymbol{\tau}_{j})\prod_{r=1}^{j}s(t-\tau_{r})d\tau_{r}$
as in equation \ref{eq:univariate Volterra series; time domain},
for a Volterra series in the time domain. Where we do, then when $j=0$
it is to be understood that $\tau_{0}$ is the unique point in the
domain $\mathbb{R}^{0}$, $v_{0}:\mathbb{R}^{0}\rightarrow\mathbb{R}$
is a constant function at the point, and the product $\stackrel[r=1]{0}{\prod}$on
the right-hand side evaluates to $1$ (it is the product of zero factors).}. The functions $v_{j}\in S'(\mathbb{R}^{j})$ are often called \emph{Volterra
kernels}, but we will refer to them here as Volterra kernel functions
(VKFs). They are often assumed to be symmetric - a feature which we
discuss in section (1.1.3) - and their support reflects the system's
\emph{memory}\footnote{The \emph{support} of a function is the subset of its domain that
is not mapped to zero.}. The signal output by the homogeneous Volterra operator at order
$j$ is written $y_{j}(t)=V_{j}[s](t)$, and the output of the entire
series is the sum of these homogeneous outputs: $y(t)=\sum_{j=0}^{\infty}y_{j}(t)$.

Two remarks should be made about the notation in (\ref{eq:univariate Volterra series; time domain}).
First, although a single input signal is processed, $j-$many copies
of the input signal are generated at each order $j$ and used to define
a $j-$dimensional function via the product of their amplitudes sampled
at various delays. This $j-$dimensional function is, in fact, the
$j^{\text{th}}-$tensor power of the input signal.\footnote{We will rewrite (\ref{eq:univariate Volterra series; time domain})
using the tensor product in section 1.2.1.}

Second, although the output, $V_{j}[s](t)$, of each order $j$ is
indexed by a single (time) variable $t$, at any order $j$ the value
indexed by $t$ in fact corresponds to the diagonal slice, $[\underbrace{t,t,\dots,t}_{j}]$,
of the $j-$dimensional output of the $j-$dimensional convolution
integral. This indexing along the diagonal allows for the outputs
of the different orders to be summed as ($1-$dimensional) signals,
resulting in a $1-$dimensional output from the entire system. The
classical, univariate VS can thus be seen as an extension of the LTI
convolution integral to multiple dimensions, since, at each order
$j$, the $j^{\text{th}}-$order homogeneous Volterra operator (VO)
convolves a product of $j-$many copies of the input signal by a
$j-$dimensional VKF. In particular, at order $1$, the homogeneous
Volterra operator $V_{1}$ is defined by a one-dimensional convolutional
integral, which characterizes linear, time-invariant (LTI) systems
\[
V_{1}(s)(t)=\int_{\tau\in\mathbb{R}}v(\tau)s(t-\tau)d\tau.
\]
At orders $j$ greater than one, however, nonlinearity enters via
the products $\prod_{i}^{j}s(t-\tau_{i})$ of the input signal with
itself.

\subsection{Frequency domain representation}

The convolution theorem in Fourier analysis states that, for any Fourier
pair (--say, time and frequency) convolution in one domain is equivalent
to pointwise multiplication in the other. As the Volterra series is
a series sum of convolution-type operators, it also admits (assuming
the Fourier integral converges) a frequency-domain representation,
dual to (\ref{eq:univariate Volterra series; time domain}). This
means that integral/analytic expressions can be transformed into algebraic
ones, and vice-versa. In studying nonlinear systems, where higher-order
effects emerge from complex interactions between components, working
in the frequency domain can help to ground understanding, and is,
moreover, practically useful, since the Fourier transform converts,
at each order $j$, the operation of convolution of $j-$dimensional
signals into a pointwise multiplication of their spectra.

For a Volterra series to be representable in the frequency domain,
it must have VKFs with convergent Fourier transforms\footnote{This requirement is the primary reason why we work with signals and
spectra lying in the space of tempered distributions, which includes
the Schwartz space; see Appendix A.}. Recall that the $n-$dimensional Fourier transform $F(s)=\hat{s}$
can be written (for an element $\text{s \ensuremath{\in}}S(\mathbb{R}^{n})$
of the Schwartz space of functions with all derivatives rapidly decreasing
at infinity) as
\begin{equation}
F(s)(\boldsymbol{\Omega})=\hat{s}(\boldsymbol{\Omega})=\int_{\boldsymbol{t}\in\mathbb{R}^{n}}e^{-i\boldsymbol{\Omega}\cdot\boldsymbol{t}}s(\boldsymbol{t})d\boldsymbol{t}\label{eq:multidimensional Fourier integral}
\end{equation}
where $[\omega_{1},\omega_{2},\dots,\omega_{n}]=\boldsymbol{\Omega}_{n}\in\mathbb{R}^{n}$
is a vector of frequency variables, $[t_{1},t_{2},\dots,t_{n}]=\boldsymbol{t}\in\mathbb{R}^{n}$
is a vector of time variables, and the dot product $\boldsymbol{\Omega}\cdot\boldsymbol{t}$
is defined $\boldsymbol{\Omega}\cdot\boldsymbol{t}=\sum_{r=1}^{n}\omega_{r}t_{r}$.
Then at each order $j$, the time-domain output $y_{j}(t)$ is 
\[
y_{j}(t)=\int_{\boldsymbol{\Omega}_{j}\in\mathbb{R}^{j}}e^{i\Sigma\boldsymbol{\Omega}_{j}t}\cdot\hat{v}_{j}(\boldsymbol{\Omega}_{j})\prod_{q=1}^{j}\hat{s}(\omega_{q})d\omega_{q},
\]
where $\sum\boldsymbol{\Omega}_{j}=(\omega_{1}+\dots+\omega_{j})$
and the function $\hat{v}_{j}=F(v_{j})$ is called the \emph{$j^{\text{th}}-$}order
\emph{Volterra frequency response function }(VFRF).

By the projection-slice theorem in Fourier analysis\footnote{The projection-slice theorem and closely-related Radon transform are central to the mathematics of Volterra series, and while we do not treat them explicitly with any real depth in this work, their implications run through its entire course.}, the \emph{$j^{\text{th}}-$order output spectrum}, $\hat{y}_{j}$,
is then given by integrating along the subspace perpindicular to the diagonal in $j-$dimensions, i.e., along the hyperplane of
vectors $\boldsymbol{\Omega_j}$ such that $\Sigma\boldsymbol{\Omega}_{j}=\omega$ --
\[
\hat{y}_{j}(\omega)=\int_{\boldsymbol{\Omega}_{j}\in\mathbb{R}^{j}\,|\,\Sigma\boldsymbol{\Omega}_{j}=\omega}\hat{v}_{j}(\boldsymbol{\Omega}_{j})\prod_{q=1}^{j}\hat{s}(\omega_{q})d\omega_{q}
\]
\[
=\int_{t \in \mathbb{R}}e^{-i \omega t}y_j(t)dt.
\]
The frequency-domain output $\hat{y}_{0}(\omega)$ of the zeroth-order
Volterra operator, $V_{0}$ is written
\[
\hat{y}_{0}(\omega)=\frac{1}{2\pi}\hat{v}_{0}\delta(\omega),
\]
where $\delta$ is the delta distribution.

\subsection{Kernel symmetry}

We will assume, throughout this work, that the VKFs of a Volterra series are symmetric.
Given a non-symmetric $j-$th order VKF, a unique symmetrization can be generated, whose
Fourier transform, i.e. whose corresponding VFRF, $\hat{v}_{j}^{\text{sym}}$, is given at the input frequency
vector $\Omega_{j}$ by the normalized sum of the values of $\hat{v}_j$ at all possible
permutations of $\Omega_{j}$, 
\begin{equation}
\hat{v}_{j}^{\text{sym}}(\Omega_{j})=\frac{1}{j!}\sum_{\text{\ensuremath{\sigma\in S_{j}}}}\hat{v}_{j}(\text{\ensuremath{\sigma}(}\Omega_{j})),\label{eq:symmetric VFRF (without multiplicities)}
\end{equation}
where $\sigma$ is an element of the symmetric group, $S_{j}$, on
$j$ symbols. Although equation (\ref{eq:symmetric VFRF (without multiplicities)})
defines a symmetric function, it does not take into account the possibility
that the vector $\Omega_{j}$ contains frequency variables that have
multiplicities greater than one; a better formula is 
\begin{equation}
\hat{v}_{j}^{\text{sym}}(\Omega_{j})=\frac{1}{n^{*}(\Omega_{j})}\sum_{\sigma\in S_{j}}\hat{v}_{j}(\text{\ensuremath{\sigma(}}\Omega_{j})),\label{eq:symmetric VFRF (with multiplicities)}
\end{equation}
where $n^{*}(\Omega_{j})$ is the multinomial coefficient indexed
by the multicombination (or combination taken with repetition) with
elements $\omega_{1},\dots,\omega_{j}$; i.e., 
\[
n^{*}(\Omega_{j})=\binom{j}{n_{\iota}(\omega_{1}),\dots,n_{\iota}(\omega_{j})}=\frac{j!}{n_{\iota}(\omega_{1})!\cdot\dots\cdot n_{\iota}(\omega_{j})!},
\]
where $n_{\iota}(\omega_{i})$ counts the number of occurences of
each $\omega_{i}$ in the frequency vector $\Omega_{j}$, and $\iota:j\rightarrow\Omega_{j}$
is a function from $j$ to the underlying set of $\Omega_{j}$, which
chooses $j$ many frequency variables, possibly with repetition. Thus,
$n_{\iota}(\omega_{i})$ is given by the cardinality of the fiber,
or preimage, of $\iota$ over the element $\omega_{i}$,
\[
n_{\iota}(\omega_{i})=|\iota^{-1}(\omega_{i})|.
\]
When all of the frequency variables are different, then the multinomial
factor is equal to $j!$, as above; and in polar contrast, when all
of the frequency variables are the same, then it is equal to one.

We will see that the multinomial structure of the Volterra kernel
functions, expressed in (\ref{eq:symmetric VFRF (with multiplicities)}),
reflects the combinatorics of the Volterra series itself. But to understand
this, it must first be understood how, in the general, multivariate
case, the different inputs to a Volterra series are combined using
the tensor product to form the inputs to the various homogeneous operators
of the series. Before we come to this, though, we briefly review the
notion of parametric Volterra series.

\subsection{Parametricity}

As classically defined, the Volterra series is a nonlinear,\emph{
time-invariant operator}, since it is a series sum of convolutions
and so commutes with translations (which we show in Chapter 2). It
has, however, been extended to model nonlinear time\emph{-varying}
phenomena, as well; this was done, e.g., in \cite{nam_volterra_2003},
to model the time-frequency shift covariant distributions of Cohen's
class, by defining the double Volterra series with parameterized kernel
functions, as in
\[
V(s)(t,\theta)=y(t,\theta)=\int_{\boldsymbol{\tau}_{2}\in\mathbb{R}^{2}}v_{2,\theta}(\tau_{1},\tau_{2})\cdot x_{\tau_{1}}(t-\tau_{1})x_{\tau_{2}}(t-\tau_{2})d\tau_{1}\,d\tau_{2},
\]
where $\theta$ is a scalar (frequency) variable upon which the kernel
depends\footnote{For example, the Volterra series representation of the \emph{Wigner-Ville
Distribution}, a core object in time-frequency analysis, has the parameterized
kernel function $v_{2,\theta}(\tau_{1},\tau_{2})=\delta(\tau_{1}+\tau_{2})e^{-2\pi i\theta(\tau_{1}+\tau_{2})}$.
We cover this in detail in Chapter 5.}. In Chapter 2, we relate parameterized Volterra series to morphisms
of Volterra series, and in Chapter 5 we again encounter parameterized
Volterra series within the context of time-frequency analysis.

\section{Multilinear maps and the tensor product}

In this section, we review the tensor product operation. It describes
the interactions of an input signal with itself, or of many input
signals with one another, and is used to construct the spaces in which
the inputs to the homogeneous Volterra operators lie. See (\cite{geroch_mathematical_physics},
Chp. 14) for a category-theoretic and physics-based introduction to
the tensor product.

Recall that the tensor product of a vector space $V$ and a vector
space $W$ over the same field $K$ is the vector space $V\otimes W$
defined by the universal property that is expressed in the following
commutative diagram,\begin{center}
\[\begin{tikzcd} {V \times W} && {V\otimes W} \\ \\ && Z \arrow["b", from=1-1, to=1-3] \arrow["{\tilde h}"', from=1-1, to=3-3] \arrow["\exists ! h", dashed, from=1-3, to=3-3] \end{tikzcd}\]\label{Universal property of the tensor product}
Fig. 1.2 Universal property of the tensor product.
\par\end{center}where $b:V\times W\rightarrow V\otimes W$ is bilinear. The diagram
says that $V\otimes W$ is the vector space (unique up to natural
isomorphism) characterized by the property that, given any other vector
space, $Z$, and any bilinear map $\tilde{h}:V\times W\rightarrow Z$,
there is a \emph{unique }linear map $h$ such that the diagram commutes,
i.e. $\tilde{h}=b\circ h$. Given a vector space $V$, the \emph{$k-$}th
iterated \emph{tensor power }of $V$ is the space $T^{k}(V)=\overbrace{V\otimes\dots\otimes V}^{k}$.
Note that if $V$ is of dimension $n$, then the dimension of $T^{k}(V)$
is $n^{k}.$ In our context, we will take $V=W=S(\mathbb{R})$ to
be the signal space.

The map $b$ is often written $-\otimes-:V\times W\rightarrow V\otimes W$,
using the same symbol to refer to the operation on the elements of
the vector spaces as is used to denote the operation on the vector
spaces themselves. Given a $k-$tensor $f\in T^{k}(V)$, and an $l-$tensor,
$g\in T^{l}(V)$, their tensor product $(f\otimes g)\in T^{k+l}(V)$
is defined as
\[
(f\otimes g)(x_{1},\dots,x_{k+l})=f(x_{1},\dots,x_{k})g(x_{k+1},\dots,x_{k+l}),
\]
where $x_{1},\dots,x_{k+l}\in V$. An element $(f\otimes g)\in T^{k+l}(V)$
is often called an \emph{elementary tensor}, or \emph{decomposable
tensor}\footnote{Not every tensor is elementary, but the elementary tensors span the
tensor product space $T^{k}(V)\otimes T^{l}(V)$. In particular, if
$K$ is a basis for $V$, and $L$ is a basis for $W$, then $V\otimes W$
is the free vector space on the Cartesian product $K\times L$.}.

The universal property expressed in the diagram of Fig. \ref{Universal property of the tensor product}
naturally extends to \emph{multilinear maps} $m:S(\mathbb{R})^{j}\rightarrow T^{j}(S(\mathbb{R}))$.
Recall that a map $f:S(\mathbb{R})^{j}\rightarrow S(\mathbb{R}^{j})$
from the $j-$fold Cartesian product of $S(\mathbb{R})$ with itself,
to $S(\mathbb{R}^{j})$, the space of $j-$dimensional signals, is
multilinear iff it is separately linear in each of its arguments,
i.e.,
\[
f(x_{1}+ax_{1}^{'},x_{2},\dots,x_{j})=f(x_{1},\dots,x_{n})+af(x_{1}^{'},\dots,x_{j})
\]
\[
f(x_{1},x_{2}+ax_{2}^{'},\dots,x_{j})=f(x_{1},x_{2},\dots,x_{j})+af(x_{1},x_{2}^{'},\dots,x_{j})
\]
\[
f(x_{1},x_{2},\dots,x_{n}+ax_{j}^{'})=f(x_{1},\dots,x_{j})+af(x_{1},\dots,x_{j}^{'})
\]
for any scalar $a\in\mathbb{R}$ and $x_{i},x_{i}'\in V$, for $1\le i\le j$.
The universal property of Fig. \ref{Universal property of the tensor product}
states that multilinear maps are completely characterized by the tensor
product.

Using the tensor product, we can rewrite the classical form of the
Volterra series, \ref{eq:univariate Volterra series; time domain},
as
\[
\sum_{j=0}^{\infty}\int_{\boldsymbol{\tau}_{j}\in\mathbb{R}^{j}}v_{j}(\boldsymbol{\tau}_{j})s^{\otimes j}(t\boldsymbol{1}_{j}-\boldsymbol{\tau}_{j})d\boldsymbol{\tau}_{j}
\]
where $s^{\otimes j}$ denotes the $j^{\text{th}}-$tensor power of
the signal $s$, and $t\boldsymbol{1}_{j}\in\mathbb{R}^{j}$ is the
vector of all $t$s. This shows that each homogeneous VO is a \emph{multilinear
operator}, on the set of input signals, that acts as the \emph{composition
}of a linear operation (convolution) applied to an input \emph{signal
tensor}. In the case of a univariate series, e.g., the input tensor
to the $j^{\text{th}}-$VO is an element of the subspace $T_{\text{sym}}^{j}(S(\mathbb{R}))\subset T^{j}(S(\mathbb{R}))$
of symmetric tensors that lies within the tensor power $T^{j}(S(\mathbb{R}))$
of the signal space $S(\mathbb{R}).$ The inputs are necessarily symmetric
in this case, because each is of the form $\prod_{i=1}^{j}u$, for
some input $u$ (i.e., where {$u_{1}=u_{2}=...=u_{j}=u$}).
In the case of multivariate series, the inputs lie more generally
in the space $T^{j}(S(\mathbb{R})$), as will be explained.

Each multilinear homogeneous VO thus factors as the composition of
an elementary nonlinear operation (forming the iterated tensor product
out of the Cartesian product) followed by a linear operation (a convolution
of the signal tensor). At order $j$, this is represented by the diagram\[\begin{tikzcd} {S(\mathbb R )^j} && {T^j (S(\mathbb R ))} \\ \\ {S(\mathbb R)} && {T^j (S(\mathbb R ))} \arrow["{\otimes j}", from=1-1, to=1-3] \arrow["{*_{v_j}}", from=1-3, to=3-3] \arrow["{\tilde{h}}"', from=1-1, to=3-3] \arrow["{(-)^j}", from=3-1, to=1-1] \arrow["{\text {diag} (-) }", from=3-3, to=3-1] \end{tikzcd}\]where
$(-)^{j}$ is the operation of forming the $j-$fold Cartesian product;
$\otimes j$ is the multilinear operation of forming, out of the $j^{\text{th}}-$fold
Cartesian product, the $j-$fold signal tensor; $*_{v_{j}}$ is the
linear operation of convolution by the Volterra kernel function of
order $j$; $\text{diag}(-)$ is the operation of indexing along the
diagonal; and $\tilde{h}$ is an arbitrary multilinear map.\\

\subsection{Tensor algebra, $T(S_{\mathbb{R}})$}

As a Volterra series consists of a (in general infinite) series of
homogeneous Volterra operators, and as the input to the linear part
of each homogeneous VO of order $j$ lies in a tensor power, $T^{j}(S(\mathbb{R}))$,
of the signal space $S(\mathbb{R})$, the action of the entire series
of operators factors through the direct sum of these tensor powers.
This space, $T(S(\mathbb{R})=\bigoplus_{j=1}^{\infty}T^{j}(S(\mathbb{R}))$,
is known as the \emph{tensor algebra} of the vector space $S(\mathbb{R})$;
its algebra multiplication is given by the tensor product. The tensor
algebra $T(V)$ of a vector space $V$ is referred to as a \emph{graded
space}, meaning that it decomposes as a direct sum of spaces of increasing
dimension.

Thus, a univariate VS is an operator $S(\mathbb{R})\rightarrow S(\mathbb{R})$
that can be viewed in more detail as a composition of operators, as
shown in the following diagram:

\[\begin{tikzcd} {S(\mathbb R )} && {T_{\text{sym}} (S(\mathbb R ))} \\ \\ && {T_{\text{sym}}(S(\mathbb R ))} \arrow["{T(-)}", from=1-1, to=1-3] \arrow["{*_{v}}", from=1-3, to=3-3] \arrow["{\sum_j \text{diag}(-)}", from=3-3, to=1-1] \end{tikzcd}\]where
$T(-)$ is the operation that, given an element $s\in S(\mathbb{R})$,
generates the infinite tower of tensors $[s,s\otimes s,s\otimes s\otimes s,\dots]$;
$*_{v}$ is the operation of \emph{level-wise }or \emph{graded }convolution
by the VKFs, $v_{j}$ for $1\le j\le\infty$; and $\sum_{j}\text{diag}(-)$
is the operation of summing the diagonal elements of the component
tensors in the tensor tower (i.e., the element of $T(S(\mathbb{R}))$)
that is obtained as the direct sum of the multidimensional input tensors
convolved by the VKFs of their corresponding orders.

In the multivariate case, for a Volterra series which processes $B-$many
input signals, a different version of the above diagram holds:

\[\begin{tikzcd} {S(\mathbb R )^B} && {\oplus_{j=1}^{\infty}\oplus^{B^{j}}T^{j}(S(\mathbb{R}))} \\ \\ && {\oplus_{j=1}^{\infty}\oplus^{B^{j}}T^{j}(S(\mathbb{R}))} \arrow["{T'(-)}", from=1-1, to=1-3] \arrow["{*_{v}}", from=1-3, to=3-3] \arrow["{\sum_j \sum^{B^j} \text{diag}(-)}", from=3-3, to=1-1] \end{tikzcd}\]wherein
the tensors are no longer symmetric and, at each order $j$, there
are in general $B^{j}-$many input tensors that are processed, as
we next explain.

\section{Multivariate Volterra series}

A multivariate VS is a model of a nonlinear system that processes
sets of signals; it is a mapping 
\[
V:S(\mathbb{R})^{B}\rightarrow S(\mathbb{R})^{A}
\]
from the cartesian product $S(\mathbb{R})^{B}=\prod_{b\in B}S(\mathbb{R})$
of $B-$many copies of the space of signals, $S(\mathbb{R})$, to
the product $S(\mathbb{R})^{A}=\prod_{a\in A}S(\mathbb{R})$, of $A-$many
copies of the same, where $B$ is the size of the set $U$ of input
signals ($|U|=B$) and $A$ is the size of the set $Y$ of output
signals ($|Y|=A$). If $B=1$, then the VS is referred to as single-input
multiple output (SIMO), and if $A=1,$ then it is referred to as a
multiple-input single-output (MISO). Unlike in the univariate case,
in the multivariate case the homogeneous outputs receive contributions
from many homogeneous operators, which process different combinations
of the inputs, taken with repetition.

Specifically, the $a^{\text{th}}-$output of the series is given by
the sum of the $a-$indexed outputs at each order

\[
V^{(a)}[U](t)=y^{(a)}(t)=\sum_{j}^{\infty}y_{j}^{(a)}(t)
\]
where the $a^{\text{th}}-$output at each order is
\[
y_{j}^{(a)}(t)=\sum_{\tilde{f}\in\nicefrac{U^{j}}{S_{j}}}\binom{j}{n_{f}(u_{1}),\dots,n_{f}(u_{B})}\int_{\boldsymbol{\tau}_{j}\in\mathbb{R}^{j}}v_{j,a}^{\text{sym},\tilde{f}}(\boldsymbol{\tau}_{j})\prod_{i=1}^{j}u_{f(i)}(t-\tau_{i})d\tau_{i},
\]
where $S_{j}$ denotes the symmetric group on $j$ symbols, $U^{j}$
is the set of functions from $j$ to the set, $U$, of input signals,
$\binom{j}{b_{1},\dots,b_{j}}$ denotes the multinomial coefficient
corresponding to the collection of indices $b_{i},1\le i\le j$ (i.e.,
such that $\sum_{i}b_{i}=j$), and $f$ is the canonical representative
of the equivalence class of functions, $U^{j}/S_{j}$, corresponding
to the identity permutation\footnote{Recall that the set of functions from a set $B$ to a set $A$ is
isomorphic to the set $A^{B}$; we refer interchangably to the the
set of functions and the exponential object.}.The symmetric kernel functions $v_{j,a}^{\text{sym},\tilde{f}}:\mathbb{R}^{j}\rightarrow\mathbb{\mathbb{R}}$
are defined by 
\begin{equation}
v_{j,a}^{\text{sym},\tilde{f}}(\boldsymbol{\tau}_{j})=n^{*}(\boldsymbol{\tau}_{j})\sum_{\sigma\in S_{j}}v_{j,a}^{f}(\tau_{\sigma(1)},\dots,\tau_{\sigma(j)}),\label{eq:symmetric kernel, multivariate series}
\end{equation}
and are indexed by functions $f:j\rightarrow\{\tau_{i}\}_{i=1}^{j}$
which are distinct up to a permutation of the delay variables\footnote{Recall that the symmetric group, $S_{j}$, on $j$ elements
is of order $j!$.}. The distinguishing feature of (\ref{eq:symmetric kernel, multivariate series}),
compared to univariate case (\ref{eq:symmetric VFRF (with multiplicities)}),
is that the kernel function is additionally indexed by an equivalence
class of functions, which target a subset of the input signals.

Each homogeneous Volterra operator in a multivariate VS thus processes
a certain set of multicombinations of signals from the input set $U$,
where each multicombination defines a different pathway through the
nonlinear system. It is furthermore specified by the particular output,
$(a)$, to which it contributes. In total, for each output there are
$B^{j}$ possible input pathways at order $j$ of nonlinearity, and
so $\sum_{j=1}^{\infty}B^{j}$-many pathways through the entire system.
Summing the multinomial coefficients associated to the different multicombinations
at a given order, we recover the full set of functions: 
\begin{equation}
\sum_{\tilde{f}\in\nicefrac{U^{j}}{S_{j}}}\binom{j}{n_{f}(u_{1}),\dots,n_{f}(u_{B})}=B^{j}.\label{eq:multinomial sum identity}
\end{equation}
In light of the combinatorics, a multivariate Volterra series is thus
a complete model of all the different ways that a nonlinear system
can respond to input signals received at its ports - a fact which
its usual definition as an operator, without regards to the tensor algebra,
does not really reflect.

Although many of the following results will be stated implicitly or
explicitly for univariate Volterra series, we will see multivariate
series return at two important points: when discussing the series
composition of Volterra series, in Chapter 4; and when considering
the Volterra series representation of time-frequency distributions,
in Chapter 5.

\section{Volterra series representations of some simple systems}

It is instructive to build an inventory of some elementary Volterra
series, both linear and nonlinear. In general, different systems may
be simpler to represent in one or the other of the time and frequency
domains. See \cite{carassale_modeling_2010} for further details.\\

The \emph{delay operator }$T_{\tau}(s)(t)=s(t-\tau)$ is linear and
time-invariant, and so its VS representation is simply given by the
$1^{\text{st}}-$order homogeneous operator 
\[
T_{\tau}(s)(t)=V_{1}(s)(t)
\]
\[
=\int_{\tau}\delta(\tau)s(t-\tau)d\tau
\]
\[
=\int_{\omega}e^{i\omega t}\cdot e^{-i\omega\tau}\hat{s}(\omega)d\omega
\]
\[
=\int_{\omega}e^{i\omega(t-\tau)}\cdot\hat{s}(\omega)d\omega
\]
with VFRF $\hat{v}_{1}(\omega)=e^{i\omega\tau}$.

Similarly, the \emph{integro-differential operator} $D^{r}(s)=\frac{d^{r}}{dt^{r}}s$
is given by the $1^{\text{st}}-$order homogeneous operator
\[
D^{r}(s)(t)=V_{1}(s)(t)
\]
\[
=\frac{d^{r}(\int e^{i\omega t}\hat{s}(\omega)d\omega)}{dt^{r}}
\]
\[
=\int(i\omega)^{r}e^{i\omega t}\hat{s}(\omega)d\omega
\]
with VFRF $\hat{v}_{1}(\omega)=(i\omega)^{r}$.

The simplest nonlinear operator is the \emph{memoryless polynomial},
$P[s]=\sum_{j=0}^{\infty}a_{j}s^{j}$. For a univariate Volterra series
$V$, and representing the input signal using the inverse Fourier
transform, $s(t)=\frac{1}{2\pi}\int e^{i\omega t}\hat{s}(\omega)d\omega$,
we have that 
\[
V[s](t)=\sum_{j=0}^{\infty}a_{j}(\frac{1}{2\pi}\int e^{i\omega t}\hat{s}(\omega)d\omega)^{j}
\]
\[
=\sum_{j=0}^{\infty}a_{j}(\frac{1}{2\pi}\int_{\Omega_{j}\in\mathbb{R}^{j}}e^{i\Sigma\Omega_{j}t}\prod_{q=1}^{j}\hat{s}(\omega_{q})d\Omega_{j})
\]
\[
=\sum_{j=0}^{\infty}(\frac{1}{2\pi}\int_{\Omega_{j}\in\mathbb{R}^{j}}a_{j}\cdot e^{i\Sigma\Omega_{j}t}\prod_{q=1}^{j}\hat{s}(\omega_{q})d\Omega_{j})
\]
which shows that the corresponding VFRF for the memoryless polynomial
operator is 
\[
\hat{v}_{j}(\Omega_{j})=a_{j},
\]
i.e., a constant which does not depend on the frequency.

\typeout{} 
\chapter{The Volterra series as a functor}

In this chapter, we begin the work of reframing the Volterra series
in category-theoretic terms. Our motive for thinking about Volterra
series categorically is to develop a language\emph{ }for speaking about nonlinear systems,
not as they exist in isolation, but as they are functionally related to one another.
Indeed, our eventual aim is to describe \emph{morphisms of Volterra series},
which, categorically-speaking, are natural transformations between them, and model ways that
nonlinear systems change.

To define morphisms of Volterra series categorically, we must treat
the Volterra series not as a type of operator acting on isolated elements
of a vector space, but rather as a type of \emph{functor}, which is
a structure-preserving maps between categories, one that preserves the compositional
structure of the morphisms\footnote{This kind of conceptual move, of going from viewing things as elements of a set
to objects of a category, is idiomatic to the process of \emph{categorification},
as described in \cite{baez_categorification_1998}.}. That is, not only will the Volterra series
by our definition process signals, but it will also transform the elementary linear maps between signals.
But to begin to see the Volterra series as a kind of functor,
we must first define the category on which it acts.

\section{Categories and functors}

Since category theory may not yet be familiar to a signal processing
context, we first review the definitions of its two most basic concepts:
category and functor. See \cite{riehl_category_2017,spivak_category_2014}
for an introduction to the subject, or \cite{geroch_mathematical_physics}
for a physics-inspired treatment.

\paragraph{Categories}

A \emph{category}, $C$, consists of

a collection, $C_{0}$, of \emph{objects},

for each pair $x,y\in C_{0}$, a collection $C_{1}(x,y)$ of \emph{morphisms
from x to y};

for each pair of morphisms $f\in C(x,y)$ and $g\in C(y,z)$, a \emph{composite
morphism }$g\circ f\in C(x,z)$;

for each object $x$, an \emph{identity morphism}, $\text{1}_{x}\in C(x,x)$;

such that:
\begin{itemize}
\item composition satisfies the left and right unit laws: given a morphism
$f\in C(x,y)$, it holds that $f\circ1_{x}=f$ and $1_{y}\circ f=f$;
and
\item composition is associative: for any composable triple of morphisms,
$f\in C(a,b)$, $g\in C(b,c)$, and $h\in C(c,d)$, the two possible
composites are equal, $f\circ(g\circ h)=(f\circ g)\circ h$.
\end{itemize}
One also writes $f:x\rightarrow y$ to denote a morphism $f$ in $C_{1}(x,y$)
-- and refers to $x$ as the\emph{ domain}, or \emph{source} of $f$,
and to $y$ as its \emph{codomain}, or \emph{target}. One may also
write $C(x,y)$ or $\text{Hom}_{C}(x,y$) for $C_{1}(x,y)$, where
the latter option is pronounced `the hom-set' between $x$ and $y$,
where `hom' stands for homomorphism.

An equivalent, graph-theoretic definition of a category is that it
is the \emph{reflexive, transitive closure }of its underlying graph\footnote{The underlying graph of a category is the graph having vertices the
objects in the category, and directed edges the non-composite morphisms.}. That is, given any graph, one can form the unique category corresponding
to it by formally adjoining: (1) a self-loop at every vertex; (2)
for any path (sequence of composable arrows) in the graph, a composite
arrow; such that (3) the composition of arrows is associative.

\paragraph{Example: the category Set\protect \\
}

The category, Set, of sets and functions has, as objects, sets, and
as morphisms, functions between sets. It is a category because, for
any set $x$, there is an identity function $1_{x}$; for any pair
of functions $g:b\rightarrow c$ and $f:a\rightarrow b$ there is
a composite $(f\circ g)(-)=g(f(-))$; and the composition of functions
is associative.\footnote{The category Set is important to the story of Volterra series
because it plays, in relation to polynomial functors--which are functors Set $\rightarrow$ Set that are,
in a sense, to sets what Volterra series are to signals--the role played in relation to Volterra series
by the category $S'(\mathbb{R})$, which we present in sect. 2.2.}\\

Just as there are morphisms between objects in a category, there are
maps between categories. But whereas the objects of a category may
be treated as elemental, or having no internal structure, categories
are imbued with the structure of their morphisms. It is natural to
define a notion of a map between categories that is \emph{structure-preserving};
such a map is called a functor.

\paragraph{Functors}

\begin{figure}[h]
\centering{}\includegraphics[scale=0.33]{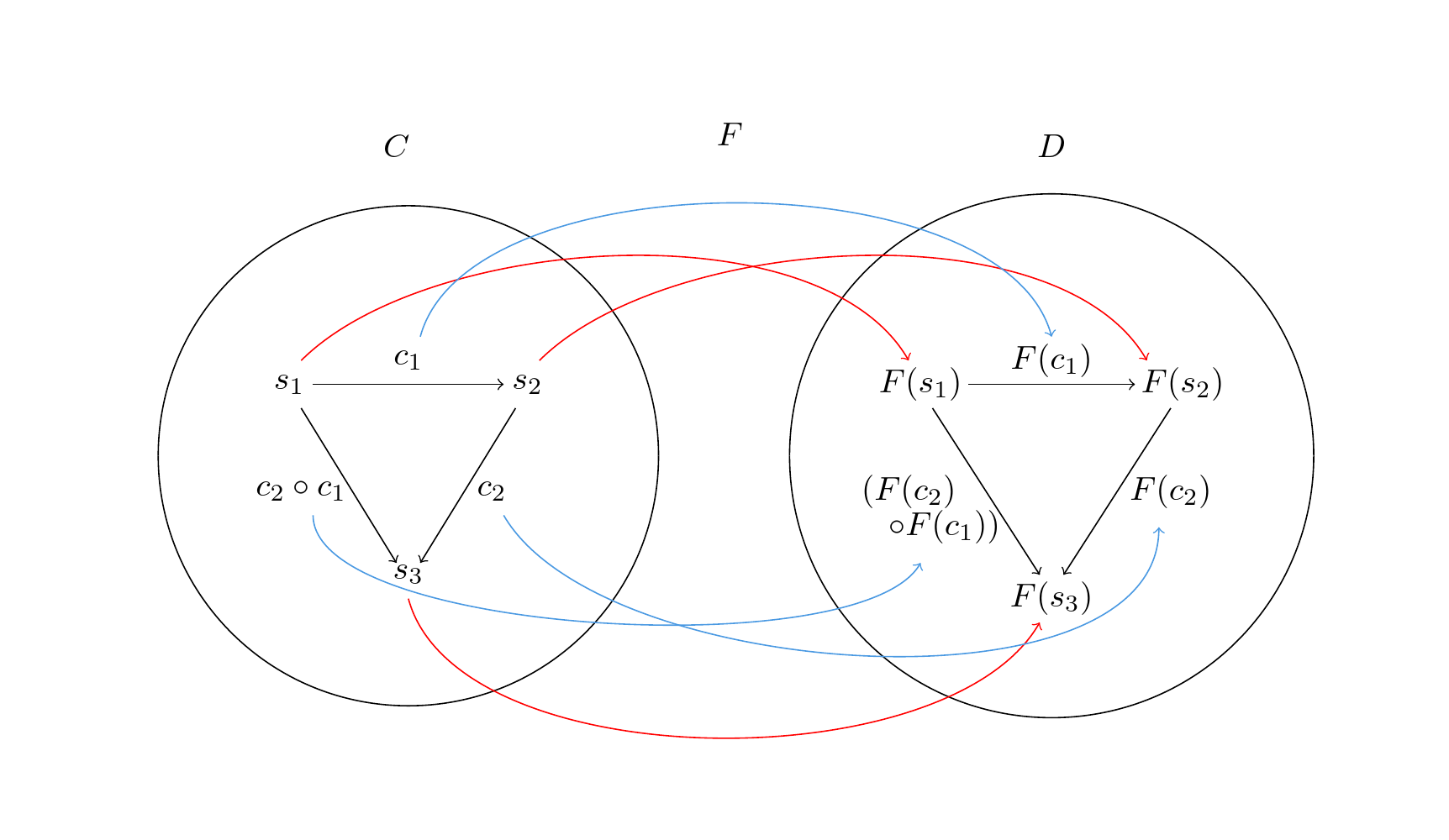}\label{Action of a functor on a commuting diagram}\caption{The action of a functor on a commuting diagram.}
\end{figure}

A\emph{ functor} $F:C\rightarrow D$ from a category $C$ to a category
$D$ is a map sending each object $x\in C$ to an object $F(x)\in D$
and each morphism $f:x\rightarrow y$ in $C$ to a morphism $F(f):F(x)\rightarrow F(y)$
in $D$, such that
\begin{itemize}
\item $F$ preserves identity morphisms: for each object $x\in C$, $F(1_{x})=1_{F(x)}$,
\item $F$ preserves composition: for any composite $g\circ f$ in $C$,
$F(g\circ f)=F(g)\circ F(f)$ in $D.$
\end{itemize}
These conditions are equivalent to requiring that a functor preserve
commuting diagrams. That is, a functor $F:C\rightarrow D$ is a mapping
such that, for all commutative diagrams in $C$ of the form shown
on the left in Fig. \ref{Action of a functor on a commuting diagram},
we have a commutative diagram in $D$ as shown on the right.

\paragraph{Example: Representable functors\protect \\
}

The most fundamental type of functor is a \emph{representable functor}.
A functor $F:C\rightarrow\text{Set}$ from a category $C$ to $\text{Set}$
is \emph{representable} if it is naturally isomorphic to the (co-
or contravariant) hom-functor, $C(c,-):C\rightarrow\text{Set}$, for
some object $c$ (in which case $c$ is called the \emph{representing
object}). The hom-functor based at an object $c$ of the category
$C$ is defined as follows: for any object $c'\in C$, it returns
the set $C(c,c')$, of morphisms from $c$ to $c'$; and for any morphism
$g:x\rightarrow y$, it returns the function $C(c,g):C(c,x)\rightarrow C(c,y)$
between the sets $C(c,x)$ and $C(c,y)$, given by sending each element
$f\in C(c,x)$ to the composite, $g\circ f$. Varying the object $c$
provides a host of perspectives on the structure of the ambient category
$C$, each from the vantage point of the representing object.

\section{The category $S'(\mathbb{R})$}

We next introduce the `base category', $S'(\mathbb{R})$, over which
the Volterra series will be defined as a type of endofunctor\footnote{An endofunctor is a functor whose domain and codomain are the same
category.}. We define $S'(\mathbb{R})$, also denoted $S'_{\mathbb{R}}$, as the category
having:
\begin{itemize}
\item as objects, tempered distributions (including the Schwartz functions;
see Appendix A for a review);
\item as morphisms, convolutors between them (see Appendix A).
\end{itemize}
The objects of $S'(\mathbb{R})$ are signals that are
`well-behaved', in the sense of having convergent Fourier transforms.
The morphisms are similarly well-behaved; the convolutors
are precisely those maps that take Schwartz functions to Schwartz functions
under convolution.\footnote{In fact, a more precise approach would be to define a pair of categories, which could be denoted
by $S'_c(\mathbb{R})$ and $S'_m(\hat{\mathbb{R}})$, where as morphisms in the first we include only convolutors between
signals, and in the second only multipliers between spectra. The Fourier transform and
its inverse then act bijectively on these spaces, taking each to the other, and in a sense computing their opposites as categories. In fact, one could then
dualize again, considering convolutions of spectra and modulations of time-domain signals, obtaining four categories in total.}

\paragraph{Comparison of $S'(\mathbb{R})$ and Set\protect \\
}

The following comparison may be useful to draw between $S'(\mathbb{R})$
and the category Set. Whereas a set has elements, of which it is the
(disjoint) union, a signal $s\in S'(\mathbb{R})$
is a linear combination of its components
\begin{equation}
s=\sum_{r}a_{r}\boldsymbol{c}_{r},\label{eq:signal decomposition onto a frame}
\end{equation}
where the vectors $\{\boldsymbol{c}_{r}\}_{r\in R}$ are elements
of a frame\footnote{A frame for a vector space $A$ is a collection of vectors that spans
$A$. The vectors of a frame need not, in contrast to a basis, be
linearly independent.} for $S'(\mathbb{R})$ (and if $\boldsymbol{c}_{r}(t)=e^{2\pi irt}$,
then of the Fourier basis, in which case \ref{eq:signal decomposition onto a frame}
is called a Fourier series). Whereas a function between sets maps
elements to elements, a morphism $f:s\rightarrow s'$ of signals maps components of $s$ to
components of $s'$. Canonically, we consider the set of components of a signal to be its
Fourier or frequency components. Then, a multiplier maps between signals by weighting each frequency:

\[
m(\hat s)(\omega)=\hat s(\omega)\gamma(\omega),
\]
where $\gamma$ is the weight function.

Now, consider the collection\footnote{Beyond forming a mere set or collection, the morphisms
between two objects can be equipped with additional structure.
For example, in a \emph{linear category}, each hom-set has the structure
of a vector space, meaning that morphisms can be added and scaled
to obtain new morphisms that are also in the set. In such cases, we
speak of \emph{hom-objects} instead of hom-sets, and are in the domain
of\emph{ enriched category theory}.} $S'_{\mathbb{R}}(s_{b},s_{c})$ of morphisms between two objects,
$s_{b},s_{c}\in\text{Ob}(S'_{\mathbb{R}})$. Just as any function
between sets is either injective, surjective, or bijective (i.e.,
both), a morphism $f\in S'_{\mathbb{R}}(s_{b},s_{c})$ is either:
\begin{itemize}
\item a monomorphism: for any other object $s_{a}$ and pair of parallel
arrows $g_{1},g_{2}:s_{a}\rightarrow s_{b},$ $f\circ g_{1}=f\circ g_{2}\implies g_{1}=g_{2}$
\item an epimorphism: for any other object $s_{d}$ and pair of parallel
arrows $g_{1},g_{2}:s_{c}\rightarrow s_{d},$ $g_{1}\circ f=g_{2}\circ f\implies g_{1}=g_{2}$
\item or both (an isomorphism): $\exists f^{-1}:s_{c}\rightarrow s_{b}|f^{-1}\circ f=\text{id}_{s_{b}}$
and $f\circ f^{-1}=\text{id}_{s_{c}}.$
\end{itemize}
Restricting our attention to the epimorphisms, for example, we see
that if $\text{supp}(\hat{s'})\subset\text{supp}(\hat{s})$ (where
$\text{supp}(x)$ denotes the support of $x$) then there are no
convolutions in the time-domain taking $s'$ fully to $s$; however,
there are \emph{infinitely many }such morphisms from $s$ to
$s'$ (since any number may multiply the amplitude of any zero-amplitude
frequency component without affecting the result). That said, there
is a unique\footnote{I.e., unique up to an integer winding of the phase factor.}
\emph{minimal }spectral multiplier $m:\hat{s}\rightarrow\hat{s'}$,
defined by 
\[
m(\hat{s})[f]=\begin{cases}
\frac{\hat{s'}(f)}{\hat{s}(f)} & \hat{s'}(f),\hat{s}(f)\ne0\\
0 & \hat{s'}(f)=0
\end{cases};
\]
and if the situation is reversed--i.e. if $\text{supp}(\hat{s})\subset\text{supp}(\hat{s'})$--then
there is a unique such minimal modulation in the time-domain (equiv.
convolution in the spectral domain).

\section{The Volterra series as a functor}

We next define the Volterra series as a kind of endofunctor,
$V:S'(\mathbb{R})\rightarrow S'(\mathbb{R})$. To do this, we must define its action on morphisms and show that
it preserves identities and composites, equivalently commuting diagrams in $S'(\mathbb{R})$.
The action of a Volterra series $V$ on a morphism $l:s \rightarrow s'$ is depicted

\[\begin{tikzcd}[ampersand replacement=\&] 	s \& {s'} \\ 	{V(s)} \& {V(s')} 	\arrow["l", from=1-1, to=1-2] 	\arrow[maps to, "V", from=1-2, to=2-2] 	\arrow[maps to, "V"', from=1-1, to=2-1] 	\arrow["{V(l)}"', from=2-1, to=2-2] \end{tikzcd}\]
Thus, $V$ must send
a morphism $l:s\rightarrow s'$ in $S'(\mathbb{R})$ to a linear transformation
$V(l):V(s)\rightarrow V(s')$, also in $S'(\mathbb{R})$, that maps components of $V(s)$
to components of $V(s')$. It must furthermore do this such that $V(\text{id}_s) = \text{id}_{V(s)}$ and $V(m \circ l) = V(m) \circ V(l)$ for each object $s$ and pair of composeable morphisms $l,m$.

Recall that, for a homogeneous Volterra operator $V_j$, the spectrum $V_j(s)(\omega)$ at a frequency $\omega$ is given by an integral over all frequency vectors $\boldsymbol {\Omega}_j \in \mathbb R^j$ whose components sum to $\omega$, $\Sigma \boldsymbol{\Omega}_j = \omega$:

\[
V_j(\hat s)(\omega)=\int_{\boldsymbol{\Omega}_{j}\in\mathbb{R}^{j}\,|\,\Sigma\boldsymbol{\Omega}_{j}=\omega} \hat{\,v_{j}}(\boldsymbol{\Omega}_{j})\,\stackrel[q=1]{j}{\prod}\hat{s}(\omega_{q})d\boldsymbol{\Omega}_{j}
\]
\[
=\int_{\boldsymbol{\Omega}_{j}\in\mathbb{R}^{j}\,|\,\Sigma\boldsymbol{\Omega}_{j}=\omega} (\hat{\,v_{j}} \odot \hat{s}^{\otimes i})(\boldsymbol{\Omega}_j)d\boldsymbol{\Omega}_{j}
\]
We thus define the action of a Volterra series on a convolutor $l: s \rightarrow s'$, corresponding to a multiplier $m$ with weight function $\gamma$, in the spectral domain as

\[
V(m)(\hat s)(\omega)=\stackrel[j=0]{\infty}{\sum}\,\int_{\boldsymbol{\Omega}_{j}\in\mathbb{R}^{j}\,|\,\Sigma\boldsymbol{\Omega}_{j}=\omega} \hat{\,v_{j}}(\boldsymbol{\Omega}_{j})\,\stackrel[q=1]{j}{\prod} m(\hat s)(\omega_{q})d\boldsymbol{\Omega}_{j}
\]

\[
=\stackrel[j=0]{\infty}{\sum}\,\int_{\boldsymbol{\Omega}_{j}\in\mathbb{R}^{j}\,|\,\Sigma\boldsymbol{\Omega}_{j}=\omega} \hat{\,v_{j}}(\boldsymbol{\Omega}_{j})\,\stackrel[q=1]{j}{\prod}\gamma (\omega_{q})\hat{s}(\omega_{q})d\boldsymbol{\Omega}_{j}
\]

\[
=\stackrel[j=0]{\infty}{\sum}\,\int_{\boldsymbol{\Omega}_{j}\in\mathbb{R}^{j}\,|\,\Sigma\boldsymbol{\Omega}_{j}=\omega} (\hat{\,v_{j}} \odot \hat{s}^{\otimes i} \odot \gamma ^{\otimes i})(\boldsymbol{\Omega}_j)d\boldsymbol{\Omega}_{j},
\]
or equivalently in the time domain by

\[
V(l)(s)(t)=\stackrel[j=0]{\infty}{\sum}\int_{\boldsymbol{\tau}_{j}\in\mathbb{R}^{j}}v_{j}(\boldsymbol{\tau}_{j})\stackrel[r=1]{j}{\prod}l(s)(t-\tau_{r})d\boldsymbol{\tau}_{j}
\]
\[
=\stackrel[j=0]{\infty}{\sum}\int_{\boldsymbol{\tau}_{j}\in\mathbb{R}^{j}}v_{j}(\boldsymbol{\tau}_{j})\stackrel[r=1]{j}{\prod}\int_{\tau'\in\mathbb{R}}l(\tau')s(t-\tau_{r}-\tau')d\tau'd\boldsymbol{\tau}_{j}.
\]
In other words, we simply `post-compose' by $l$, or apply it to each instance of $s$ occuring in the Volterra integral expansion. This weights each element, $\hat{s}^{\otimes i}(\boldsymbol{\Omega}_j)$, of the tensor power of $s$ by an element, $\gamma ^{\otimes j} (\boldsymbol {\Omega}_j)$, of the tensor power of the weight function $\gamma$ at the corresponding frequency vector, $\boldsymbol {\Omega}_j \in \mathbb{R}^j$.


We can see this definition of the action of $V$ on morphisms of signals in analogy with that of representable Set-valued endofunctors on functions between sets, as follows. Given two sets, $A,B\in \text{Set}$, and a function $f:A\rightarrow B$, the functor $y^X$ represented by the object (set) $X$ acts on $f$ by sending each function $g\in A^X$ to a function in $B^X$, via postcomposition with $f$
\[
f^X : A^X \rightarrow B^X
\]
\[
f^X(g) = f\circ g.
\]
Now, just as an element of the exponential object $A^X$ is a function from $X$ to $A$, equivalently an $|X|-$tuple of elements in $A$, so too does a frequency component of $V_i(s)$ correspond to a (weighted) element of the $i-$fold tensor product of $\hat s$, i.e., a product of $i-$many frequency components in $s$, and so to a function $i \rightarrow \hat s$. And just as $f^X(g)$ is the function in $B^X$ obtained by mapping each element $g(x)\in A$, $x\in X$, to an element $f(g(x) \in B$, so too is the spectrum $V_i(f)(\hat s)$ obtained by mapping each frequency component
\[
V_i(\hat s)(\omega)=\int_{\boldsymbol{\Omega}_{i}\in\mathbb{R}^{i}\,|\,\Sigma\boldsymbol{\Omega}_{i}=\omega} \hat{\,v_{i}}(\boldsymbol{\Omega}_{i})\,\stackrel[q=1]{i}{\prod}\hat{s}(\omega_{q})d\boldsymbol{\Omega}_{i}
\]
in $V_i(\hat s)$ to a frequency component
\[V(l)(\hat s)(\omega)=\stackrel[i=0]{\infty}{\sum}\,\int_{\boldsymbol{\Omega}_{i}\in\mathbb{R}^{i}\,|\,\Sigma\boldsymbol{\Omega}_{i}=\omega} \hat{\,v_{i}}(\boldsymbol{\Omega}_{i})\,\stackrel[q=1]{i}{\prod}l(\hat s)(\omega_{q})d\boldsymbol{\Omega}_{i}
\]
in $V(\hat{s}')$.

We can see by the above definition that identities are preserved, namely: $V(\delta)(s) = V(\delta (s)) = V(s)$, since $\hat \delta (\omega) = \boldsymbol 1$. Likewise, composites are preserved, since the weight function of a composition of two multipliers is simply the pointwise product of their respective weight functions (or the product of two diagonal matrices):
\[
V(g \circ f) =\stackrel[j=0]{\infty}{\sum}\,\int_{\boldsymbol{\Omega}_{j}\in\mathbb{R}^{j}\,|\,\Sigma\boldsymbol{\Omega}_{j}=\omega} (\hat{\,v_{j}} \odot \hat{s}^{\otimes i} \odot  \widehat {g\circ f}^{\otimes i})(\boldsymbol{\Omega}_j)d\boldsymbol{\Omega}_{j},
\],
\[
=\stackrel[j=0]{\infty}{\sum}\,\int_{\boldsymbol{\Omega}_{j}\in\mathbb{R}^{j}\,|\,\Sigma\boldsymbol{\Omega}_{j}=\omega} (\hat{\,v_{j}} \odot \hat{s}^{\otimes i} \odot  \hat{g}^{\otimes i} \odot \hat{f}^{\otimes i})(\boldsymbol{\Omega}_j)d\boldsymbol{\Omega}_{j},
\]
\[
=V(g)V(f)(s).
\]
Thus, as we have defined it, the Volterra series is functorial.  

However, there is a potential obstruction to the definition given above, which has to do with whether the image $V(l)(s)$ lies in the codomain $V(s')$. Recall that the support of a signal is weakly-contracting in the spectral domain under convolution, i.e., its image has a spectrum whose support is smaller than or equal to that of its argument. The image of such a filtered signal under the action of a Volterra series will tend to be harmonically simpler than that of the original signal, since fewer or the same amount, but never more frequencies are being intermodulated and combined, but since the intermodulated products of freqencies are being summed, it is possible for cancellations to occur due to destructive interference. Similarly, $V(l)$ may cause frequencies that were absent in $V(s)$ to reappear, by shifting the relative phases of their intermodulation components. 
  
We elide this subtlety and its implications throughout the rest of the paper. If no cancellations occur, then the spectrum of the kernel of $V(l)$ - i.e., by abuse of notation, of the distribution by which $V(s)$ is linearly convolved - can be written

\[
V(\hat l)(\omega)=\frac {V(\widehat{l(s)})(\omega)} {V(\hat s)(\omega)}.
\]

\section{Examples of the action of a VS on a linear transformation}

It is therefore useful to catalogue some of the ways that the Volterra series acts on elementary
linear transformations, from which actions more complex transformations
can then be assembled. In this section we survey four: translation, modulation, periodization, and sampling \footnote{Of course, modulation and sampling
are not time-invariant operations. We consider in this section also a base category in which there exist multiplication-type linear maps as morphisms between signals or spectra, per the footnote
in section 2.2}.

\subsection{Translation (time shift)}

Let $c:s\rightarrow s'$ be the translation-by-$l$ map, given in
the time-domain as $c(s)(t)=(s*\delta_{l})(t)=\int_{\tau}s(\tau)\delta(t-l-\tau)d\tau$.
Then $c$ commutes with the action of a Volterra series; i.e., $V(c):V(s)\rightarrow V(s')$
is defined by
\[
V(c)V(s)(t)=V(c(s))(t)
\]
\[
=\sum_{j=0}^{\infty}\int_{\omega\in\mathbb{R}}e^{i\omega t}\int_{\boldsymbol{\Omega}_{j}\in\mathbb{R}^{j}|\Sigma\boldsymbol{\Omega}_{j}=\omega}\hat{v}_{j}(\boldsymbol{\Omega}_{j})\prod_{r=1}^{j}e^{-i\omega_{r}\tau}\hat{s}(\omega_{r})d\boldsymbol{\Omega}_{j}d\omega
\]
\[
=\sum_{j=0}^{\infty}\int_{\omega\in\mathbb{R}}e^{i\omega t}\int_{\boldsymbol{\Omega}_{j}\in\mathbb{R}^{j}|\Sigma\boldsymbol{\Omega}_{j}=\omega}e^{-i(\omega_{1}+\dots+\omega_{j})\tau}\hat{v}_{j}(\boldsymbol{\Omega}_{j})\hat{s}^{\otimes j}(\boldsymbol{\Omega}_{j})d\boldsymbol{\Omega}_{j}d\omega
\]
\[
=\sum_{j=0}^{\infty}\int_{\omega\in\mathbb{R}}e^{i\omega(t-\tau)}\hat{y}_{j}(\omega)d\omega
\]
\[
=V(s)(t-\tau),
\]
where $\hat{y}_{j}$ is the spectrum of the output of the $j^{\text{th}}-$order
operator, $V_{j}$.

\subsection{Modulation (frequency shift)}

Let $m:s\rightarrow s'$, $m(s)(t)=e^{i\xi t}s(t)$ be the modulation-by-$\xi$
map. Then $V(m):V(s)\rightarrow V(s')$ is written in the frequency domain as
\[
\widehat{V(m)V(s)}(\omega)
\]
\[
=\stackrel[j=0]{\infty}{\sum}\int_{\boldsymbol{\Omega}_{j}\in\mathbb{R}^{j}\,|\,\Sigma\boldsymbol{\Omega}_{j}=\omega}\hat{v}_{j}(\boldsymbol{\Omega}_{j})\stackrel[q=1]{j}{\prod}(\delta_{\xi} * \hat{s})(\omega_q)d\boldsymbol{\Omega}_{j}
\]
\[
=\stackrel[j=0]{\infty}{\sum}\int_{\boldsymbol{\Omega}_{j}\in\mathbb{R}^{j}\,|\,\Sigma\boldsymbol{\Omega}_{j}=\omega}\hat{v}_{j}(\boldsymbol{\Omega}_{j})\stackrel[q=1]{j}{\prod}\hat{s}(\omega_q - \xi)d\boldsymbol{\Omega}_{j}
\]
\begin{equation}
=\stackrel[j=0]{\infty}{\sum}\int_{\boldsymbol{\Omega}_{j}\in\mathbb{R}^{j}\,|\,\Sigma\boldsymbol{\Omega}_{j}=\omega}\hat{v}_{j}(\boldsymbol{\Omega}_{j})\hat{s}^{\otimes j}(\boldsymbol{\Omega}_j - \xi \boldsymbol{1})d\boldsymbol{\Omega}_{j}.\label{eq:VS action on modulation}
\end{equation}


If the input $s$ is the constant unity signal $\boldsymbol{1}$,
then (\ref{eq:VS action on modulation}) reduces to the
response of a Volterra series to a pure complex exponential input
(see: \cite{rugh_nonlinear_1981}, Section 2.4; or \cite{mathews_polynomial_2000}):
\[
V(m)V(\boldsymbol{1})(t)=\sum_{j}e^{ijt\xi}\,\hat{v}_{j}(\xi\boldsymbol{1}).
\]

\subsection{Periodization}

See (Appendix A, Sect. A.2) for background on the Dirac comb and its
tensor products.\\

Let $c:s\rightarrow s'$ be the operation of convolution against the
Dirac comb with period $T$: $c(s)(t)=(\text{Ш}_{T}*s)(t)$; such
an operation `periodizes' the signal $s$. Then $V(c):V(s)\rightarrow V(s')$
is defined by
\[
V(c)V(s)(t)=V(c(s))(t)
\]
\[
=\sum_{j=0}^{\infty}\int_{\omega\in\mathbb{R}}e^{i\omega t}\int_{\boldsymbol{\Omega}_{j}\in\mathbb{R}^{j}|\Sigma\boldsymbol{\Omega}_{j}=\omega}\hat{v_{j}}(\boldsymbol{\Omega}_{j})\prod_{r=1}^{j}\text{Ш}_{\frac{1}{T}}(\omega_{r})\hat{s}(\omega_{r})d\boldsymbol{\Omega}_{j}d\omega
\]
\[
=\sum_{j=0}^{\infty}\int_{\omega\in\mathbb{R}}e^{i\omega t}\int_{\boldsymbol{\Omega}_{j}\in\mathbb{R}^{j}|\Sigma\boldsymbol{\Omega}_{j}=\omega}\hat{v_{j}}(\boldsymbol{\Omega}_{j})\,\text{Ш}_{\frac{1}{T}}{}^{\otimes j}(\boldsymbol{\Omega}_{j})\hat{s}^{\otimes j}(\boldsymbol{\Omega}_{r})d\boldsymbol{\Omega}_{j}d\omega
\]
\[
=\sum_{j=0}^{\infty}\int_{\omega\in\mathbb{R}}e^{i\omega t}\sum_{\boldsymbol{k}\in\mathbb{Z}^{j}\,|\,\Sigma T^{-1}\boldsymbol{k}=\omega}\hat{v_{j}}(T^{-1}\boldsymbol{k})\hat{s}^{\otimes j}(T^{-1}\boldsymbol{k})d\omega
\]
\\
which, collecting multicombinations of integer indices $k_{i}\in\mathbb{Z}$
making up the vector $\boldsymbol{k}=[\overbrace{k_{1},\dots,k_{1}}^{n(k_{1})};\dots;\overbrace{k_{p},\dots,k_{p}}^{n(k_{p})}]$,
and counting permutations of these multicombinations, can be rewritten
\\
\begin{equation}
=\sum_{j=0}^{\infty}\int_{\omega\in\mathbb{R}}e^{i\omega t}\sum_{\boldsymbol{k}|\Sigma T^{-1}\boldsymbol{k}=\omega}\binom{j}{n(k_{1}),\dots,n(k_{p})}\,\hat{v_{j}}(T^{-1}\boldsymbol{k})\,\hat{s}^{\otimes j}(T^{-1}\boldsymbol{k})d\omega.\label{eq:VS response to periodization}
\end{equation}

\subsection{Sampling}

Let $m:s\rightarrow s'$ be the operation of multiplication against
the Dirac comb with period $T$: $m(s)(t)=(\text{Ш}_{T}\cdot s)(t)$
; such an operation `samples' the signal $s$. Then $V(m):V(s)\rightarrow V(s')$
is defined 
\[
V(m)V(s)(t)=V(m(s))(t)
\]
\[
=\stackrel[j=0]{\infty}{\sum}\int_{\boldsymbol{\tau}_{j}\in\mathbb{R}^{j}}v_{j}(\boldsymbol{\tau}_{j})\stackrel[r=1]{j}{\prod}\text{Ш}_{T}(t-\tau_{r})s(t-\tau_{r})d\boldsymbol{\tau}_{j}
\]
\[
=\stackrel[j=0]{\infty}{\sum}\int_{\boldsymbol{\tau}_{j}\in\mathbb{R}^{j}}v_{j}(\boldsymbol{\tau}_{j})\,\text{Ш}_{T}{}^{\otimes j}(t\boldsymbol{1}_{j}-\boldsymbol{\tau}_{j})\,s^{\otimes j}(t\boldsymbol{1}_{j}-\boldsymbol{\tau}_{j})d\boldsymbol{\tau}_{j}
\]
\begin{equation}
=\sum_{j=0}^{\infty}\,\,\sum_{\boldsymbol{k}\in\mathbb{Z}^{j}|\,\Sigma n(k_{i})=j}\binom{j}{n(k_{1}),\dots,n(k_{p})}\,v_{j}(t\boldsymbol{1}-T\boldsymbol{k})\,s^{\otimes j}(t\boldsymbol{1}-T\boldsymbol{k})\label{eq:VS response to sampling}
\end{equation}
where $\boldsymbol{k}=[\overbrace{k_{1},\dots,k_{1}}^{n(k_{1})};\dots;\overbrace{k_{p},\dots,k_{p}}^{n(k_{p})}]$,
as in example 2.4.3.

If the input is the constant unity signal, $\boldsymbol{1}$, then
(\ref{eq:VS response to sampling}) reduces to the result known for
the response of a Volterra series to a combination of impulsive inputs
(\cite{rugh_nonlinear_1981}, Section 5.1), specialized to the case
where the inputs are spaced at regular intervals of length $T$:\\
\[
V(\text{Ш}_{T})(t)=\sum_{j=0}^{\infty}\,\,\sum_{\boldsymbol{k}\in\mathbb{Z}^{j}|\,\Sigma n(k_{i})=j}\binom{j}{n(k_{1}),\dots,n(k_{p})}\,v_{j}(t\boldsymbol{1}-T\boldsymbol{k}).
\]
\\


\typeout{} 
\chapter{The category\emph{ }of Volterra series}

No system in the physical world remains forever the same; rather, its identity, the particular way that it processes signals,
inevitably changes with time. 
Such a system is called \emph{nonstationary},
in contrast to an idealized one whose response pattern is invariant
with respect to its inputs and past. For example, a linear translation-invariant
system is nonstationary if its impulse response is time-varying. Likewise,
we call a nonlinear system represented by a Volterra series nonstationary
if any of its VKFs or higher-order impulse response functions are.

Nonstationarity, as a concept, provides a context for questions such
as: which parts of a system are changing, and in what ways, and how
fast? The stability of identity becomes a property to be measured and controlled.
Like nonlinearity, however, nonstationarity is only a negativistic
notion. Knowing, in principle that a system changes does not tell
us \emph{how} it does so. For example, in section (1.3), we briefly
reviewed a form of parametricity for Volterra series that was achieved apparently
by simply formally adjoining a scalar variable upon which the kernel functions of
the Volterra series were then made to depend; yet, the mechanism by which those
dependencies had been introduced had not, itself, been made clear.

In this chapter, we introduce the notion of a \emph{morphism of Volterra
series}, which models a change in the response pattern of a nonlinear
system. We do this by defining a Volterra series morphism $\phi:V\rightarrow W$ as a kind of 
\emph{lens map}, which consists of a pair, $(\phi_{1},\phi^{\#})$,
of functions: a function $\phi_{1}$ between the index-sets
of the series; and, for each pair of indices $(i,\phi_{1}(i))$, a
\emph{dependent }linear map $\phi_{i}^{\#}:S'(\mathbb{R}^{[\phi_{1}(i)]})\rightarrow S'(\mathbb{R}^{[i]})$
between the function spaces of their respective kernels. We further show that, under appropriate restrictions
on the map $\phi_{i}^\#$, such morphisms are actually natural transformations between the Volterra series considered as functors.
Equipped with morphisms of Volterra series, we then define a category thereof, the category of
Volterra series. We begin by recalling the notion of natural transformation.

\section{Natural transformations}

Just as functors are structure-preserving maps between categories,
natural transformations are structure-preserving maps between functors.
Given two functors $F,G:C\rightarrow D$, a natural transformation
$\alpha:F\rightarrow G$ consists of a collection of \emph{components},
$\alpha_{(-)}\in\text{Mor}(D),$ which are functions indexed by the objects of $C$. In
order to comprise a natural transformation, these components must satisfying the following
condition: for any morphism $f:x\rightarrow y$ in $C$, the square

\[\begin{tikzcd}[ampersand replacement=\&] 	F(x) \& {F(y)} \\ 	{G(x)} \& {G(y)} 	\arrow["F(f)", from=1-1, to=1-2] 	\arrow["\alpha_{y}", from=1-2, to=2-2] 	\arrow["\alpha_{x}"', from=1-1, to=2-1] 	\arrow["{G(f)}"', from=2-1, to=2-2] \end{tikzcd}\]\\
commutes in $D$; i.e., $\alpha_{y}(F(f)(F(x)))=G(f)(\alpha_{x}(F(x))$.
A natural transformation $\alpha:F\rightarrow G$ is also denoted\[\begin{tikzcd}[ampersand replacement=\&] 	{C} \&\& {D} 	\arrow[""{name=0, anchor=center, inner sep=0}, "G"', curve={height=20pt}, from=1-1, to=1-3] 	\arrow[""{name=1, anchor=center, inner sep=0}, "F", curve={height=-20pt}, from=1-1, to=1-3] 	\arrow["\alpha"', shorten <=8pt, shorten >=8pt, Rightarrow, from=1, to=0] \end{tikzcd}\]to
emphasize the fact that, being a morphism between functors,
it is `two-dimensional', or a $2-$morphism in a category whose $1-$morphisms are functors\footnote{In\emph{ }the theory of higher categories, morphisms may be distinguished
by their level. For example, there is a $2-$category Cat, the category
of categories, in which the objects are, themselves, categories, the
$1-$morphisms are functors between them, and the $2-$morphisms are
natural transformations between the functors.}.

\paragraph{Example: Natural Transformation between Representable functors $\text{Set}\rightarrow\text{Set}$}

Recall the notion of representable functor from Section 2.1. Let $y^{s},y^{z}:\text{Set}\rightarrow\text{Set}$
be two representable endfunctors on the category Set, represented by the
sets $s$ and $z$, respectively\footnote{Recall that the exponential object, notated $a^{b}$, denotes the
space of maps from \textbf{$b$ }to $a$.}. Given any morphism $q:z \rightarrow s$, there is an induced function
$\alpha_s:a^s \rightarrow a^z$ between Hom-sets, given by precomposition with $q$; i.e., given any morphism $g:s\rightarrow a$
in $a^s$, we obtain a function $\alpha_s(g) \in a^z$ where $\alpha_s(g)=q;g$. The following diagram then commutes

\[\begin{tikzcd}[ampersand replacement=\&] 	a^s \& b^s \\ 	a^{z} \& b^{z} 	\arrow["f^s", from=1-1, to=1-2] 	\arrow["\alpha_{b}", from=1-2, to=2-2] 	\arrow["\alpha_{a}"', from=1-1, to=2-1] 	\arrow["f^z"', from=2-1, to=2-2] \end{tikzcd}\]\\
(where $f^s$ represents the operation of postcomposition with $f$), since $(q;g);f = q;(g;f)$, by the associativity of function composition.\\

\section{Morphisms of Volterra series}

In this section, we define the concept of a morphism between Volterra series and exhibit it as a kind of lens map. We show that it our definition satisfies the property of being a natural transformation. Before we do, however, we briefly introduce a notation that will allow us to distinguish the index of a homogeneous Volterra operator from its order of nonlinearity.

\paragraph{Notational aside: index sets and nonlinear orders\protect \\
}

A Volterra series is often assumed to consist of a single homogeneous
operator at each \emph{order of nonlinearity}, as reflected by the indexing
variable $j\in\mathbb{N}$ in the left-most sum of the defining equation
(\ref{eq:univariate Volterra series; time domain}). However, in Chapter
4 we will describe ways of combining Volterra series that result in
series which contain multiple homogeneous operators of the same order
of nonlinearity; we therefore need to be able to distinguish homogeneous
operators of similar order.

Let $I_{V}$ or just $I\in\text{Ob(Set)}$ denote the index-set of
the homogeneous operators in a Volterra series $V$, and let $i\in I$
be such an index. Then, in certain contexts that will be made clear,
we write $[i]$ for the order of nonlinearity of the homogeneous
operator indexed by $i$. Thus, for example, if $[i]=j$, then the
VKF at at index $i$ has type $v_{i}:\mathbb{R}^{j}\rightarrow\mathbb{R}$
(modulo symmetry and any other constraints).

In the special case that the magnitude of the $0^{\text{th}}-$frequency
component of every homogeneous operator in a Volterra series $V$
is $1$, then an elegant way of generating the index-set of $V$ is
to have $V$ operate on the constant input $\boldsymbol{1}$: $I_{V}\cong V(\boldsymbol{1})$.\footnote{This is analagous to the notation used for polynomial endofunctors
on Set, as described in \cite{poly_book}; there, the index set, or
set of \emph{positions},\emph{ }$I_{p}$, of a polynomial $p$ is
isomorphic to $p$ evaluated at the number $1$: $I_{p}\cong p(\text{1)}$.} We will make this assumption (that $I_{V}\cong V(\boldsymbol{1})$)
at times in what follows, and use the notation $V(\boldsymbol{1})$
to refer to the index set of the homogeneous operators in a series.\\

\subsection{Morphisms as lens maps}

The definition we are about to give of a morphism of Volterra series
is an example of what is known in category theory and computer science
as a \emph{lens}. In the category Set, a lens $\binom{\phi_{1}}{\phi_{(-)}^{\#}}$$:(A\times B)\rightarrow(C\times D)$
is comprised of: a map $\phi_{1}:A\rightarrow C$, and for each $a\in A$,
a function $\phi_{a}^{\#}:A\times D\rightarrow B$.
This construction is used in \cite{poly_book} to define morphisms
between \emph{polynomial functors}, which are functors Set$\rightarrow$Set
that are given as sums (coproducts) of representable functors. In what follows,
we adapt the lens construction to work for Volterra series.

The general shape of a morphism between Volterra series comprises the following: a map \emph{forwards},
identifying, for each homogeneous operator in the source series, a homogeneous operator in the target one;
and for each such pair, a continuous linear map \emph{backwards}, from the function space of the kernel function of the
target to that of the source homogeneous operator. However,
the precise definition that we next give is somewhat more restricted, for reasons that will be explained. 

\paragraph{Definition:}

A \emph{morphism $\phi:V\rightarrow W$ of Volterra series }is comprised
of the following data:
\begin{itemize}
\item a function $\phi_{1}:I_{V}\rightarrow I_{W}$ between index-sets;
\item for each pair $(i,\phi_{1}(i))$, with $i\in I_{V}$ and $\phi_{1}(i)\in I_{W}$,
\begin{itemize}
\item a linear map $\phi_{i}: \mathbb{R}^{[i]} \rightarrow \mathbb{R}^{[\phi_{1}(i)]}$.
\item and a weight function, or \emph{mask}, $\psi : \mathbb{R}^{[i]} \rightarrow \mathbb{C}$
\end{itemize}
\end{itemize}
Notice that the map $\phi_i$ is going in the forwards direction. This is because we obtain from it and the weight function $\psi$ a map $\phi_{i}^{\#}: S(\mathbb{R}^{[\phi_{1}(i)]}) \rightarrow S(\mathbb{R}^{[i]})$ called the \emph{weighted pullback} from $S'(\mathbb{R}^{[\phi_{1}(i)]})$ to $S'(\mathbb{R}^{[i]})$ along $\phi_i$, which is given by precomposition with $\phi_i$ followed by weighting by $\psi$ - i.e., $\phi_i^{\#}(z)(\boldsymbol{x}) = \psi(\boldsymbol x) \cdot z(\phi_i(\boldsymbol{x}))$, where $z\in S'(\mathbb{R}^{\phi_{1}[i]})$ and $\boldsymbol{x} \in \mathbb{R}^{[i]}$. 
We furthermore require the following condition on the map $\phi_i$: that it preserve the weak compositions of frequencies, i.e. $\Sigma \Omega_i = \omega \implies \Sigma \phi_i(\Omega_i) = \omega$.\footnote{This restriction limits the form of the maps $\phi_i$ in general to that of a certain shearing-type mapping followed by a projection. Depending on the supports of the kernel functions of the series involved, other mappings may be possible.} This ensures that the two systems which are interacting do so at the same (composite) set of frequencies, and so that the image of a component of the source series always lies within the spectrum of the target.

Operationally, a morphism of Volterra series instructs us to do the following: for each homogeneous operator $V_{i}$ of the series $V$,
choose a homogeneous operator $W_{\phi_{1}(i)}$ of the series
$W$; then, for each pair $(i,j)$ where $j=\phi_1(i)$, choose a linear map
from the domain of the kernel function $v_j$ to the domain of the kernel function $w_i$, as well as a masking function over the former; this
gives a third map, the weighted pullback, going in the \emph{backwards} direction, from the signal space living over the domain to that living over the codomain of $\phi_i$. The map
$\phi_i$ acts geometrically as an interface, parameterizing the area of interaction of the two systems, and the mask $\psi$ modulates this interaction. 


We can think of a morphism between (homogeneous) Volterra series once more in analogy to representable (and polynomial) functors,
in the following sense. Recall that, given two representable functors on the category Set, whose representing
sets are $s$ and $z$, respectively, a morphism between them consists of a map $q:z \rightarrow s$ and an induced map
$\alpha_s: a^s \rightarrow a^z$, between the exponential objects, which is given by precomposition with $q$.
Now, just as exponentiating one set $a$ by another of cardinality $j$ produces the set of $j-$tuples of elements of $s$,
a homogeneous Volterra series of order $j$ produces from an input signal $s$ an output signal whose frequency components are
products of $j-$many frequencies in $s$.\footnote{This is most clearly seen in the case where the kernel function $v_j$ is the
$j-$dimensional delta distribution, $v_j=\delta^{\otimes j}$.} Additionally, just as one can distinguish between
the Hom-set, or exponential object, $a^s$, and its cardinality (a number), one can distinguish between the (graded) object $V(s)$ and the signal
to which it evaluates as a sum\footnote{The analogy here can probably
be further extended via a combinatorial argument based on partitions, by looking at the cardinalities of the fibers over various multi-frequencies in
the output spectrum.}. We can thus view the map $\phi_i:\mathbb{R}^{[i]}\rightarrow \mathbb{R}^{[\phi_1(i)]}$, along which we pullback, as similar to the map $q:z\rightarrow s$,
between representing sets, by which we precompose.

However, there are many more degrees of freedom when defining a map $S'(\mathbb{R}^{[j]}) \rightarrow S'(\mathbb{R}^{[i]})$
than there are when defining a map of sets $[j] \rightarrow [i]$. In order to ensure that the morphisms are well-defined and natural, we need to place geometric restrictions on the map $\phi_i^\#$, which also has the affect of strengthening the analogy above.

\subsection{Naturality of Volterra morphisms}

Given the above lens-type definition of a morphism, we obtain a collection of maps indexed by objects of the signal space $S'(\mathbb R)$: given a signal $s$,
we define a map $\phi_s$, called the \emph{component of $\phi$ at s}, in the spectral domain via its action
on Fourier components, as\footnote{We focus here on homogeneous Volterra operators, but our definition extends additively to Volterra series that are sums of the same.}:

\[
\phi_s(V_{i}(\hat s))(\omega)=
\]
\begin{equation}
  \int_{\genfrac{}{}{0pt}{}{\boldsymbol{\Omega}_{i}\in\mathbb{R}^{i}}{\Sigma\boldsymbol{\Omega}_{i}=\omega}} (\psi \odot (\hat{\,v_{i}} \odot \hat{s}^{\otimes [i]}) \odot \phi_{i}^{\#}(\hat{w}_j))(\boldsymbol{\Omega}_{j})d\boldsymbol{\Omega}_{j}=\label{eq:morphism of VS}
\end{equation}
\begin{equation}
  \int_{\genfrac{}{}{0pt}{}{\boldsymbol{\Omega}_{i}\in\mathbb{R}^{i}}{\Sigma\boldsymbol{\Omega}_{i}=\omega}} (\psi \odot \hat{\,v_{i}} \odot \hat{s}^{\otimes [i]})(\boldsymbol{\Omega}_{j})\hat{w}_j(\phi_{i}(\boldsymbol{\Omega}_{j}))d\boldsymbol{\Omega}_{j}
\end{equation}
where the symbol $\odot$ denotes pointwise multiplication. 

The next step is to show that these maps are natural, i.e., that all squares of the form\[\begin{tikzcd}[ampersand replacement=\&] 	V(s) \& {V(s')} \\ 	{W(s)} \& {W(s')} 	\arrow["V(f)", from=1-1, to=1-2] 	\arrow["\alpha_{s'}", from=1-2, to=2-2] 	\arrow["\alpha_{s}"', from=1-1, to=2-1] 	\arrow["{W(f)}"', from=2-1, to=2-2] \end{tikzcd}\]for
$s,s'\in S'(\mathbb{R})$ and $f:s\rightarrow s'$ commute. We immediately see that such a condition excludes those linear maps which in the spectral domain are of convolution-type, since convolution and modulation do not, in general, commute. However, our definition using the weighted pullback involves only a change of variables and additional multiplication in the spectral domain. Thus, we see that the bottom path is equal to the top path of the diagram:

\[
\int_{\genfrac{}{}{0pt}{}{\boldsymbol{\Omega}_{i}\in\mathbb{R}^{i}}{\Sigma\boldsymbol{\Omega}_{i}=\omega}} (\hat{f}^{\otimes[i]}\odot (\psi \odot (\hat{\,v_{i}} \odot \hat{s}^{\otimes [i]}) \odot \phi_i^\#(\hat{w}_j)))(\boldsymbol{\Omega}_{j})d\boldsymbol{\Omega}_{j}
\]
\[
=\int_{\genfrac{}{}{0pt}{}{\boldsymbol{\Omega}_{i}\in\mathbb{R}^{i}}{\Sigma\boldsymbol{\Omega}_{i}=\omega}} (\psi \odot ((\hat{\,v_{i}} \odot \hat{s}^{\otimes [i]} \odot \hat{f}^{\otimes[i]}) \odot \phi_i^\#(\hat{w}_j)))(\boldsymbol{\Omega}_{j})d\boldsymbol{\Omega}_{j}.
\]
$ $

An alternative route to ensuring the naturality of Volterra morphisms would be to simply impose it by fiat: a lens map of the shape given in section 3.2.1 is a valid morphism of Volterra series \emph{precisely when} all of the corresponding naturality squares in $S'(\mathbb R)$ commute.

\subsection{Examples of Volterra morphisms}

In this section we survey some elementary examples of Volterra series morphisms, including
the identity morphism at a Volterra series, and introduce the notion of
parameterized morphism.

\paragraph{{\small{}Trivial Morphism}}

The trivial morphism $\alpha:V\rightarrow(\sum_{j}(\delta^{\otimes j}))$
is the morphism with target the series whose VKF at each order $j$
is a multidimensional distribution centered at the origin, and whose
maps $\phi_{1}$ and $\phi^{\#}$ are both the identity maps.

\paragraph{{\small{}Autoconvolution}}

The autoconvolution $\text{aut}_{V}:V\rightarrow V$ is given by the
pair $(\phi_{1},\phi^{\#})$, where both $\phi_{1}$ and all of the
$\phi_{i}^{\#}$ are identity maps. This map results in the Volterra
series whose VKF at each order $i$ is the autoconvolution $(v_{i}*v_{i})$
of $v_{i}$.

\paragraph{{\small{}Identity morphism}}

The identity morphism $\text{id}_{V}:V\rightarrow V$ is given by
the pair $(\phi_{1},\phi^{\#})$, where $\phi_{1}=\text{id}_{V(\boldsymbol{1})}$
is the identity and where, for any $i\in V(\boldsymbol{1})$, $\phi_{i}^{\#}:S'(\mathbb{R}^{[i]})\rightarrow S'(\mathbb{R}^{[i]})$
is the weighted pullback along the identity on $\mathbb{R}^{[i]}$ that scales by the reciprocal of the spectrum
of $v_{i}$, i.e. $\psi(\boldsymbol{\Omega}_i)=\frac{1}{\hat{v}_{i}(\boldsymbol{\Omega}_i)}$
for $\hat{v}_{i}(\boldsymbol{\Omega}_i)\ne0$. The definition of $\phi_{i}^{\#}$
follows from the fact that the spectrum of the autoconvolution $R_{v_{i}}=(v_{i}*v_{i})$
is $\hat{R}_{v_{i}}(\boldsymbol{\Omega}_{i})=\hat{v}_{i}(\boldsymbol{\Omega}_{i})^{2}$.

\subsubsection{Parameterized morphisms}
Morphisms of Volterra series can also be parameterized by one or many variables. Each of the following examples of this corresponds
to a well-known operation in signal processing, and provides a basic
template by which a parameterized family of Volterra series can be
generated in a continuous fashion.

\paragraph{{\small{}Translation}}

Let $\boldsymbol{\tau}=[\tau_{1},\boldsymbol{\tau}_{2},\dots,\boldsymbol{\tau}_{j}]\in\mathbb{R}\times\mathbb{R}^{2}\times\dots\times\mathbb{R}^{j}$;
then the translation-by-$\boldsymbol{\tau}$ morphism is the morphism
with target the VS whose VKF at each order $i$ is a multidimensional
distribution centered at $\boldsymbol{\tau}_{i}$, with $\phi_{1}=\text{id}$
and all of the $\phi_{i}^{\#}$ also identities. The offsets can be varied and/or the morphism iterated.

\paragraph{{\small{}Sampling}}

Similarly, sampling corresponds to a morphism $V\rightarrow W$,
where the VKF of the target series at order $j$ is a Dirac comb, $w_{j}(\boldsymbol{\tau}_{j})=\text{\textcyr{\CYRSH}}(\boldsymbol{\tau}_{j})$.
The period $T_{j}$ of the comb at order $j$ can be varied.

\paragraph{{\small{}Smoothing}}

Smoothing corresponds, in one example, to a morphism $V\rightarrow(\sum_{j}W_{j})$
where the VKFs of the target series are normal distributions, i.e.
$w_{j}(\boldsymbol{\tau}_{j})=e^{-\boldsymbol{\tau}_{j}^{\top}C_{j}\boldsymbol{\tau}_{j}}$
and $C_{j}$ is a positive-definite $j\times j$ matrix. The morphism
can be iterated. Similarly, a target series whose VKFs are the stencils
of Laplacian operators can be used to drive diffusion.\\  

Other morphisms, which might, for example, affect transformations such as periodization or shearing, can be similarly defined -
and one might as well go further, defining families of Volterra series generated by morphisms arising, for example, under the orbits of 
various groups.

\section{Volterra series form a category}

Given a category $C,$ and an endofunctor $F:C\rightarrow C$,
one can construct the identity natural transformation, denoted $1_{F}$. Furthermore,
natural transformations compose associatively. Thus, for any category
$C$, there is a \emph{functor category }whose objects are endofunctors
on $C$ and whose morphisms are natural transformations between them.

In the preceding sections,
we defined the Volterra series as a kind a functor, and defined
morphisms of Volterra series as lens maps, which we showed were natural. We now show that Volterra series and these
morphisms assemble into a functor category.
Recall from 3.2.3 that for any Volterra series $V$, we have an identity morphism $\text{id}_V : V \rightarrow V$. It remains
to address the composition of Volterra series morphisms. For any three
Volterra series $V,W,Z$, and morphisms $f,g,h$ between them as shown,
the diagram:

\[\begin{tikzcd} V && W \\ \\ && X \arrow["f", from=1-1, to=1-3] \arrow["g", from=1-3, to=3-3] \arrow["h"', from=1-1, to=3-3] \end{tikzcd}\]
commutes in the sense that the following two diagrams, describing the corresponding
lens maps, do:

\[\begin{tikzcd} {V(\bf1)} && {W(\bf1)} && {S(\mathbb{R} ^{[i]})} && {S(\mathbb{R} ^{[f_1(i)]})} \\ \\ && {X(\bf1)} &&&& {S(\mathbb{R} ^{[h_1(i)]})} \arrow["{f_1}", from=1-1, to=1-3] \arrow["{g_1}", from=1-3, to=3-3] \arrow["{h_1}"', from=1-1, to=3-3] \arrow["{f_{i}^{\#}}"', from=1-7, to=1-5] \arrow["{g_{f_{1}(i)}^{\#}}"', from=3-7, to=1-7] \arrow["{h_{i}^{\#}}", from=3-7, to=1-5] \end{tikzcd}\] where $h_{1}=g_{1}\circ f_{1}$ and $h_{i}^{\#}=g_{f_{1}(i)}^{\#}\circ f_{i}^{\#}$, $\forall i \in V(\boldsymbol 1)$. We can now
give the following definition:

\paragraph{Definition:}

The category, \emph{Volt}, of Volterra series is the category having,
as objects, Volterra series, and as morphisms, natural transformations between them.\\


In Chapter 4, we study three ways of combining Volterra series, each
of which has the form of a monoidal product, $(-\diamond-):\text{\emph{Volt}\ensuremath{\times}\emph{Volt}\ensuremath{\rightarrow\emph{Volt}}}$.
By working not, merely, with the set of all Volterra series, but rather
with Volterra series and their morphisms in concert, it becomes possible
and, indeed, natural to study the associated universal properties
of these operations. This is fundamentally because morphisms of Volterra
series provide a means of describing how Volterra series transform,
not in isolation, but functionally and in relation to one another.

\typeout{} 
\chapter{\emph{Volt }as a monoidal category}

The modeling of complex systems as interconnections and compositions
of simpler ones, be it via a synthetic or an analytic approach to
the problem, is a hallmark of scientific methodology. However, representing
a complex system merely as an interconnection may bring little benefit
if the rules for interconnection and the properties of those rules
are not well understood. This understanding constitutes the difference
between mere modularity and compositionality: whereas, in a modular
paradigm, one is able to arrange systems and interconnect them but
might not know how the various interconnections are related, in a
compositional framework one has access, in the form of equivalences
or isomorphisms, to knowledge of this kind. Mathematically,
such a framework has the structure of a monoidal category.

The need for a compositional theory is especially poignent
when it comes to modeling nonlinear systems, since the complexity of a nonlinear system
grows extremely quickly with its order of nonlinearity. For example, in nonlinear
system identification, harmonic probing methods for the estimation
of Volterra series parameters have an algorithmic complexity that,
in their naive formulation, is exponential in the system memory
and order of nonlinearity, limiting their usefullness for identifying
Volterra series that are represented monolithically\footnote{See \cite{rugh_nonlinear_1981,mathews_polynomial_2000} for introductions
to Volterra series parameter estimation and \cite{cheng_volterra-series-based_2017}
for a recent overview of approaches to kernel estimation in both the
time and frequency domains.}. Modeling a complex nonlinear system instead as a composition of
lower-order Volterra series in theory enables the parameters of the
simpler series to first be estimated separately, and then the results
to be combined using the interconnection rules. In such cases, beyond
the benefit brought by mere modularity, compositionality further provides
the ability to flexibly rearrange the component subsystems -- for
example, to re-associate them in series -- without requiring the
reestimation of their parameters. Compositionality stands, dually,
to bring the same kind of benefit to the synthesis, rather than the
analysis, of complex nonlinear systems.

Modularity and compositionality are furthermore idiomatic to the domain
of audio. From the design of a modular synthesizer, to the orchestration
of an ensemble, to the composition of a musical score, to the performance
of a musical piece, musicians, composers, instrument designers, and
audio engineers rearrange and interconnect nonlinear systems in
a rich variety of ways. What is more, these musical practices and acts are often
carried out \emph{live} and in \emph{real-time}.

Our goal in this chapter is to place upon category-theoretic foundations
a compositional theory of nonlinear audio systems and their transformations,
as modeled by Volterra series and their morphisms. We do this by studying
three core operations - following closely the analogous parts of the exposition in \cite{poly_book} -
each of which has the form of a bifunctor on
the category Volt, and show how each satisfies a corresponding universal
property. As the culmination of our efforts, we show that \emph{Volt
}is a monoidal category under the operation of series composition.
We begin by recalling the concept of monoidal category.

\section{Monoidal categories}

A \emph{monoidal category} is a category $C$ equipped with:
\begin{itemize}
\item a bifunctor
\[
\diamond:C\times C\rightarrow C,
\]
also called the monoidal (or tensor) product\footnote{Monoidal categories are sometimes called \emph{tensor categories}, but the notion of monoidal product is more general than any particular tensor product, such as that of vector spaces. The
product we use in this chapter to exhibit \emph{Volt
}as a monoidal category is the series composition product, which we denote by $\triangleleft$.}, that is associative up to a natural isomorphism;
\item an object, $I$, called the \emph{monoidal unit}, that acts as both
a left and right identity for $\diamond$, up to natural isomorphism;
\item and natural isomorphisms:
\begin{itemize}
\item the associator: $\mathfrak{a}:((-)\diamond(-))\diamond(-)\stackrel{\cong}{\rightarrow}(-)\diamond((-)\diamond(-)),\,\mathfrak{a}_{x,y,z}:(x\diamond y)\diamond z\rightarrow x\diamond(y\diamond z)$
\item the left and right unitors:
\end{itemize}
\end{itemize}
\begin{center}
$\begin{array}{cccc}
\lambda:(1\diamond(-))\stackrel{\cong}{\rightarrow}(-) &  &  & \rho:((-)\diamond1)\stackrel{\cong}{\rightarrow}(-)\\
\lambda_{x}:1\diamond x\rightarrow x &  &  & \rho_{x}:x\diamond1\rightarrow x
\end{array}$\\
\par\end{center}

\begin{flushleft}
such that the following diagrams commute: \emph{}\\
\emph{triangle identity}
\par\end{flushleft}

\emph{}\emph{\[\begin{tikzcd}[ampersand replacement=\&] 	\& {x\diamond y} \\ 	\\ 	{(x\diamond 1)\diamond y} \&\& {x\diamond (1\diamond y)} 	\arrow["{\rho_{x}\diamond 1_{y}}"', from=3-1, to=1-2] 	\arrow["{\alpha_{x,1,y}}"', from=3-1, to=3-3] 	\arrow["{1_{x}\diamond\lambda_{x} }", from=3-3, to=1-2] \end{tikzcd}\]}\emph{pentagon
identity}

\[\begin{tikzcd}[ampersand replacement=\&,column sep=tiny,row sep=huge] 	\&\& {(w\diamond x) \diamond (y\diamond z)} \\ 	\\ 	{((w\diamond x)\diamond y)\diamond z} \&\&\&\& {w\diamond(x\diamond y(\diamond z))} \\ 	\\ 	\\ 	\& {(w\diamond(x\diamond y))\diamond z} \&\& {w\diamond((x\diamond y)\diamond z)} 	\arrow["{\alpha_{w \diamond x,y,z}}", from=3-1, to=1-3] 	\arrow["{\alpha_{w,x,y\diamond z}}", from=1-3, to=3-5] 	\arrow["{\alpha_{w,x\diamond y,z}}", from=6-2, to=6-4] 	\arrow["{\lambda _w \diamond \alpha_{x,y,z}}", from=6-4, to=3-5] 	\arrow["{\alpha_{w,x,y}\diamond \rho_z}"{description}, from=3-1, to=6-2] \end{tikzcd}\]\\

The notion of monoidal category is a categorification of that of \emph{monoid},
which is a category having only one object; see \cite{fong_seven_2018}
for an introduction to monoidal category theory and a tour of various
examples of monoidal categories. Examples of monoidal categories include
the category Vect, of vector spaces and linear maps, with monoidal
product the ordinary tensor product; and the category Set, equipped
with either the disjoint union or the Cartesian product as the monoidal
product\footnote{A category may admit multiple monoidal structures.}.

To any monoidal category there is furthermore associated a diagrammatic
language, whose terms are known as \emph{string diagrams}, in which
applying the tensor product corresponds to putting morphisms in parallel,
and regular composition to putting them in series. Such a language
is resemblant of wiring diagrams in electrical engineering and signal
flow graphs in signal processing, but is fully formal, meaning that
topological deformations or manipulations of the diagrams correspond to known algebraic
operations; see \cite{selinger_graphical_languages} for a comprehensive
overview of the various string diagrammatic languages corresponding
to different monoidal categories.

\section{Sum and product of Volterra series}

\paragraph{Sum (coproduct)\protect \\
}

What happens when the outputs of two Volterra series that process
the same input signal are added? Can we represent the result using
a single series? Indeed, the sum $V+W$ of two Volterra series is
defined level-wise by the addition of their homogeneous operators;
equivalently, of the VKFs of the same order
\[
(V+W)(s)(t)
\]
\[
=\sum_{j}(V+W)_{j}(s)(t)=\sum_{j}V_{j}(s)(t)+W_{j}(s)(t)
\]
\[
=\int_{\boldsymbol{\tau}_{j}\in\mathbb{R}^{j}}(v_{j}(\boldsymbol{\tau}_{j})+(w_{j}(\boldsymbol{\tau}_{j}))\prod_{q=0}^{j}s(t-\tau_{q})d\boldsymbol{\tau}_{j}.
\]

While this operation is well-known in the Volterra series literature,
we would like to show that it is, category-theoretically, the \emph{coproduct}
in the category \emph{Volt} - i.e., that it satisfies the following
universal property. Let $\iota,\kappa$ denote the inclusions of $V$
and $W$ into $V+W$. Then given any other Volterra series $X$ and
any pair of morphisms $f,g$ where $f:V\rightarrow X$ and $g:W\rightarrow X$,
there is a unique morphism $h:V+W\rightarrow X$ making the following
diagram commute:\[\begin{tikzcd} V && {V+W} && W \\ \\ && X \arrow["\iota", from=1-1, to=1-3] \arrow["f"', from=1-1, to=3-3] \arrow["\kappa"', from=1-5, to=1-3] \arrow["h", dashed, from=1-3, to=3-3] \arrow["g", from=1-5, to=3-3] \end{tikzcd}\]Recall
from section 3.3 that, by representing the morphisms $f,g,\iota,\kappa$,
and $h$ as lens maps, this is equivalent to requiring the commutation
both of: (1) a diagram of maps between the index sets of the homogeneous
operators of the series, and; (2) a diagram of maps between their
respective function spaces. The first of these conditions states that
the diagram

\[\begin{tikzcd} {V(\bf 1)} && {V({\bf{1}}) + W({\bf 1})} && {W(\bf 1)} \\ \\ \\ && {X(\bf 1)} \arrow["{\iota_1}", from=1-1, to=1-3] \arrow["{f_1}"', from=1-1, to=4-3] \arrow["{\kappa_1}"', from=1-5, to=1-3] \arrow["{h_1}", dashed, from=1-3, to=4-3] \arrow["{g_1}", from=1-5, to=4-3] \end{tikzcd}\]must
commute; i.e., $h_{1}$ is the function which sends each $(1,i)\in V(\boldsymbol{1})+W(\boldsymbol{1})$,
where $i\in V(\boldsymbol{1})$, to $f_{1}(i)$, and each $(2,j)\in V(\boldsymbol{1})+W(\boldsymbol{1})$,
where $j\in W(\boldsymbol{1})$, to $g_{1}(j)$\footnote{Here, $(1,-)$ indexes into the part of the disjoint union tagged
by $V$, and $(2,-)$ indexes into the part tagged by $W$.}. This diagram is the standard one for the coproduct of sets.

For the diagram of the maps between function spaces, we obtain:

\[\begin{tikzcd} {S({\mathbb R}^{[i]})} && {S({\mathbb R}^{[(1,i)]})} && {S({\mathbb R} ^{[(2,j)]})} && {S({\mathbb R}^{[j]})} \\ \\ && {S({\mathbb R}^{[f_{1}(i)]})} && {S({\mathbb R}^{[g_{1}(j)]})} \arrow["{\iota_{i}^\#}"', from=1-3, to=1-1] \arrow["{f_{i} ^\#}", from=3-3, to=1-1] \arrow["{h_{(1,i)}^\#}"', dashed, from=3-3, to=1-3] \arrow["{h_{(2,j)} ^\#}", dashed, from=3-5, to=1-5] \arrow["{\kappa_{j}^\#}", from=1-5, to=1-7] \arrow["{g_{j} ^\#}"', from=3-5, to=1-7] \end{tikzcd}\]But both
$\iota_{i}^{\#}$ and $\kappa_{j}^{\#}$ are identities, so we must
have that $h_{(1,i)}^{\#}=f_{i}^{\#}$ and $h_{(2,j)}^{\#}=g_{j}^{\#}.$
Since both sets of diagrams commute, so does the original.

\paragraph{Product\protect \\
}

\begin{figure}[H]
\begin{centering}
\includegraphics[scale=0.2]{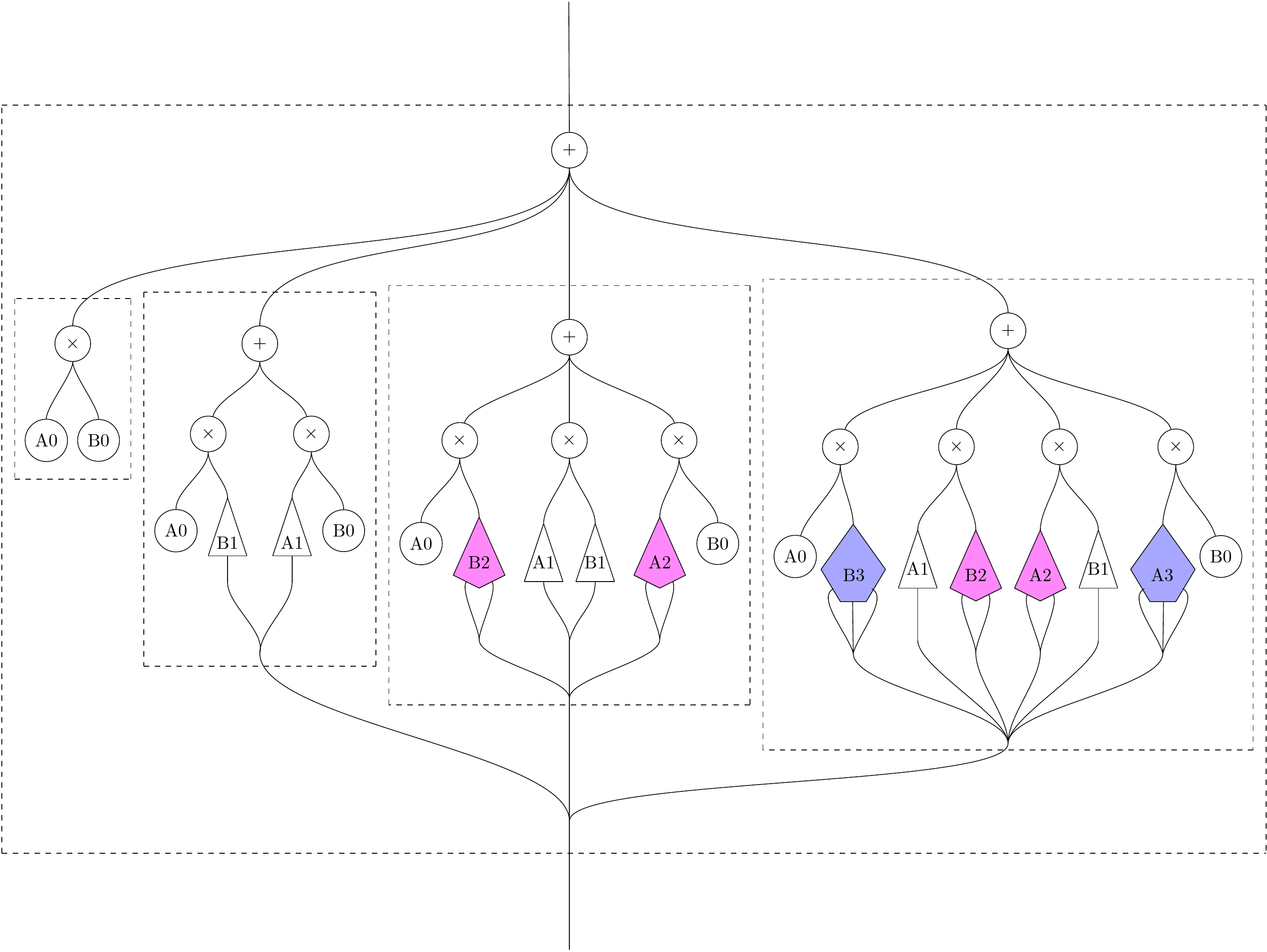}
\par\end{centering}
\centering{}\caption{Product of Volterra series. The interior dashed boxes represent the
composite homogeneous Volterra operators of the product series $A\times B$.
\label{fig:Product wiring diagram}}
\end{figure}

What happens if the outputs of two Volterra series that process the
same signal are multiplied? Can we express the result as the output
of a single series? Indeed, the product $V=A\times B$ of two Volterra
series $A$ and $B$ is given by the formula \cite{carassale_modeling_2010,rugh_nonlinear_1981}
\[
V(s)(t)=(A\times B)(s)(t)
\]
\begin{equation}
=\sum_{j}\sum_{k=0}^{j}(A_{k}(s)B_{j-k}(s))(t)\label{eq:product of VS}
\end{equation}
which can be rewritten
\begin{equation}
\begin{array}{c}
V_{j}(s)(t)=\sum_{k=0}^{j}\sideset{}{_{\Omega_{k}\in\mathbb{R}^{k}}}\int e^{i\Sigma\Omega_{k}t}\hat{a}_{k}(\Omega_{k})\stackrel[p=0]{k}{\prod}\hat{s}(\omega_{p})d\omega_{p}\\
\times\sideset{}{_{\Omega_{j-k}\in\mathbb{R}^{j-k}}}\int e^{i\Sigma\Omega_{j-k}t}\hat{b}_{j-k}(\Omega_{j-k})\stackrel[q=0]{j-k}{\prod}\hat{s}(\omega_{q})d\omega_{q}.
\end{array}\label{eq:product of VS expanded}
\end{equation}
Generalizing to products of arbitrary arity, the above can be further rewritten under a single integral using an
index notation similar to that from section $1.2$. Let $\theta_{1}^{p},\theta_{2}^{p},...,\theta_{k}^{p}$,
with $\theta_{i}^{p}\in\mathbb{R}^{\alpha_{i}^{p}}$, be a sequence
of vectors that partition (into parts of possibly zero size) the vectorial
frequency variable $\Omega_{j}$, where the superscript $p$ indicates
a given $k-$multicomposition of $j$; i.e., whose lengths 
\[
\alpha_{i}^{p}\ge0,\,\,(i=1,2)
\]
are such that 
\[
\stackrel[i=1]{k}{\sum}\alpha_{i}^{p}=j.
\]
Then in the case of the binary product, i.e. where $k=2$, eq. (\ref{eq:product of VS expanded})
can be rewritten as
\[
V_{j}(s)(t)=\sum_{p=1}^{\binom{j+1}{j}}\sideset{}{_{\Omega_{j}\in\mathbb{R}^{j}}}\int e^{i\Sigma\Omega_{j}t}\,\widehat{a}_{\alpha_{1}^{p}}(\theta_{1}^{p})\widehat{b}_{\alpha_{2}^{p}}(\theta_{2}^{p})\prod_{r=1}^{j}\widehat{s}(\omega_{r})d\omega_{r}
\]
where the binomial coefficient, $\binom{j+1}{j}=j$, counts the number
of possible binary partitions of $\Omega_{j}$. Thus, the $j^{\text{th}}-$order
VFRF for the composite series is
\begin{equation}
\widehat{v}_{j}(\Omega_{j})=\sum_{p=1}^{j}\widehat{a}_{\alpha_{1}^{p}}(\theta_{1}^{p})\,\,\widehat{b}_{\alpha_{2}^{p}}(\theta_{2}^{p}).\label{eq:kernel of product}
\end{equation}
The binomial structure of the product can be seen reflected in the
wiring diagram of Fig. \ref{fig:Product wiring diagram}, of a product
of two series up to order $3$.

Just as we showed that the sum of Volterra series is indeed a categorical
coproduct in \emph{Volt}, so would we like to show that the product
of Volterra series is the categorical product. First, let $\pi:A\times B\rightarrow A$
and $\varphi:A\times B\rightarrow B$ denote the projections out of
the product and onto the factors. Then, in order for $A\times B$
to be the categorical product, the following diagrams must commute:

\[\begin{tikzcd} && Y \\ \\ A && {A\times B} && B \arrow["\pi", from=3-3, to=3-1] \arrow["\varphi"', from=3-3, to=3-5] \arrow["g", from=1-3, to=3-5] \arrow["f"', from=1-3, to=3-1] \arrow["h"{description}, dashed, from=1-3, to=3-3] \end{tikzcd}\]

As with the coproduct, the commutation of a diagram of morphisms of
Volterra series as natural transformations is equivalent to that of
a pair of diagrams in which the morphisms are represented as lens
maps. Thus, the index-level diagram

\[\begin{tikzcd} {Y(\bf 1)} && {B(\bf 1)} \\ \\ {A(\bf 1)} && {{A(\bf 1)} \times {B(\bf 1)}} \arrow["{g_1}", from=1-1, to=1-3] \arrow["{f_1}"', from=1-1, to=3-1] \arrow["{\varphi_1}"', from=3-3, to=1-3] \arrow["{\pi_1}", from=3-3, to=3-1] \arrow["{h_1}", dashed, from=1-1, to=3-3] \end{tikzcd}\]must
commute. Since $\pi_{1}$ and $\varphi_{1}$ are projections, we must
have that $h_{1}(k)=(f_{1}(k),g_{1}(k))$. Similarly, the diagram

\[\begin{tikzcd} {S'(\mathbb R ^{[k]})} && {S'(\mathbb R ^{[g_{1}(k)]})} \\ \\ {S'(\mathbb R ^{[f_{1}(k)]})} && { S'(\mathbb R ^{[f_{1}(k)]}) \oplus S'(\mathbb R ^{[g_{1}(k)]})} \arrow["{g_{k}^\#}"', from=1-3, to=1-1] \arrow["{f_{k}^\#}", from=3-1, to=1-1] \arrow["{\varphi_{(f_1 (k),g_1 (k))}^\#}", from=1-3, to=3-3] \arrow["{\pi_{(f_1 (k),g_1 (k))} ^\#}"', from=3-1, to=3-3] \arrow["h_{k}^\#"{description}, dashed, from=3-3, to=1-1] \end{tikzcd}\]
of linear transformations between function spaces must commute. But
as $\pi_{(f_{1}(k),g_{1}(k))}^{\#}:S'(\mathbb{R}^{[f_{1}(k)]})\rightarrow S'(\mathbb{R}^{[f_{1}(k)]})\oplus S'(\mathbb{R}^{[g_{1}(k)]})\,$
and $\,\varphi_{(f_{1}(k),g_{1}(k))}^{\#}:S'(\mathbb{R}^{[g_{1}(k)]})\rightarrow S'(\mathbb{R}^{[f_{1}(k)]})\oplus S'(\mathbb{R}^{[g_{1}(k)]})$
are inclusions into the coproduct (disjoint union) of function spaces,
it follows that there is a unique \emph{bilinear} such map $h_{k}^{\#}$.
Noting that the expression of the kernel function
for the $j^{\text{th}}-$homogeneous operator in (\ref{eq:kernel of product})
is exactly the tensor product of the kernel functions of the factors,
it thus follows that there is a unique linear map out of the tensor
product, $h_{k}^{' \#}:S'(\mathbb{R}^{[f_{1}(k)]})\otimes S'(\mathbb{R}^{[g_{1}(k)]})\rightarrow S'(\mathbb{R}^{[k]})$.

\section{Series Composition}

The series composition, $V=B\triangleleft A$, of two Volterra series
$A$ and $B$, in which the output of $A$ flows to the input of $B$
is defined on a signal $u$ by summing the outputs of all the multilinear
Volterra operators of $B$ whose $k-$many inputs have orders summing
to $j$; i.e., 
\begin{equation}
V(u)=\stackrel[k=0]{n_{B}}{\sum}\underset{p|\stackrel[r=1]{k}{\sum}\alpha_{r}^{p}=j}{\sum}B_{k}[y_{\alpha_{1}^{p}},\dots,y_{\alpha_{k}^{p}}]\label{Series composition (multivariate inputs from A to B)}
\end{equation}
where: $n_{B}$ is the maximum order of the series $B$; the inner
sum runs over all $k-$multi\-compositions of $j$; and $y_{r}$
denotes the $r^{\text{th}}-$order output of $A$. Note that the second
series in the composition, denoted $B$ in equation (\ref{Series composition (multivariate inputs from A to B)}),
is multivariate since, in general, many outputs from amongst the homogeneous
operators of $A$ are used to compose the signals of appropriate order
that are fed as inputs to the homogeneous operators of $B$.

A composite series $V=B\triangleleft A$ is given in terms of its
Volterra frequency response functions\cite{carassale_modeling_2010,carassale_synthesis_2014,rugh_nonlinear_1981}
by 
\begin{equation}
v_{j}(\boldsymbol{\Omega}_{j})=\stackrel[k=0]{n_{B}}{\sum}\underset{\{p|\stackrel[r=1]{k}{\sum}\alpha_{r}^{p}=j\}}{\sum}\widehat{b}_{k}(S_{p}^{(j,k)}\boldsymbol{\Omega}_{j})\stackrel[r=1]{k}{\prod}\widehat{a}_{\alpha_{r}^{p}}(\boldsymbol{\theta}_{r}^{p})\label{binary composition (kernels form)}
\end{equation}
where $S_{p}^{(j,k)}$ is the $j\times k$ matrix:
\[
\begin{bmatrix}\overbrace{1,1,...,1}^{\alpha_{1}^{p}} &  &  & 0\\
 & \overbrace{1,1,...,1}^{\alpha_{2}^{p}}\\
 &  & \ddots\\
0 &  &  & \overbrace{1,1,...,1}^{\alpha_{k}^{p}}
\end{bmatrix},
\]
$\boldsymbol{\Omega}_{j}\in\mathbb{R}^{j}$ is the vector of frequency
variables, and $\alpha_{r}^{p}$ are the numbers representing the
lengths of the corresponding vectors in a partition, $[\theta_{1}^{p},\theta_{2}^{p},...,\theta_{k}^{p}]$
of $\boldsymbol{\Omega_{j}}$.

\begin{figure}[H]
\begin{centering}
\includegraphics[scale=0.2]{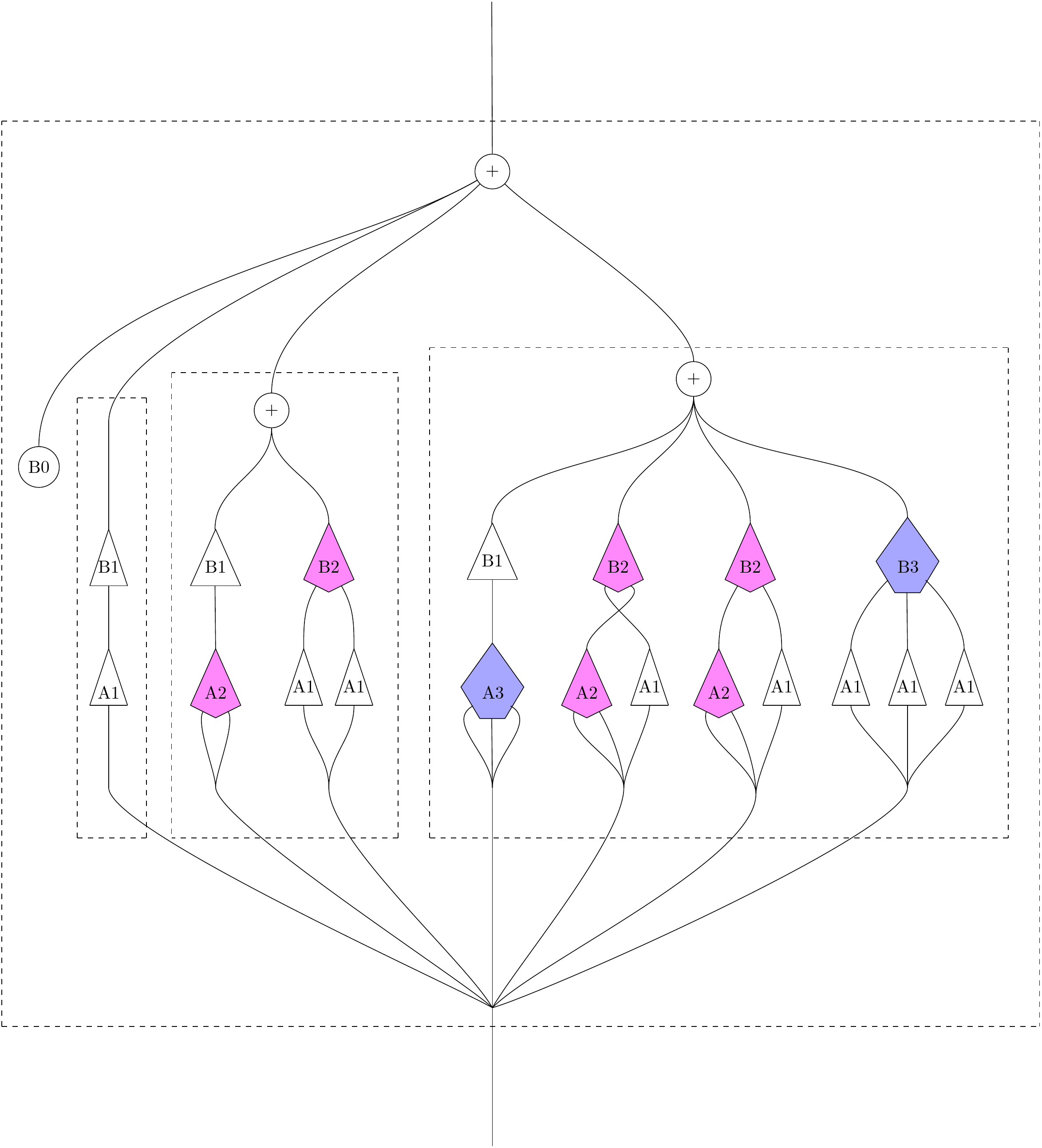}
\par\end{centering}
\centering{}\caption{Binary series composition of Volterra series. The interior dashed
boxes represent the composite homogeneous Volterra operators of the
series composition $B\triangleleft A$.\label{Binary series comp wiring diagram}}
\end{figure}

The composition operator, $\triangleleft$, is noncommutative, so
it induces a temporal ordering of its operands. The binary composition
of two series, up to order $3$ and with the $0^{\text{th}}-$order
operator $A_{0}$ assumed equal to $0$, is shown in Fig. \ref{Binary series comp wiring diagram}.\\

\subsection{Associativity of $\triangleleft$}

\begin{figure}[H]
\begin{centering}
\includegraphics[scale=0.17]{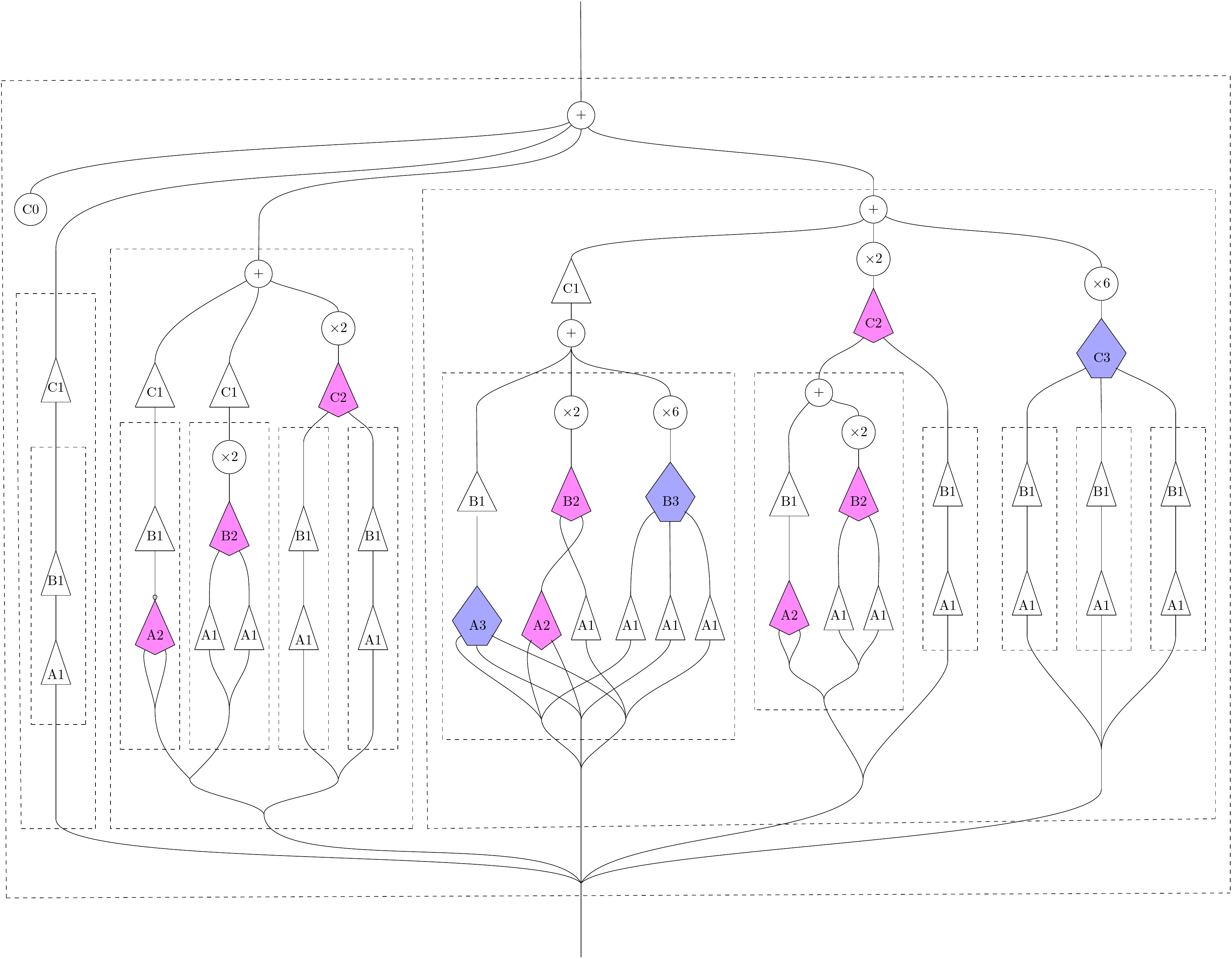}
\par\end{centering}
\centering{}\caption{Ternary Series Composition, type 1: $C\triangleleft(B\triangleleft A)$.
The interior-most dashed boxes represent composite homogeneous operators
of $B\triangleleft A$, which are composed first from the operators
of $B$ and $A$. The dashed boxes of the middle layer represent composite
homogeneous operators of $C\triangleleft(B\triangleleft A)$, which
are composed second from the operators of $C$ and $(B\triangleleft A)$.\label{Ternary-Series-Composition-Type1}}
\end{figure}

\begin{figure}[H]
\begin{centering}
\includegraphics[scale=0.15]{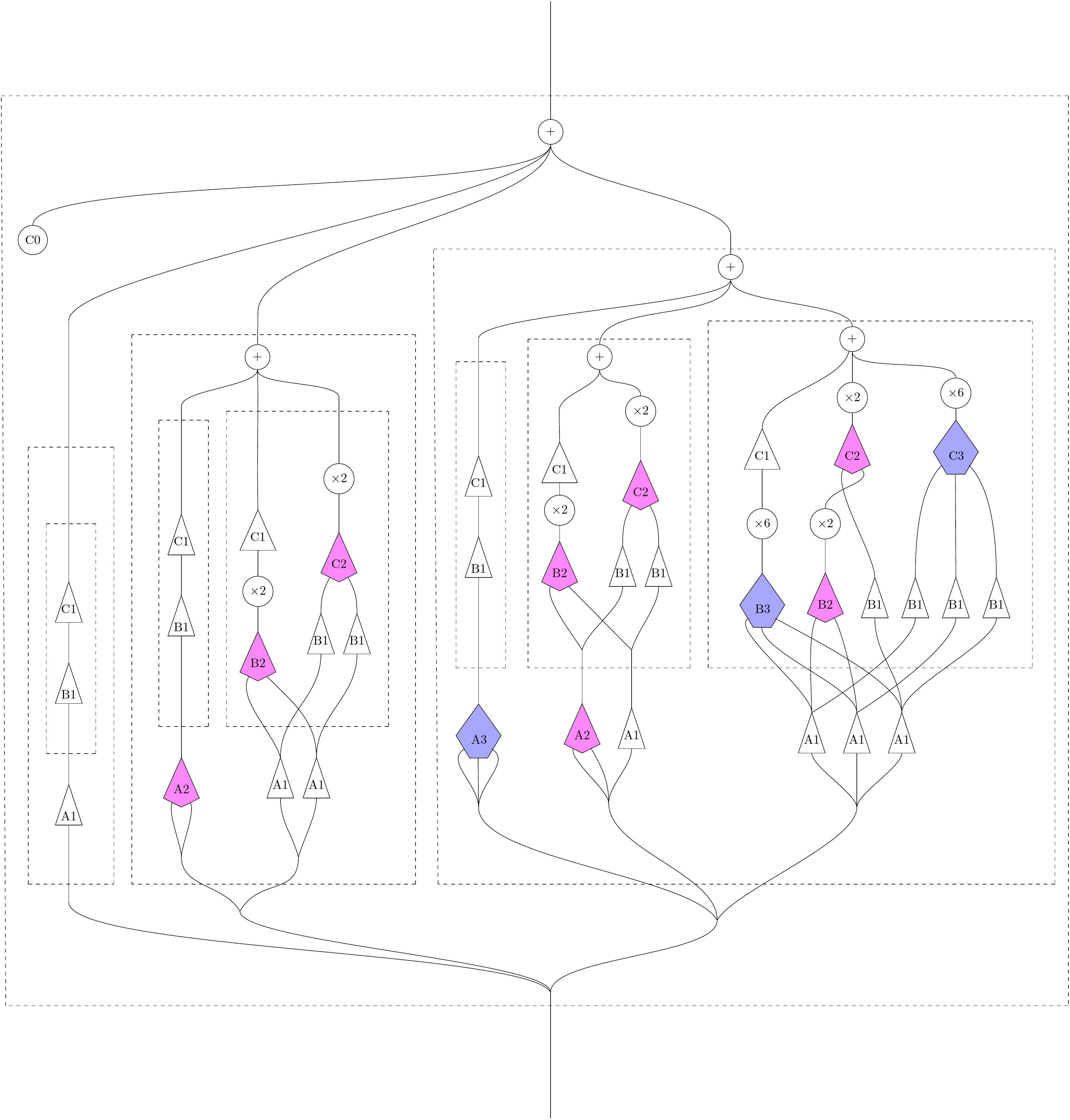}
\par\end{centering}
\centering{}\caption{Ternary Series Composition, type 2: $(C\triangleleft B)\triangleleft A$.
The interior-most dashed boxes represent the composite homogeneous
operators of $C\triangleleft B$, which are composed first from the
operators of $C$ and $B$. The dashed boxes of the middle layer represent
the composite homogeneous operators of ($C\triangleleft B)\triangleleft A$,
which are composed second from the operators of $(C\triangleleft B)$
and $A$.\label{Ternary-Series-Composition-Type2}}
\end{figure}

Given three Volterra series, $C$, $B$, and $A$, there are two possible
ways to use the binary composition rule to form the ternary composite
of the series, such that the output of $A$ flows as the input to
$B$ and the output of $B$ flows as the input to $C$; they are
\begin{equation}
C\triangleleft(B\triangleleft A)\label{eq:(type 1)}
\end{equation}
 and
\begin{equation}
(C\triangleleft B)\triangleleft A.\label{eq:(type 2)}
\end{equation}
Equationally, these two types of ternary composition correspond to
substituting eq. (\ref{binary composition (kernels form)}) into itself
either for the operators of $A$ (i.e., at the first~vertical level
in Fig. 4.2) or for the operators of $B$ (at the second vertical
level). The first of these forms of substitution is depicted in Fig.
\ref{Ternary-Series-Composition-Type1}, and the second in Fig. \ref{Ternary-Series-Composition-Type2}.

Associativity implies that Figs. \ref{Ternary-Series-Composition-Type1}
and \ref{Ternary-Series-Composition-Type2} must have the same number
of boxes of each order (color), interconnected in the same pattern
up to rearrangement; and yet, at first glance, they exhibit differences!
For example, they have different numbers of order 1 (white, triangular)
boxes. However, the following three assumptions and abuses of notation
have been made in the design of these diagrams, which account for
these variations. Firstly, constant multipliers appear wherever it
is possible to permute the inputs to a colored box that is receiving
its inputs from other homogeneous operators (i.e., excluding the first
level boxes, whose inputs are inputs to the entire system). Secondly,
certain wires in the second diagram split - a choice of visual presentation,
which was made to avoid clutter. Thirdly, the zeroth order terms of
the series $A$ and $B$ are assumed to be zero. This last simplifies
the combinatorics, changing from weak compositions to merely compositions,
and vastly reduces the number of non-trivial terms. Accounting for
these permutations and conflations, one arrives, as a quick sanity
check, at the following tally: that there are 90 white (unary) boxes;
17 magenta (dyadic) boxes; and 13 purple (triadic) boxes.

\medskip{}
In Appendix B, we prove that the two possible ways of composing three
Volterra series using $\triangleleft$ are equivalent. That is, we
prove\\
\\
\textbf{Theorem} 1: The series composition, $\triangleleft$, of Volterra
series is associative. I.e., 
\[
(c\triangleleft(b\triangleleft a))_{j}\cong((c\triangleleft b)\triangleleft a)_{j}
\]
for all $0\le j\le\infty$. \textbf{}\\
\textbf{}\\
\textbf{Proof}: See Appendix B.\\

At a high level, the proof of Theorem 1 proceeds by constructing a
bijection between:
\begin{enumerate}
\item The set of ways to compose the multivariate inputs, to each of the
homogeneous operators of $C$, from the outputs of the homogeneous
operators of the composite series $(B\triangleleft A)$;
\item The set of ways to compose the multivariate inputs, to each of the
homogeneous operators of the composite series $(C\triangleleft B)$,
from the outputs of the homogeneous operators of $A$.
\end{enumerate}
One can think of these two possibilities for composing Volterra series
in terms of stacking trees, as follows. Each $k^{\text{th}}-$order
homogeneous operator is like a tree with $k$ branches. In each of
the two cases, we are stacking trees (attaching the roots of some
trees to the leaves of others) so as to form a composite tree of height
(or depth) three that has $j-$many leaves\footnote{Thus, in the first of the two types of ternary composition, we are
stacking $A-$trees upon $B-$trees, and then stacking the $(B\triangleleft A)-$trees
upon $C-$trees; whereas, in the second type, we are stacking $B-$trees
upon $C-$trees, and then stacking $A-$trees upon the $(C\triangleleft B)-$trees.}. Theorem $1$ states that, for every order $j$, there is exactly
the same number of such trees that can be formed through the first
method above as there is through the second. The explicit bijection
we construct is the content of equations (\ref{bijection of the bs})
and (\ref{bijection of the as}), as well as of equation (\ref{equivalence of ternary identities}).
It defines, essentially, the action of the associator of Volt with
monoidal product the series composition product. The corresponding
monoidal unit is the (linear) identity Volterra series\emph{ }that
was described in Section $3.2.1.$, which has no effect when composed
with any other series.

There are many other monoidal products on \emph{Volt} that we did
not study. Nonetheless, each of the three products covered
in this chapter is fundamental, and together they allow us to describe
interconnected nonlinear systems in which the outputs of subsystems
are summed, multiplied, or fed into others, or back into themselves, as
inputs.

\typeout{} 
\chapter{Time-frequency signal processing using Volterra series}

The Fourier transform bridges between time and frequency, exchanging
a temporal representation, based on elements of an instantaneous
nature, for a spectral one, whose elements are eternal,
as they are periodic. Yet, we experience signals in the real world
as having content that is at once temporal and frequential. For example,
music and speech signals are perceived to have temporally-localized
spectral data, as in the frequencies of a tone that rings out for a finite
time. Another type of signal that contains local spectral information is
a chirp\footnote{A chirp is a signal of the form $ae^{i\phi(t)}$ where $\phi'(t)$
is everywhere nonnegative or everywhere nonpositive; i.e., whose instantaneous
frequency varies monotonically.}, whose Fourier transform informs us that, like an impulse, it contains
energy at every frequency (if it sweeps the entire frequency spectrum);
however, the information about \emph{when},
and for \emph{how long} each frequency is present, is encoded by the phase information, which
may appear difficult to interpret or even random. In signal processing terms, the
Fourier transform provides us with a picture of the signal that is
\emph{stationary}, but real signals have frequential\emph{ }content\emph{
}that is time-varying. They are evidently the products of nonstationary
processes.

The field of time-frequency analysis arose to address the problem
of the representation of signals jointly in time and frequency. Its
core methods allow one to identify where, jointly in time and frequency,
information in a signal is concentrated. Its modern formulation has
its origins both in signal analysis and quantum mechanics \cite{wigner_quantum_1932},
and is at the heart of audio signal processing. For a comprehensive
account of (bilinear) time-frequency analysis, see \cite{flandrin_time-frequencytime_1999,flandrin_explorations_2018,grochenig_foundations_2001}
and for a classical treatment \cite{cohen_time-frequency_1995}.

In this chapter, we introduce the bilinear time-frequency distributions
along with their higher-order generalizations, and show how they can
be embedded into \emph{Volt}, thus enabling their future treatment
using the results described in chapters 2, 3, and 4.

\section{Time and frequency}

Consider a (complex-valued) signal of the form 
\begin{equation}
x(t)=A(t)e^{i\varphi(t)},\label{eq:analytic signal}
\end{equation}
which is the product of a time-varying amplitude, $A(t)$, called
the carrier, and a modulation term, $e^{i\varphi(t)},$ whose argument,
$\varphi(t)$, is the phase\emph{.} Such a signal is called
\emph{analytic} if its energy is distributed purely in the positive
half of the frequency spectrum; this happens when $A(t)$ varies much
more slowly than $\varphi(t)$, in which case the derivative of the
phase can be interpreted as the \emph{instantaneous frequency }of
$x$ at time $t$ \cite{j_theorie_1948}:
\begin{equation}
f_{\text{inst}}(t)=\frac{1}{2\pi}\varphi^{'}(t).\label{eq:instantaneous frequency}
\end{equation}

Though, in practice, signals are real-valued, if $s$ is a real signal
then its complex analytic form can be obtained using the Hilbert transform,
$H$, as 
\begin{equation}
x(t)=s(t)+iH(s)(t)\label{eq:analytic signal via Hilbert transform}
\end{equation}
where $H(x)(t)=\frac{1}{\pi}\text{p.v.}\int_{\tau\in\mathbb{R}}\frac{u(\tau)}{t-\tau}d\tau$
and where $\text{p.v.}$ denotes the Cauchy principal value. Alternatively,
$x(t)$ can be defined in the spectral domain, as 
\[
\hat{x}(f)=\begin{cases}
2\cdot\hat{s}(f) & f>0\\
\hat{s}(f) & f=0\\
0 & f<0
\end{cases}.
\]

A natural way to represent the phase, $\varphi(t)$, is to let it
be a polynomial function of time,
\begin{equation}
\varphi(t)=\sum_{p=0}^{P}a_{p}t^{p}.\label{eq:polynomial phase}
\end{equation}
Thus, for example, if $\varphi(t)=\beta t+c$, then the frequency
is constant and the term $A(t)$ is modulated by a pure frequency
and shifted by a constant phase term. If, instead, $\varphi(t)=\alpha t^{2}+\beta t+c$,
then the signal includes a linear frequency modulation; and if $A(t)$
is a constant function, then the signal $x(t)$ is referred to as
a (unit-amplitude, linearly frequency-modulated, or quadratic phase)
chirp.\\

\paragraph{Energy distributions and representations\protect \\
}

Of all the possible types of transformations, a representation is
one that preserves the identity of any signal which it processes,
but which may present it in an equivalent form. The Fourier transform
is but one such example. A core property, important for physical interpretability,
that such a transformation must satisfy is that it preserve the signal
\emph{energy}. Parseval's theorem in Fourier analysis states that
this property holds for the Fourier transform, i.e.:
\begin{equation}
E_{x}=\int_{t\in\mathbb{R}}|x(t)|^{2}dt=\int_{f\in\mathbb{R}}|X(f)|^{2}df\label{eq:Energy of signal}
\end{equation}
where $E_{x}$ is the energy of $x$. In the context of time-frequency
distributions, which are \emph{bilinear}, we search more generally
for distributions of the form 
\[
\rho_{x}(t,f)=\frac{\partial^{2}E_{x}}{\partial t\partial f},
\]
and so, for energies 
\begin{equation}
E_{x}=\int_{t}\int_{f}\rho_{x}(t,f)dtdf\label{eq:Integration of the local enegy distribution}
\end{equation}
where $\rho_{x}$ is now the \emph{energy distribution }of $x$.\footnote{More generally, an energy distribution may be of the form 
\[
E_{x}=\int\int\rho_{x}(t,f)d\mu_{G}(t,f)
\]
where $d\mu_{G}$ is a measure associated to a group $G$ which acts
on the kernel of the transformation.} Such a distribution measures the energy density in both time \emph{and}
frequency, making it possible to locate where energy is present in
the signal as it is represented in the \emph{time-frequency plane}.

Distributions that preserve the signal energy may lose other forms
of information. The spectrogram, for example, is a quadratic energy-preserving
distribution that is obtained as the squared-magnitude of the STFT,
which completely loses the spectral phase information. The Wigner-Ville
distribution, which is in many respects the most important time-frequency
distribution, in contrast preserves the signal energy and is invertible
up to a constant phase factor\footnote{For the purpose of the developments in this chapter, a transformation
is a representation if what it produces is equivalent to its input
\emph{up to} a (linear, frequency-dependent) phase shift. This is
not too egregious, however, if the identity of a signal
is not meaningfully changed by such a transformation.}, but is corrupted by interferences for multi-component signals. Atomic
decompositions, which are linear representations, and amongst which
the STFT is fundamental, are representations in the stronger sense
of strict equality, yet they do not provide a measure of the local
energy density of the signal jointly in time and frequency (but such
densities can be obtained from them - as, e.g., the spectrogram is
from the STFT).

We search, therefore, for representations akin to a musical score,
capable of representing sequences of harmonies that are localized
to particular intervals of time. However, just as a musical score
provides a description of a piece of music in terms of information
that is localized discretely, in units of duration and notes in a
scale (or set of pitch classes\footnote{Consider the hypothetical case of a glissando that is written in a
musical score with no discrete measure of time; there a complete ambiguity.}), so, too, a time-frequency representation must navigate the fundamental
tradeoff of time- and frequency-resolution. This tradeoff can be stated
in a number of ways, but a clear description is given in a signal
processing context by the Balian-Low theorem\footnote{This statement, of the fundamental character of inverse proportionality
relating the variables of a Fourier pair (here, time and frequency),
is related to analogous statements about any other Fourier pair; in
particular, to position and momentum variables in quantum mechanics,
and it is thus connected to the so-called uncertainty principle. An
analog to the actual inequality that gives a quantitative measure
of the Heisenberg uncertainty principle in physics can be obtained
by taking the minimal possible time-frequency bandwidth of the product
of a signal with itself represented in the time and frequency domains.}. This theorem states that any family of elementary waveforms (called
atoms) capable of representing the time-varying spectral structure
of a signal, must have either infinite temporal or infinite spectral
bandwidth. Formally, for any family of time-frequency atoms $h(t-nt_{0})e^{i2\pi mt\nu_{0}}$
generated from a finite-energy window function $h\in L^{2}(\mathbb{R})$,
either\\
\begin{equation}
\int t^{2}|h(t)|^{2}dt=\infty\,\,\,\,\,\,\,\,\,\,\,\,\,\,\text{or \,\,\,\,\,\,\,\,\,\,\,\,\,\,\ensuremath{\int\nu^{2}|H(\nu)|^{2}d\nu=\infty}}.\label{eq:Temporal and spectral bandwidths}
\end{equation}
An example of this principle is furnished by taking the family of
atoms to the characteristic functions over finite interals, $h(t)=t_{0}^{-1/2}\boldsymbol{1}_{[0,t_{0}]}(t)$,
whose Fourier transforms are the cardinal sine functions
\[
\hat{h}(f)=t_{0}^{-1/2}e^{-\pi ift_{0}}\text{sinc}_{t_{0}}(t)
\]
\[
=t_{0}^{-1/2}e^{-\pi ift_{0}}\frac{\sin(\pi ft_{0})}{\pi ft_{0}};
\]
these atoms provide a finite temporal bandwidth, but their bandwidth
is analytic in the spectral domain. More generally, it is a fundamental
fact in Fourier analysis that for any Fourier pair, the bandwidths
in the two domains (here time and frequency) are inversely proportional
(see \cite{flandrin_time-frequencytime_1999} for a more in-depth
account).

Given the fundamental obstruction that prevents the simultaneous attainment
of optimal concentration in both time and frequency, much progress
in time-frequency analysis has revolved around the developement of
distributions that capture certain features of interest, but perhaps
not others; for example, which are positive (allowing for interpretability
as a probability density function), or optimally concentrated along
the instantaneous frequency for some class of signals\footnote{Various TFDs are classified in \cite{flandrin_time-frequencytime_1999}
according to these and various other properties that they do or do
not satisfy.}.

In section 5.1.1, we discuss the general form of time-frequency distributions,
and focus on a central member of this class, the Wigner-Ville Distribution,
which achieves optimal concentration along the instantaneous frequency
of linearly frequency-modulated signals. We then describe, in section
5.1.2, how to represent these bilinear TFDs as Volterra series. We
next show that the bilinear transformations cannot perfectly represent
the time-varying spectral information of higher-order signals (i.e.,
signals whose instantaneous frequency varies nonlinearly, such as
chirps of cubic or higher phase). This motivates the lift to so-called
polynomial time-frequency distributions (PTFD), which we study in
section 5.2.

\subsection{Bilinear time-frequency distributions}

The following is a review of the theory of bilinear time-frequency
distributions. It largely follows the treatment in \cite{flandrin_time-frequencytime_1999};
see there for more details.

In equation (\ref{eq:Integration of the local enegy distribution})
we defined the energy of a signal in terms of a distribution $\rho(t,\nu)$,
of two variables, one time and the other frequency. A general form
of the distribution $\rho$ for a signal $x$ is 
\begin{equation}
\rho_{x}(t,f)=\int\int K(u,u^{'};t,f)x(u)x^{*}(u^{'})du\,du^{'}\label{distribution of two variables}
\end{equation}
which is parameterized by a kernel, \emph{$K:\mathbb{R}^{4}\rightarrow\mathbb{R}$},
called the \emph{kernel }of the analysis. Amongst all distributions
of the form in (\ref{distribution of two variables}), the \emph{time-frequency
distributions} as those that are covariant with respect to time-frequency
shifts. A time-frequency shift is a linear transformation of the form
\begin{equation}
M_{\nu}T_{\tau}x(t)=x(t-\tau)e^{2\pi i\nu t}\label{eq:time-frequency shift}
\end{equation}
where the operator $T$ shifts the signal by a constant in time, and
the operator $M$ modulates (pointwise multiplies) it in the time
domain by a constant frequency--which is equivalent to convolution
by a delta function in the frequency domain (i.e., to a frequency
shift). It is a fundamental fact that time and frequency shifts do
not commute for all arguments $\tau$ and $\nu$
\begin{equation}
M_{\nu}T_{\tau}=e^{-2\pi i\tau\nu}T_{\tau}M_{\nu}.\label{eq:non-commutativity of time-frequency shifts}
\end{equation}
However, they \emph{do} commute when the product $\tau\nu$ is an
integer. The time-frequency shifts generate the non-commutative group
known as the \emph{Heisenberg-Weyl group}.

Imposing that $\rho$ be covariant with respect to time-frequency
shifts amounts to requiring that, for all signals $x$ and time-frequency
shifts $T_{\tau}M_{\nu}$, the diagram in Fig. 5 commutes,
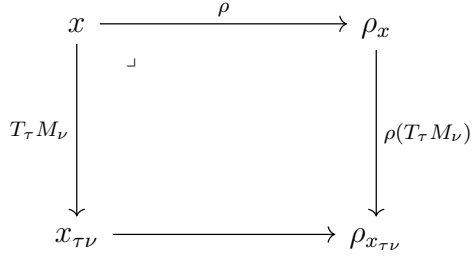
\begin{figure}[h]
\begin{center}
\[\begin{tikzcd}[ampersand replacement=\&] 	x \&\&\& {\rho_x} \\ 	\\ 	\\ 	{x_{\tau\nu}} \&\&\& {\rho_{x_{\tau\nu}}} 	\arrow["\rho", from=1-1, to=1-4] 	\arrow["{T_\tau M_\nu }"', from=1-1, to=4-1] 	\arrow[from=4-1, to=4-4] 	\arrow["{\rho(T_\tau M_\nu )}", from=1-4, to=4-4] 	\arrow["\lrcorner"{anchor=center, pos=0.125}, draw=none, from=1-1, to=4-4] \end{tikzcd}\]
\par\end{center}\caption{Covariance of time-frequency distributions with respect to time-frequency
shifts.}
\end{figure}
which fact we denote by the right angle symbol in the upper left-hand
corner\footnote{Note the similarity of this covariance diagram to the one in Section
2.3, depicting the condition of functoriality for Volterra series.
Indeed, we will see in Section 5.1.2 that TFDs are Volterra series.}. The action of the operator $\rho$ on the linear transformation
of time-frequency shifting must equivalently be given by precomposition,
$\rho(T_{\tau}M_{\nu})=\rho\circ T_{\tau}M_{\nu}$, which amounts
in this case to shifting the kernel of the distribution. Thus, we
have the following set of equivalences \cite{flandrin_time-frequencytime_1999}
\[
((\rho\circ T_{t^{'}}M_{f^{'}})(x))(t,f)=\rho_{x_{t^{'}f^{'}}}(t,f)=\rho(T_{t^{'}}M_{f^{'}})\rho_{x}(t,f)=\rho_{x}(t-t^{'},f-f^{'})
\]
where 
\[
\rho_{x_{t^{'}f^{'}}}(t,f)=\int\int K(u+t^{'},u^{'}+t^{'};t,f)e^{2\pi if^{'}(u-u^{'})}x(u)x^{*}(u^{'})du\,du^{'}
\]
and
\[
\rho_{x}(t-t^{'},f-f^{'})=\int\int K(u,u^{'};t-t^{'},f-f^{'})x(u)x^{*}(u^{'})du\,du^{'}
\]
and so, the representing kernel must satisfy
\[
K(u,u^{'};t-t^{'},f-f^{'})=K(u+t^{'},u^{'}+t^{'};t,f)e^{2\pi if^{'}(u-u^{'})}.
\]
Reorganizing, and performing a change of variables, one obtains
\begin{equation}
K(u,u^{'};t,f)=K(u+t,u^{'}+t;0,0)e^{2\pi if(u-u^{'})}.\label{bilinear kernel, time-frequency shift covariant}
\end{equation}
This says that time-frequency shifting the kernel function along both
the time and frequency axes and then evaluating at the origin is equivalent
to evaluating it at the point which is indexed by the two extents
of the shift. Equation \ref{distribution of two variables} for the
time-frequency shift-covariant distributions can now be rewritten
as
\[
\rho_{x}(t,f)=\int\int K(s-t+\frac{\tau}{2},s-t-\frac{\tau}{2};0,0)x(s+\frac{\tau}{2})x^{*}(s-\frac{\tau}{2})e^{-i2\pi f\tau}d\tau\,ds
\]
\begin{equation}
\rho_{x}(t,f)=\int\int\Pi(s-t,\xi-f)W_{x}(s,\xi)ds\,d\xi\label{eq:TFD as 2D smoothing of WVD}
\end{equation}
where 
\begin{equation}
W_{x}(s,f)=\int x(s+\frac{\tau}{2})x^{*}(s-\frac{\tau}{2})e^{-i2\pi f\tau}d\tau\label{WVD}
\end{equation}
is known as the \emph{Wigner-Ville Distribution }(WVD) of the signal
$x$, and 
\[
\Pi(t,f)=\int K(t+\frac{\tau}{2},t-\frac{\tau}{2};0,0)e^{2\pi if\tau}d\tau
\]
is the inverse Fourier transform of the kernel.

A time-frequency distribution is thus a $2-$D convolution of the
WVD of the signal with the inverse Fourier transform of the distribution
kernel. The general form (\ref{eq:TFD as 2D smoothing of WVD}) can
be rewritten more explicitly as 
\begin{equation}
C_{x}(t,f;g)=\int\int\int e^{i2\pi\xi(s-t)}\phi(\xi,\tau)x(s+\frac{\tau}{2})x^{*}(s-\frac{\tau}{2})e^{-i2\pi f\tau}d\xi\,ds\,d\tau\label{eq:Cohen's class}
\end{equation}
where 
\[
\phi(\xi,\tau)=\int\int\Pi(t,f)e^{-i2\pi(f\tau+\xi t)}dt\,df
\]
is an arbitrary parameter function. The form (\ref{eq:Cohen's class})
is known as Cohen's class of time-frequency distributions, and gives
precisely the class of all bilinear distributions that are covariant
with respect to time-frequency shifts. Further requiring conservation
of energy 
\[
\int\int K(u-t,u^{'}-t;0,0)e^{-i2\pi f(u-u^{'})}dt\,df=\delta(u-u^{'}),
\]
implies that
\[
\int\int\Pi(t,f)dt\,df=1,
\]
and so $\phi(0,0)=1$.

In (\ref{eq:Cohen's class}), the parameter function $\phi$ determines
what, if any, additional features there are of the distribution kernel.
If $\phi(\xi,\tau)=1,$ then we recover the WVD. If, instead, $\phi(\xi,\tau)=e^{2\pi i\xi\tau}$,
then we have the \emph{Rihaczek }distribution.\footnote{The Rihaczek distribution has been used as the starting point of a
time-frequency analogous approach to graph signal processing (GSP)
called vertex-frequency graph signal processing (VFGSP), as well as
in the time-frequency analysis of non-commutative groups \cite{turunen_time-frequency_2020}.} The spectrogram, probably the most widely used TFD and the only one
which is positive-definite (and so interpretable as a probability
density function), is recovered by setting $\phi(\xi,\tau)=A_{h}^{*}(\xi,\tau)$,
where $A_{h}(\xi,\tau)$ is the ambiguity function \cite{flandrin_time-frequencytime_1999}
of the short-time Fourier transform window $h$,
\begin{equation}
A_{h}(\xi,\tau)=\int h(s+\frac{\tau}{2})h^{*}(s-\frac{\tau}{2})e^{-i2\pi\xi s}ds.\label{Ambiguity function of STFT window}
\end{equation}

We next turn to the WVD, which serves as a kind of prototype TFD from
which all the others can be derived.

\subsubsection{The role and properties of the Wigner-Ville Distribution}

The WVD was originally developed in a quantum mechanical context by
Wigner, who sought a joint position-momentum distribution to use in
order to calculate deviations from equilibrium of a virial gas \cite{wigner_quantum_1932}.
Although intended to function as a probability density function, it
exhibits a feature which prevents its interpretation as such; namely,
oscillatory interferences that are generated between the components
of any multicomponent signal. Thus, the WVD is not, in general, everywhere
positive\footnote{The WVD \emph{is} everywhere positive, however, if (and only if) the
input is a generalized Gaussian \cite{hudson_when_1974}.}. However, in other respects the WVD is ideal. In particular, it exhibits
the property of being optimally concentrated around the instantaneous
frequency for linearly FM signals; i.e., for any pair $(t,f)$, $W_{z}(t,f)=\delta(f-f_{\text{inst}}(t))$.\footnote{Note that it is essential here that $z$ be analytic; see the article
\cite{boashash_note_1988} for justification of this requirement,
and \cite{cohen_time-frequency_1995,j_theorie_1948} for comprehensive
introductions to the concept of analytic, or complex, signal.}

As shown, given a signal $z$, the WVD is the Fourier transform, with
respect to $\tau$, of the time-shifted product of $z$ and its complex
conjugate. Its mechanism of action can be understood as follows \cite{boashash_higher-order_1995}.
Let $z(t)$ be an analytic signal of unit magnitude, $z(t)=e^{i\varphi(t)}$.
Then 
\[
W_{z}(t,f)=f\overset{F}{\leftrightarrow}\tau(K_{z}(t,\tau)),
\]
where $f\overset{F}{\leftrightarrow}\tau$ denotes the partial Fourier
transform which exchanges the $f,\tau$ variables, and $K_{z}(t,\tau)=z(t+\frac{\tau}{2})z^{*}(t-\frac{\tau}{2})=e^{i(\varphi(t+\frac{\tau}{2})-\varphi(t-\frac{\tau}{2}))}$.
Recall that the instantaneous frequency (IF) of an analytic signal
$e^{i\varphi(t)}$ is $f_{\text{inst}}(t)=\frac{1}{2\pi}\varphi^{'}(t)$,
which can be written 
\[
f_{\text{inst}}(t)=\frac{1}{2\pi}\lim_{\tau\rightarrow0}\frac{\varphi(t+\tau/2)-\varphi(t-\tau/2)}{\tau},
\]
where 
\begin{equation}
2\pi\tau\hat{f}_{\text{inst}}(t,\tau)=\varphi(t+\tau/2)-\varphi(t-\tau/2)\label{CFD IF estimator}
\end{equation}
is the (linear) \emph{central finite difference }(CFD)\emph{ }approximation
to the instantaneous frequency. We can then note that the values which
the WVD takes at points $(t,f)$ in the time-frequency plane are computed
using CFD estimates to the phase, scaled by the squared magnitude
of the signal.

Because the CFD is a linear estimator, we expect the WVD to accurately
measure the IF of signals whose frequency is varying linearly. Indeed,
letting $\varphi(t)=\alpha t^{2}+\beta t$, we have that 
\[
W_{z}(t,f)=\int e^{i(\alpha(t+\tau/2)^{2}+\beta(t+\tau/2))}e^{-i(\alpha(t-\tau/2)^{2}+\beta(t-\tau/2))}e^{-i2\pi\tau f}d\tau
\]
\[
=\int e^{i(2\alpha t\tau+\beta\tau)}e^{-i2\pi\tau f}d\tau
\]
\[
=\int e^{-i2\pi\tau(f-\frac{1}{2\pi}(2\alpha t+\beta))}d\tau
\]
\begin{equation}
=\delta(f-\frac{1}{2\pi}(2\alpha t+\beta)).\label{WVD of quadratic-phase chirp}
\end{equation}

As mentioned, the WVD of a linear combination of signals is not the
superposition of the component signals, oweing to its bilinear nature.
For a signal $z=z_{1}+z_{2}$ which is the sum of two terms, $W_{z}=W_{11}+W_{12}+W_{21}+W_{22}$,
where 
\begin{equation}
W_{12}(t,f)=\int z_{1}^{*}(t-\frac{1}{2}\tau)z_{2}(t+\frac{1}{2}\tau)e^{-2\pi i\tau f}d\tau\label{cross WVD}
\end{equation}
is called the \emph{cross Wigner distribution}. Though complex-valued,
by Hermitian symmetry we have that $W_{12}=W_{21}^{*}$, and so 
\[
W_{z}=W_{11}+W_{22}+2\Re\{W_{12}\}.
\]
More generally, for a signal $b$ which is a linear combination of
signals, $b(t)=\stackrel[n=1]{N}{\sum}b_{n}x_{n}(t)$, the WVD is
of the form 
\begin{equation}
W_{b}(t,f)=\stackrel[n=1]{N}{\sum}|b_{n}|^{2}W_{x_{n}}(t,f)+2\stackrel[n=1]{N-1}{\sum}\stackrel[k=n+1]{N}{\sum}\Re\{b_{n}b_{k}^{*}W_{x_{n}x_{k}}(t,f)\}.\label{WVD of atomic decomposition}
\end{equation}

Much work has gone into reducing the interference terms of the WVD.
For example, the parameter function $\phi$ in Cohen's class may be
used as an averaging kernel in order to smooth out the interferences
that are oscillatory \cite{flandrin_time-frequencytime_1999}. However,
although the interference terms ruin the viability of interpretation
of the WVD as a probability density function, they exhibit remarkable
combinatorial and geometric structure. In particular, they are located
exactly half-way along the line connecting the pairs of interacting
components in the input signal, are oriented in the direction perpendicular
to the line, and have a frequency depending on the distance between
the components.

\subsection{Volterra series representation of Cohen's class}

In \cite{nam_volterra_2003}, Powers and Nam showed that a class of
double second-order Volterra series are equivalent to Cohen's class
of bilinear TFDs. Their result was motivated by the following observations:
\begin{enumerate}
\item Cohen's class of time-frequency distributions has the form of a bilinear
transformation wherein the input signal appears twice; and the $2^{\text{nd}}$-order
(multivariate) Volterra series is a bilinear transformation of the
same form.
\item The expression of Cohen's class is parameterized in terms of a kernel
function of two variables; and the $2^{\text{nd}}$-order Volterra
series is also parameterized in terms of such a kernel.
\item All the members of Cohen's class of bilinear distributions can be
obtained from the WVD via a 2D convolution; and the $2^{\text{nd}}$-order
Volterra series performs a 2D convolution.
\end{enumerate}
We next recapitulate their result.

\paragraph{{\small{}Second-order double Volterra series}}

Let $V$ be a Volterra series with two inputs, $x_{a}$ and $x_{b}$
, and all kernel functions zero except for $h_{2}$, 
\begin{equation}
y(t)=H_{2}[x_{u},x_{v}]=\int\int h_{2}(u,v)\cdot x_{a}(t-u)x_{b}(t-v)du\,dv.\label{eq:double, second-order Volterra series (non-parameterized kernel)}
\end{equation}
This was referred to in \cite{nam_volterra_2003,rice_volterra_1973}
as a \emph{double, second-order VS}, and further generalized by allowing
the kernel function to be parameterized by a scalar variable $\theta$,
as in
\begin{equation}
y(t,\theta)=\int\int h_{2,\theta}(u,v)\cdot x_{a}(t-u)x_{b}(t-v)du\,dv.\label{double Volterra series, parameterized kernel}
\end{equation}
From the perspective of our framework, this could also be modeled
as a family of Volterra series that are generated by a morphism which
is \emph{parameterized }by the variable $\theta$; i.e., $y(t,\theta)=\alpha_{\theta}(H_{2}(x_{u},x_{v}))(t)$,
where $\alpha$ is a frequency-shift morphism, as described in section
3.2.3. (However, for the sake of consistency with respect to the notation
of the articles we're referring to, we preserve the original notation.)

By the time-shift properties of the Fourier transform, (\ref{double Volterra series, parameterized kernel})
can be rewritten 
\[
y(t,\theta)=\int\int\int\int h_{2,\theta}(u,v)\cdot e^{2\pi i(f_{1}+f_{2})t}\cdot e^{-2\pi i(f_{1}u+f_{2}v)}\cdot X_{a}(f_{1})X_{b}(f_{2})df_{1}df_{2}dudv
\]
\begin{equation}
=\int\int H_{2,\theta}(f_{1},f_{2})\cdot e^{2\pi i(f_{1}+f_{2})t}\cdot X_{a}(f_{1})X_{b}(f_{2})df_{1}df_{2}\label{eq:40}
\end{equation}
where $H_{2,\theta}$ is the $2-$D Fourier transform of the kernel
function $h_{2,\theta}$. In the case where the two inputs $x_{a},x_{b}$
are given by
\[
\begin{array}{c}
x_{a}(t-u)=s^{*}(t-u)\\
\\
x_{b}(t-v)=s(t-v),
\end{array}
\]
then we have that $X_{a}(f_{1})=S^{*}(-f_{1})$ and $X_{b}(f_{2})=S(f_{2})$,
so (\ref{eq:40}) becomes 
\[
y(t,\theta)=\int\int H_{2,\theta}(f_{1},f_{2})\cdot e^{2\pi i(f_{1}+f_{2})t}\cdot S^{*}(-f_{1})S(f_{2})df_{1}df_{2}.
\]

Now, recall Cohen's class of TFDs. Using once again the time-shift
properties of the Fourier transform, (\ref{eq:Cohen's class}) can
be rewritten 
\[
C(t,f)=\int\int\int\delta(\xi+f_{1}+f_{2})\cdot e^{-2\pi i\xi t}\cdot\Phi_{2}(\xi,(f+\frac{1}{2}f_{1}-\frac{1}{2}f_{2}))d\xi\cdot S^{*}(-f_{1})S(f_{2})df_{1}df_{2}
\]
\[
=\int\int\Phi_{2}(-f_{1}-f_{2},(f+\frac{1}{2}f_{1}-\frac{1}{2}f_{2}))\cdot e^{2\pi i(f_{1}+f_{2})t}\cdot S^{*}(-f_{1})S(f_{2})df_{1}df_{2},
\]
where $\Phi_{2}(\xi,\alpha)=\int\phi(\xi,\tau)\cdot e^{-2\pi i\alpha\tau}d\tau$,
and $\phi$ is an arbitrary parameter function, as in (\ref{eq:Cohen's class}).
It can therefore be seen that Cohen's class is precisely the class
of double, $2^{\text{nd}}-$order Volterra series with kernel function
of the form
\begin{equation}
\begin{array}{c}
H_{2,f}(f_{1},f_{2})\\
\\
=\Phi_{2}(-f_{1}-f_{2},(f+\frac{1}{2}f_{1}-\frac{1}{2}f_{2}))\\
\\
=\int\phi(-f_{1}-f_{2},\tau)\cdot e^{-2\pi i(f+(1/2)f_{1}-(1/2)f_{2})\tau}d\tau.
\end{array}\label{41}
\end{equation}

Note further that equation (\ref{41}) is equivalent to
\[
H_{2,f}(f_{1},f_{2})=\int\int\delta(\xi+f_{1}+f_{2})\phi(\xi,\tau)\cdot e^{-2\pi i(f+(1/2)f_{1}-(1/2)f_{2})\tau}d\tau d\xi.
\]
Thus, $h_{2,f}(u,v)$ can be obtained via a $2-$D inverse Fourier
transform, as 
\[
h_{2,f}(u,v)=\int\int\phi(\xi,\tau)\cdot e^{-2\pi if\tau}\cdot\int\int\delta(\xi+f_{1}+f_{2})\cdot e^{2\pi i[(u-\frac{1}{2}\tau)f_{1}+(v+\frac{1}{2}\tau)f_{2}]}df_{1}df_{2}d\tau d\xi
\]
\[
=\int\int\phi(\xi,\tau)\cdot e^{-2\pi if\tau}\cdot\delta(u-v-\tau)\cdot e^{-2\pi i(u-\frac{1}{2}\tau)\xi}d\tau d\xi
\]
\begin{equation}
=e^{-2\pi if(u-v)}\cdot\int\phi(\xi,u-v)\cdot e^{\pi i(u+v)\xi}d\xi.\label{eq:42}
\end{equation}
Cohen's class can thus be described in Volterra series form as
\[
C(t,f)=\int\int[h_{2,f}(u,v)]\cdot s^{*}(t-u)\cdot s(t-v)dudv
\]
\[
\int\int H_{2,f}(f_{1},f_{2})\cdot e^{2\pi i(f_{1}+f_{2})t}\cdot S^{*}(-f_{1})S(f_{2})df_{1}df_{2}
\]
 where $H_{2,f}(f_{1},f_{2})$ and $h_{2,f}(u,v)$ are as given in
(\ref{41}) and (\ref{eq:42}), respectively.

The WVD is a core member of Cohen's class; it is obtained when $\phi(\xi,\tau)=1$,
in which case the time- and frequency-domain kernels become 
\[
h_{2,f}(u,v)=e^{-2\pi if(u-v)}\delta(\frac{u+v}{2})=2e^{-4\pi ifu}\delta(u+v)
\]
and 
\[
H_{2,f}(f_{1},f_{2})=\delta(f+\frac{1}{2}f_{1}-\frac{1}{2}f_{2}),
\]
respectively. The Volterra series representation of the WVD can, in
a sense, be read off from the standard definition by recognizing that
the kernel function $h_{2,f}$ must comprise both the Fourier kernel
and a delta functions that annihilates all but the symmetrically time-shifted
copies of the input. The Volterra series representation of the WVD
is thus 
\begin{equation}
W(t,f)=\int\int2e^{-2\pi if(u-v)}\delta(u+v)\cdot s^{*}(t-u)\cdot s(t-v)dudv\label{eq:VS representation of the WVD}
\end{equation}
\[
=\int\int\delta(f+\frac{1}{2}f_{1}-\frac{1}{2}f_{2})\cdot e^{2\pi i(f_{1}+f_{2})t}\cdot S^{*}(-f_{1})S(f_{2})df_{1}df_{2}.
\]

\section{Higher-order and polynomial time-frequency distributions}

If the phase of a signal of the form (\ref{eq:analytic signal}) is
not a polynomial of order two or lower, then the WVD is not optimally
concentrated along the signal's instantaneous frequency law; this
is because the CFD estimate (\ref{CFD IF estimator}) of the phase
is linear. For example, in the case of a cubic chirp signal, $s(t)=e^{i\varphi(t)}$,
with $\varphi(t)=\gamma t^{3}/3+\beta t^{2}/2+\omega_{0}t$ and so
$\varphi^{'}(t)=\gamma t^{2}+\beta t+\omega_{o}$ , calculation of
the WVD kernel results (\cite{cohen_time-frequency_1995}, example
8.6) in 
\[
-\varphi(t-\frac{1}{2}\tau)+\varphi(t+\frac{1}{2}\tau)=\gamma\tau^{3}/12+(\gamma t^{2}+\beta t+\omega_{0})\tau
\]
\[
=\gamma\tau^{3}/12+\varphi^{'}(t)\tau
\]
from which it follows that 
\[
\begin{array}{c}
W_{z}(t,f)=\frac{1}{2\pi}\int e^{i\gamma\tau^{3}/12+i(\varphi^{'}(t)-\omega)\tau}d\tau\\
\\
=\frac{2}{2\pi}(\frac{4}{\gamma})^{1/3}\int\cos[\gamma\tau^{3}/12+(\varphi^{'}(t)-\omega)\tau]d\tau\\
\\
=(\frac{4}{\gamma})^{1/3}\text{Ai}((\frac{4}{\gamma})^{1/3}[\varphi^{'}(t)-\omega])
\end{array}
\]
where $\text{Ai}(x)$ is the Airy function $\text{Ai}(x)=\frac{1}{\pi}\int_{0}^{\infty}\cos(u^{3}/3+xu)du$.
What is more, no bilinear TFD can overcome this limitation, since
they are all obtained from the WVD by the operation of convolution.
This motivates the search for higher-order time-frequency representations,
capable of modeling nonstationary signals of a more strongly nonlinear
nature.

\subsection{The Polynomial Wigner-Ville Distribution}

In \cite{boashash_higher-order_1995,boashash_polynomial_1994,boashash_polynomial_1996,boashash_polynomial_1998,ristic_relationship_1996},
higher-order analogues of the WVD were described, generalizing the
WVD to the `time-multifrequency' (or `multi-time, multi-frequency')
setting. A core member of these transforms is the Higher-Order Wigner-Ville
Distribution (HOWVD), defined by
\begin{equation}
W_{x}^{(k)}(t,\boldsymbol{f})=\int_{\boldsymbol{\tau}\in\mathbb{R}^{k-1}}x^{*}(t-\alpha_{k})\prod_{r=1}^{k-1}P_{r+1}\{x(t+\tau_{r}-\alpha_{k})\}e^{-i2\pi\boldsymbol{f}^{T}\boldsymbol{\tau}}d\boldsymbol{\tau}\label{eq:HOWVD}
\end{equation}
where $\boldsymbol{f}\in\mathbb{R}^{k-1}$, $P_{r}(x)=\begin{cases}
x^{*} & r\text{ odd}\\
x & \text{else}
\end{cases}$, $\boldsymbol{\tau}=\{\tau_{1},\dots,\tau_{k-1}\}$, and $\alpha_{k}=\frac{1}{k}\sum_{r=1}^{k-1}\tau_{r}$.
It is a straightforward extension of the WVD to higher dimensions.

It was shown in \cite{ristic_relationship_1996} that any HOWVD uniquely
determines a two-dimensional distribution, called the \emph{Polynomial
Wigner-Ville Distribution} (PWVD), which is of the form
\begin{equation}
W_{x}^{(k)}(t,f)=\int_{\tau}x^{*}(t+\lambda_{1}\tau)\prod_{r=1}^{k-1}P_{r+1}\{x(t+\lambda_{r+1}\tau)\}e^{-i2\pi f\tau}d\tau\label{eq:PWVD}
\end{equation}
where $k$ is even and the coefficients $\lambda_{r},$ $r=1,\dots,k$
satisfy a set of nonlinear equations. It is obtained from the HOWVD
as a projection from multifrequency to one-dimensional frequency space,
with projection axis 
\[
f=\frac{f_{r}}{\lambda_{r+1}+\sum_{l=1}^{k-1}\lambda_{l+1}},\,\,(r=1,\dots,k-1)
\]
and corresponding slice axis
\begin{equation}
\tau_{r}=(\lambda_{r+1}+\sum_{l=1}^{k-1}\lambda_{l+1})\tau\,\,(r=1,\dots,k-1).\label{slice axis, HOWVD to PWVD-1}
\end{equation}
 The PWVD was also used to define a `higher Cohen's class' of TFDs
\cite{boashash_polynomial_1994}.

Like the WVD, the PWVD is a function of time and frequency, but unlike
the WVD it exhibits optimal concentration along the IF for signals
whose phase is an arbitrary polynomial of order less than or equal
to that of the distribution\footnote{The augmented representational power comes, however, at the expense
of a proliferation of interference terms. Specifically, for a multicomponent
signal $z(t)=\sum_{r=1}^{M}z_{r}(t)$ with $M$ components, the HOWVD
and PWVD generate $C_{k}$-many interference terms, where
\[
C_{k}=\sum_{r=1}^{k-1}\binom{k}{r}=2^{k}-2
\]
Furthermore, not all of these are oscillatory, as was the case with
the WVD, and so can no longer be smoothed away.}. That is, for a constant-amplitude analytic signal of the form $z(t)=Ae^{i\varphi(t)}$,
where $\varphi(t)$ is a polynomial of order $P$, $\varphi(t)=\sum_{r=0}^{P}a_{r}t^{r}$,
then for a certain selection of the coefficients $\lambda_{r}$, we
have that 
\begin{equation}
W_{z}^{(k)}(t,f)=A^{k}\delta(f-\frac{1}{2\pi}\sum_{m=1}^{P}ma_{m}t^{m-1})\label{eq:representation of polynomial phase IF}
\end{equation}
where $A$ is the constant amplitude of the analytic signal. Note
that eq. (\ref{eq:representation of polynomial phase IF}) implies
that
\begin{equation}
\arg\{\prod_{r=1}^{k/2}z(t+\lambda_{r}\tau)z^{*}(t+\lambda_{-r}\tau)\}=\tau\sum_{m=1}^{p}ma_{m}t^{m-1}.\label{PWVD derivation of optimality}
\end{equation}
Substituting $Ae^{i\varphi(t)}$ for $z$ in the left-hand side of
(\ref{PWVD derivation of optimality}), and letting $t=0$, we arrive
at
\[
\arg\{\prod_{r=1}^{k/2}e^{i\varphi(t+\lambda_{r}\tau)-i\varphi(t+\lambda_{-r}\tau)}\}=\arg\{\prod_{r=1}^{k/2}e^{i(\sum_{m=0}^{P}a_{m}(\lambda_{l}\tau)^{m}-\sum_{m=0}^{P}a_{m}(\lambda_{-l}\tau)^{m})}\},
\]
and so 
\begin{equation}
\sum_{m=0}^{p}\sum_{r=1}^{k/2}a_{m}[\lambda_{r}^{m}+(-1)\lambda_{-r}^{m}]=0\cdot a_{0}+1\cdot a_{1}+0\cdot a_{2}+...+0\cdot a_{p}.\label{eq:equivalence for polynomial phase at t=00003D0}
\end{equation}

Equation (\ref{eq:equivalence for polynomial phase at t=00003D0})
implies that the property of concentration along the IF of a polynomial
phase signal requires that the coefficients $\lambda_{r}$ satisfy
a nonlinear set of equations, namely:
\[
\sum_{r=1}^{k/2}(\lambda_{r}-\lambda_{-r})=1
\]
and
\[
\sum_{r=1}^{k/2}(\lambda_{r}^{m}-\lambda_{-r}^{m})=0,\,\,\,m=2,...,p;
\]
or, equivalently,\\
\begin{equation}
\begin{array}{c}
\lambda_{r}=-\lambda_{r-1}\,\,\,(r=1,...,k/2)\\
\\
\stackrel[r=1]{k/2}{\sum}\lambda_{r}=1/2\\
\\
\stackrel[r=1]{k/2}{\sum}\lambda_{r}^{m}=0
\end{array}\label{constraint equations PWVD}
\end{equation}
for $m\text{ odd};m=3,...,p$.

The equations (\ref{constraint equations PWVD}) have no exact, real
solutions for $p=3$ or $p=4$ for $k\le4$, nor for $p=6,8,...,$
by the Abel-Ruffini theorem. For $p=3$ or $p=4$ with $k=6$, there
is an infinite family of solutions given by 
\begin{equation}
\begin{array}{c}
\lambda_{1}=\frac{1}{4}-\frac{\lambda_{3}}{2}\pm\frac{\sqrt{3}\sqrt{24\lambda_{3}^{3}+12\lambda_{3}^{2}-6\lambda_{3}+1}}{12\sqrt{2\lambda_{3}-1}}\\
\\
\lambda_{2}=\frac{1}{4}-\frac{\lambda_{3}}{2}\mp\frac{\sqrt{3}\sqrt{24\lambda_{3}^{3}+12\lambda_{3}^{2}-6\lambda_{3}+1}}{12\sqrt{2\lambda_{3}-1}},
\end{array}\label{crazy formulae}
\end{equation}
where, in order that $\lambda_{1}$ and $\lambda_{2}$ exist and are
real, we must have that $\lambda_{3}\ge\frac{1}{2}$ \cite{boashash_higher-order_1995}.

\paragraph{{\small{}Proof of formula (\ref{crazy formulae}) $\text{[Depalle]}$:}}

Denote by $s=\lambda_{1}+\lambda_{2}=\frac{1}{2}-\lambda_{3}$ and
$p=\lambda_{1}\lambda_{2}=\frac{1}{3}(\frac{s^{3}+\lambda_{3}^{3}}{s})$
the product and sum of the roots $\lambda_{1}$ and $\lambda_{2}$.
Then they are solutions to $x^{2}-sx+p=0$, and so are of the form
\[
s^{2}-4p
\]
\[
=s^{2}-\frac{4}{3}(\frac{s^{3}+\lambda_{3}^{3}}{s})=-\frac{1}{3}(\frac{s^{3}+4\lambda_{3}^{3}}{s})
\]
\[
=-\frac{1}{3}\left(\frac{(\frac{1}{2}-\lambda_{3})^{3}+4\lambda_{3}^{3}}{\frac{1}{2}-\lambda_{3}}\right)=-\frac{1}{3}\left(\frac{(1-2\lambda_{3})^{3}+32\lambda_{3}^{3}}{4(1-2\lambda_{3})}\right)
\]
\[
=-\frac{1}{12}\left(\frac{1-6\lambda_{3}+12\lambda_{3}^{2}-8\lambda_{3}^{3}+32\lambda_{3}^{3}}{1-2\lambda_{3}}\right)=\frac{24\lambda_{3}^{3}+12\lambda_{3}^{2}-6\lambda_{3}+1}{12(2\lambda_{3}-1)}
\]
from which it follows that
\[
\lambda_{1,2}=\frac{1}{4}-\frac{\lambda_{3}}{2}\pm\sqrt{3}\cdot\frac{\sqrt{24\lambda_{3}^{3}+12\lambda_{3}^{2}-6\lambda_{3}+1}}{12\sqrt{(2\lambda_{3}-1)}}.
\]

Some properties of the PWVD were reported in \cite{boashash_higher-order_1995};
for example, the PWVD is covariant with respect to time-frequency
shifts, and distributes over modulation. Experiments were also performed
demonstrating the benefit brought by the PWVD when compared to bilinear
distributions such as the WVD; for example, it was shown in \cite{boashash_higher-order_1995},
example 3.6.2 that for a quadratic chirp signal which is modulated
by white Gaussian noise, the $4^{\text{th}}-$order PWVD, written
\[
W_{z}^{(4)}(t,f)=\int[x(t+\tau/4)]^{2}[x(t-\tau/4)]^{2}e^{-2\pi i\tau f}d\tau,
\]
is clearly concentrated along the linear time-varying frequency component,
whereas the WVD of the same signal is unreadable.

\subsection{Volterra series representation of the Polynomial WVD}

In order to render the PWVD in Volterra series form, we need to internalize
the conditions from (\ref{constraint equations PWVD}) as components
of the Volterra kernel function, $v_{k}^{\text{PWVD}}$. We will use
the following symmetric notation for the PWVD
\begin{equation}
W_{z}^{(k)}(t,f)=\int_{\tau\in\mathbb{R}}e^{-2\pi if\tau}\prod_{l=1}^{k/2}z(t+\lambda_{l}\tau)\cdot z^{*}(t+\lambda_{-l}\tau)d\tau\label{PWVD (symmetric notation)}
\end{equation}
where the coefficients $\lambda_{l}$, $l=(1,\dots,\frac{k}{2})$,
must satisfy the set of equations (\ref{constraint equations PWVD}).
Notice that this form is nearly identical to a Volterra series with
Fourier kernel function, but for the coefficients $\lambda_{l}$.

Recall from equation \ref{slice axis, HOWVD to PWVD-1} that the variable
of integration, $\tau$, is the unique slice of a multidimensional
parameter $\boldsymbol{\tau}_{k}\in\mathbb{R}^{k}$, and is related
to the coefficients $\lambda_{l}$. Just as the Volterra series form
of the WVD involves a component that is a delta distribution, so do
we impose that $\tau$ take on only those values that satisfy the
constraint in (\ref{constraint equations PWVD}); we do this by incorporating,
into the Volterra kernel function, terms that are composed of delta
distributions which annihilate the signal energy outside the solution
set for the variable of integration, $\boldsymbol{\tau}_{k}$.

Let $v_{k}^{'}(\boldsymbol{\tau}_{k})=ke^{-2\pi if(\sum_{l=-k/2}^{k/2}\tau_{l})}$
denote the multidimensional Fourier kernel function of a Volterra
series, which we would like to exhibit the properties of the PWVD,
and let $v_{k}^{\text{PWVD}}(\boldsymbol{\tau}_{k})$ denote the true
PWVD kernel function. Then the first condition of (\ref{constraint equations PWVD})
on the coefficients $\lambda_{l}$ translates into a constraint on
the kernel function, as 
\[
\lambda_{l}=-\lambda_{l-1}\,\,\,(l=1,...,k/2)\implies v_{k}^{\text{PWVD}}(\boldsymbol{\tau}_{k})=v_{k}^{'}(\tau_{k})\prod_{l=1}^{k/2}\delta(\tau_{l}+\tau_{-l});
\]
the second, as
\[
\stackrel[l=1]{k/2}{\sum}\lambda_{l}=1/2\implies v_{k}^{\text{PWVD}}(\boldsymbol{\tau}_{k})=v_{k}^{'}(\boldsymbol{\tau}_{k})\delta(\left[\sum_{l=1}^{k/2}\tau_{l}\right]-1/2);
\]
and the third, as
\[
\stackrel[l=-k/2]{k/2}{\sum}\lambda_{l}^{m}=0\implies v_{k}^{\text{PWVD}}(\boldsymbol{\tau}_{k})=v_{k}^{'}(\boldsymbol{\tau}_{k})\delta(\sum_{l=-k/2}^{k/2}\tau_{l}^{m}).
\]
Putting these terms together results in the kernel function
\begin{equation}
v_{k}^{\text{PWVD}}(\boldsymbol{\tau}_{k})=\left[\prod_{l_{1}=1}^{k/2}\delta(\tau_{l_{1}}+\tau_{-l_{1}})\right]\cdot\delta\left(\sum_{l_{2}=-k/2}^{k/2}\tau_{l_{2}}^{m}\right)\cdot\delta\left(\left[\sum_{l_{3}=1}^{k/2}\tau_{l_{3}}\right]-1/2\right)\cdot ke^{-2\pi if\left(\sum_{l=-k/2}^{k/2}\tau_{l}\right)}.\label{PWVD Volterra series kernel function}
\end{equation}
Thus, as a $k^{\text{th}}-$order Volterra series, (\ref{PWVD (symmetric notation)})
is of the form 
\begin{equation}
V_{z}^{\text{PWVD}}(t,f)=\int_{\boldsymbol{\tau}_{k}\in\mathbb{R}^{k}}v_{k}^{\text{PWVD}}(\boldsymbol{\tau}_{k})\prod_{l=1}^{k/2}z_{l}(t+\tau_{l})\cdot z^{*}(t+\tau_{-l})d\tau_{l}d\tau_{-l},\label{Volterra series representation of the PWVD}
\end{equation}
where $v_{k}^{\text{PWVD}}(\boldsymbol{\tau}_{k})$ is as in (\ref{PWVD Volterra series kernel function}).

Following \cite{boashash_polynomial_1994}, equation (\ref{Volterra series representation of the PWVD})
can also be used to define a class of Volterra series generalizing
Cohen's class to higher dimensions: 
\[
C_{z}^{\text{(k)}}(t,f)=\int_{\boldsymbol{\Omega}_{k}\in\mathbb{R}^{k}}e^{-2\pi it\sum\Omega_{k}}\cdot\hat{v}_{k}^{\text{PWVD}}(\boldsymbol{\Omega}_{k})\cdot c_{k}(\boldsymbol{\Omega}_{k})\prod_{m=1}^{k/2}\hat{z}_{m}(\omega_{m})\cdot\hat{z^{*}}(\omega_{-m})d\omega_{m}d\omega_{-m}
\]
where $c_{k}:\mathbb{R}^{k}\rightarrow\mathbb{R}$ is the smoothing
parameter function, and $\boldsymbol{\Omega}_{k}=[\omega_{-k/2},\dots,\omega_{-1},\omega_{1},\dots,\omega_{k/2}]$
is the $k-$dimensional vector of frequency variables.


\typeout{} 
\chapter*{Conclusion and prospects}

\markboth{Conclusion}{}\addcontentsline{toc}{chapter}{\numberline{6}Conclusion and Prospects}In
this work, we have atempted to lay the foundations of a compositional approach
to nonlinear audio signal processing that is based on the Volterra
series, a universal model of nonlinear dynamics. We have done so by \emph{categorifying} the Volterra series,
by recasting it as a functor on the category $S'(\mathbb{R})$,
whose objects are signals and whose morphisms are convolutors. 
This motivated us to consider the affect that applying a linear transformation to
the input of a Volterra series has upon its output.

We then introduced the concept of morphism between Volterra series, to model how a nonlinear system changes.
We defined a morphism as a kind of lens map, and showed that it satisfies the property of being a natural transformation.
This led us to define the category of Volterra series, \emph{Volt}, a kind of universe of nonlinear systems.
We then studied three elementary monoidal products on \emph{Volt}: sum,
product, and series composition, thus exhibiting \emph{Volt} as a monoidal category, and in doing so, set on
firm theoretical ground a compositional approach to the modeling of complex nonlinear systems and their transformations.

We finally turned to the theory of bilinear time-frequency distributions
and their higher-order counterparts, in particular the polynomial TFDs. We reviewed an equivalence between
Cohen's class of TFDs and a subclass of second-order double Volterra
series, and showed how it extends to one between the polynomial TFDs
and a class of higher-order VS. Thus we provided a new bridge whereby
higher-order TFDs may be treated using Volterra series methods.

\section*{Future work}

The theory we have described could be extended or generalized in a
number of directions. For example, we did not discuss the analysis
-- or \emph{de}composition -- and identification of nonlinear systems
via probing techniques \cite{peyton_jones_new_2018}, although this
is of critical practical importance and a major subject in the VS
literature. Such techniques can also be used to solve for the parameters
of entire classes of nonlinear systems algebraically, as shown in
\cite{peyton_jones_computation_2019}. Related to identification techniques
are also \emph{order separation }methods, which are used to preprocess
Volterra series models and speed up subsequent probing-based procedures.
An interesting example of order separation is given by the phase-based
technique of Bouvier et al. \cite{bouvier_phase-based_2021}. Nonlinear
system identification methods stand, moreover, to benefit greatly
from compositionality, since if the components in a factorization
can be identified separately then the results can be combined, resulting
in multiplicative gains. In a similar vein to identification, it would
be useful to study regularization techniques, such as those presented
in \cite{dalla_libera_kernel-based_2021,libera_novel_2019}, in order
to reduce the sensitivity of system identification techniques to edge-cases
or anomolous patterns of behavior.

We only gave, moreover, a tiny sampling of the various interesting
and even fundamental transforms that are contained in \emph{Volt},
any number of which might serve in a useful and structural role, and
many of which already have known applications. For example, in higher-dimensional
signal processing, the Riesz transform plays a role analogous to that
of the Hilbert transform in classical signal processing. It has been
used, e.g., in demodulation of speech signals \cite{aragonda_demodulation_2015}
as well as fringe patterns in optics \cite{larkin_natural_2001}.
It could be fruitful to study the Volterra series representations
of known transforms, and integrate them into new nonlinear audio processing
chains within our framework. Even more fundamentally, a fully general treatment of
the Volterra series would involve at its center the relationship of the
Volterra series and the Radon transform.

Additionally, in our explorations we have only scratched the surface
of the abstract categorical structure of \emph{Volt}. A programme
for elucidating this could follow, as a roadmap, the development of
the theory of polynomial functors as described in \cite{poly_book},
since Volterra series appear to be to signals what polynomials are to sets. Continuing
the theme of categorification, one could also study higher-categorical structures than we
have done here; considering, for example, what a change in a change in a nonlinear system
looks like, and so on.

Although we have treated only Volterra series which process
inputs and produce outputs that are elements of $S'(\mathbb{R})$ and whose
kernels are elements of $S'(\mathbb{R}^{n})$, other domains can be considered
for use in constructing the signal space. In particular, one can define \emph{graph Volterra series},
which process signals that are elements of a space $S'(G)$ defined over
a graph $G$. Convolution and spectral analysis
are then carried out over powers of the underlying graph via spectral
decompositions of a shift operator, principally the graph
Laplacian. For example, given a
graph $G$, the Laplacian matrix of their Cartesian product is of
the form $L(G\times G)=L(G)\otimes I_{n}+I_{m}\otimes L(G)$, where
$\otimes$ is the Kronecker product; an associated eigenbasis is obtained
via $L(G\times G)(X_{i}\otimes Y_{j})=(\lambda_{i}+\mu_{j})(X_{i}\otimes Y_{j})$,
where $X_{i},Y_{j}$ are eigenvectors and $\lambda_{i},\mu_{j}$ are
associated eigenvalues. We expect that there would be many interesting
uses for other graph products, as well - 256 of which are reported
in \cite{barik_laplacian_2015} that are generated from the adjacency,
non-adjacency, and equality relations, alone.

Moreover, just as the field of time-frequency bridges between time
and frequency for signals indexed over $\mathbb{R}$, the recent field
of `vertex-frequency graph signal processing' plays an analogous role
for signals over graph, bridging between the graph signal and graph
spectral domains \cite{stankovic_vertex-frequency_2020}. Furthermore,
the theory of graphons \cite{beck_signal_2023,ghandehari_noncommutative_2022,morency_graphon_2021,ruiz_graphon_2021}
in principle allows for discrete methods to be obtained from continuous
ones via sampling operations, and so for a corresponding continuous
theory of graph-based Volterra series to be developed.

More generally, it should be possible to lift much if not all of the
formalism developed in this work to the setting of nonlinear signal
processing over simplicial complexes or other higher models of space,
wherein associated continuous theories would be obtained in the limit
of refinement. The inputs, outputs, and kernel functions of Volterra
series should be definable over structured spaces such
as manifolds - e.g., over the $d-$dimensional torus, $\mathbb{T}^{d}$
- and simplicial versions thereof, with morphisms of Volterra series being suitably adapted in like
kind. It would be interesting to explore the structure and uses of
morphisms between Volterra series defined over such spaces, including
potential connections with the circle maps described in \cite{essl_deforming_2022}.

\appendix
\typeout{} 
\chapter*{Appendix A: Construction of the category $S'(\mathbb{R})$}

\addcontentsline{toc}{chapter}{Appendix A}

In this appendix, we construct the \emph{base category}, $S'(\mathbb{R})$.
Its collection of objects is the union of the set of signals and the
set of spectra, and its morphisms are certain linear transformations.
It is this category over which, we conjecture, \emph{Volt }exists as a functor category.
Since the ability to work in the spectral domain is crucial to the
use and formalization of Volterra series, we are compelled to define
signals and their spectra as elements of the Schwartz space and its
dual - i.e., to functions that are rapidly decreasing at infinity
and to distributions that are slowly growing at infinity, respectively
- since it is precisely for these objects that the Fourier integral
is well-defined. We provide in this appendix a review, largely drawn
from \cite{vuojamo_timefrequency_2022}, of the relevant notions from
functional and harmonic analysis, but our recapitulation is quite
sparse; see \cite{melrose_introduction_2008}, \cite{gasquet_fourier_2013},
\cite{friedlander_introduction_1998} for a fuller treatment.

\subsection*{Linear Operators}

Consider a signal $u:\mathbb{R}\rightarrow\mathbb{C}$. A system is
linear iff it obeys the superposition and proportionality constraints:
\[
\begin{array}{c}
L(u+u')=L(u)+L(u')\\
\\
L(c\cdot u)=c\cdot L(u)
\end{array}
\]
for any other $u':\mathbb{R}\rightarrow\mathbb{C}$ and $c\in\mathbb{C}$.
Linear systems come in two important classes. To introduce them, we
need the following two elementary operators.

The\emph{ modulation }of $u$ by a frequency $\xi\in\mathbb{\hat{R}}$
, where $\hat{\mathbb{R}}$ denotes the unitary dual of $\mathbb{R}$,
i.e. the space of functions $e_{\xi}(x)=e^{2\pi ix\cdot\xi}$, by
\[
M_{\xi}u(x)=e^{2\pi ix\cdot\xi}u(x).
\]
The \emph{translation} of $u$ by $\tau\in\mathbb{R}$ is given by
\[
T_{\tau}u(x)=u(x-\tau).
\]
Any linear operator, $L$, that commutes with translations is referred
to as a \emph{linear, time-invariant }(LTI)\emph{ system}. It satisfies
the property\\
\[
y(t+\tau)=(L\circ T_{\tau})s(t)=(T_{\tau}\circ L)s(t)
\]
for all constants $\tau\in\mathbb{R}$, where $y(t+\tau)$ is the
system output at time $t+\tau$. Dually, we might call any linear
operator commuting with $M$ a linear\emph{ frequency-invariant }(LFI)
system. An important characterization of LTI systems is that they
do not increase the frequency-domain support of the signals which
they process; likewise, LFI systems do not increase the temporal support
of signals.\footnote{To see this, note that complex exponentials are the
eigenfunctions of LTI systems, so the result of processing any signal by
an LTI system is to scale each Fourier component by a complex number.
That is, LTI systems are represented by \emph{Fourier multiplier }operators.}

Recall the integral definition, given in (4), of the Fourier transform,
$F$. Modulations and translations satisfy the following set of equations
\[
\begin{array}{c}
F\circ M_{\xi}=T_{\xi}\circ F\\
\\
F\circ T_{\tau}=M_{-\tau}\circ F
\end{array}
\]
Thus, we see that translations and modulations do not, exactly, commute;
rather, we have that 
\[
T_{\tau}M_{\xi}=e^{-2\pi i\tau\cdot\xi}M_{\xi}T_{\tau}
\]
(and, so, they \emph{do} commute when the product $\tau\cdot\xi$
is an integer).

Another linear operator that has an important relationship with the
Fourier transform is complex conjugation, $z\mapsto\bar{z}$. Its
interaction with $F$ is given by 
\[
F(\iota(\bar{u}))=\overline{Fu}
\]
where $\iota u(t)=u(-t)$ denotes the \emph{reflection} of the signal
$u$ about zero. We will use this when defining the operation of convolution
by a tempered distribution. First, we recall the operation of convolution
of functions: given two absolutely integrable functions, $u,v\in\mathbb{R}\rightarrow\mathbb{C}$,
their convolution is another function, $u\ast v:\mathbb{R}\rightarrow\mathbb{C}$,
defined by the formula
\[
(u\ast v)(t)=\int u(t-\tau)v(\tau)d\tau.
\]
Convolution is associative and commutative. A central result in Fourier
analysis and signal processing is the \emph{convolution theorem},
which states that the Fourier transform interchanges convolution and
pointwise-multiplication
\[
\widehat{u\ast v}(\xi)=\hat{u}(\xi)\hat{v}(\xi).
\]

\subsection*{The spaces $\mathscr{S}(\mathbb{R})$ and $\mathscr{S}^{'}(\mathbb{R})$}

We now turn to the Schwartz space of rapidly decreasing functions
and its dual, the space of tempered distributions. Denote by $\mathscr{C}^{\infty}(\mathbb{R})$
the space of infinitely differentiable, or smooth, functions. For
$\alpha,m\in N_{0}$, and $K$ a compact subset of $\mathbb{R}$,
$\mathscr{C}^{\infty}(\mathbb{R})$ becomes a topological vector space
when equipped with the family of seminorms $\varphi\mapsto|\varphi|_{m,K}$,
where 
\[
|\varphi|_{m,K}=\sup_{|\alpha|\le m}\left(\sup_{x\in K}|\partial_{x}^{\alpha}\varphi(x)|\right).
\]
The subspace of compactly supported smooth functions is denoted $\mathscr{C}_{c}^{\infty}(\mathbb{R})$.

\paragraph{Definition \textmd{(Schwartz space)}}

The \emph{Schwartz space $\mathscr{S}(\mathbb{R})$ }of rapidly decreasing
smooth functions, or \emph{test functions}, is the subspace of functions
$\varphi\in\mathscr{C}^{\infty}(\mathbb{R})$ for which 
\[
\sup_{x\in\mathbb{R}}|x^{\beta}\partial_{x}^{\alpha}\varphi(x)|<\infty,
\]
for $\beta\in\mathbb{N}_{0}$ -- i.e., all of whose derivatives decay
faster than any polynomial in $x$, as $x\rightarrow\infty$.\\
\\
Schwartz functions are well-behaved under the operations of multiplication
and differentiation. In particular, we have the following duality:

\paragraph{Theorem:}

If $\varphi\in\mathscr{S}(\mathbb{R})$ and $p$ is a polynomial,
then the following identities hold: 
\[
\begin{array}{c}
F(p(\partial_{x})\varphi)(\xi)=p(2\pi i\xi)\cdot F\varphi(\xi)\\
\\
F(p\cdot\varphi)(\xi)=p((-2\pi i)^{-1}\partial_{\xi})F\varphi(\xi),
\end{array}
\]
where $p(\partial_{x})$ denotes the polynomial formed from the polynomial
$p(x)=\sum_{\alpha}a_{\alpha}x^{\alpha}$ by the substitution $x^{\alpha}\mapsto\partial_{x}^{\alpha}$.

\paragraph{Definition \textmd{(Tempered distributions)}}

The space $\mathscr{S}^{'}(\mathbb{R})$ of \emph{tempered distributions
}is the space of continuous linear functionals on $\mathscr{S}(\mathbb{R})$.
The evaluation of a distribution $\eta\in\mathscr{S}^{'}(\mathbb{R})$
at a Schwartz function $\varphi$ is denoted by
\[
\eta(\varphi)\eqqcolon<\eta,\varphi>_{\mathscr{S}^{'},\mathscr{S}}
\]
where the inner product is bilinear.

Any function $u\in L^{p}(\mathbb{R})$ with bounded $L^{p}-$norm
can be made into a tempered distribution $\Lambda_{u}$ by setting
\[
\Lambda_{u}(\varphi)\coloneqq\int u(x)\varphi(x)dx.
\]
However, in general, $\mathscr{S}(\mathbb{R})$ and $\mathscr{S}^{'}(\mathbb{R})$
are not in bijective correspondance; rather, we have the sequence
of relationships
\[
\mathscr{S}(\mathbb{R})\subset L^{p}(\mathbb{R})\subset\mathscr{S}^{'}(\mathbb{R}).
\]
However, the Schwartz functions are dense in the tempered distributions.
\\

\paragraph*{Example: delta function\protect \\
}

The Dirac delta `function' is the distribution, 
\[
\delta\in\mathscr{S}^{'}(\mathbb{R}),\,\,\delta(\phi)=\phi(0).
\]
In practice, $\delta$ must be approximated by a function, referred
to as a \emph{nascent delta function}. The canonical such function
is the \emph{sinc }function:
\[
\text{sinc}(x)=\frac{\sin(\pi x)}{\pi x}.
\]
We then have that
\[
\delta(x)=\lim_{a\rightarrow0}\text{\ensuremath{\frac{\sin(\frac{\pi x}{a})}{\pi x}}}.
\]

\paragraph{Example: Dirac comb function\protect \\
}

The \emph{Dirac Comb $\text{\textcyr{\CYRSH}}$} is the distribution
given by
\[
\begin{array}{c}
\text{\textcyr{\CYRSH}}_{T}\in\mathscr{S}^{'}(\mathbb{R}^{n})\\
\text{\textcyr{\CYRSH}}_{T}(\phi)=\begin{cases}
\phi(x) & x=nT,n\in\mathbb{Z}\\
0 & \text{else}
\end{cases}
\end{array}.
\]
As a periodic function, it is defined by $\text{\textcyr{\CYRSH}}_{T}(t)\coloneqq\sum_{n=-\infty}^{\infty}\delta(t-nT)$,
i.e. it is an infinite train of delta functions at evenly spaced intervals.
$\text{\textcyr{\CYRSH}}_{T}$ has Fourier series representation $\text{\textcyr{\CYRSH}}_{T}(t)=\frac{1}{T}\sum_{n=-\infty}^{\infty}e^{2\pi in\frac{t}{T}}$,
and so Fourier transform $\widehat{\lyxmathsym{\textcyr{\CYRSH}}_{T}}(f)=\begin{cases}
\frac{1}{T}, & f\in\mathbb{Z}\\
0, & \text{else}
\end{cases}$. The Dirac comb is a fundamental object in signal processing, since
multiplication of a signal against \emph{samples} the signal; whereas,
convolution against it results in \emph{periodization}.

In $n-$dimensional space, a cuboid lattice can be defined as the
tensor product of the Dirac comb with itself, $\text{\textcyr{\CYRSH}}_{(T\boldsymbol{1})}^{\text{nD}}(\boldsymbol{k}_{j})=(\text{\textcyr{\CYRSH}}_{(T\boldsymbol{1})})^{\otimes j}(\boldsymbol{k}_{j})=\begin{cases}
\frac{1}{T^{j}}, & \boldsymbol{k}_{j}\in\mathbb{Z}^{n}\\
0, & \text{else}
\end{cases}$, which lattice is regular -- i.e., has period equal in each dimension.
More generally, an $n-$dimensional lattice with not-necessarily-the-same
period in each dimension is notated $\text{\textcyr{\CYRSH}}_{\boldsymbol{T}_{n}}^{\text{nD}}(\boldsymbol{k}_{j})$,
with $\boldsymbol{T}_{n}\in\mathbb{Z}^{n}$. By abuse of notation,
we just write $\text{\textcyr{\CYRSH}}_{T}^{\text{nD}}(\boldsymbol{k}_{j})$
for $\text{\textcyr{\CYRSH}}_{(T\boldsymbol{1})}^{\text{nD}}(\boldsymbol{k}_{j})$,
or $(\text{\textcyr{\CYRSH}}_{T})^{\otimes j}(\boldsymbol{k}_{j})$
for $(\text{\textcyr{\CYRSH}}_{(T\boldsymbol{1})})^{\otimes j}(\boldsymbol{k}_{j})$,
if it is understood that the period of the comb is equal in each dimension.

\subsection*{The spaces of multipliers, $\mathscr{O}_{M}(\mathbb{R})$, and convolutors,
$\mathscr{O}_{C}^{'}(\mathbb{R})$}

Schwartz functions (resp., tempered distributions) are the objects
(signals and spectra, resp.) of our category, $S(\mathbb{R})$. In
a moment, we will define the maps that are its morphisms; but first,
recall the following two definitions:

\paragraph{Definition \textmd{(Fourier transform of tempered distributions)}}

For any $\varphi\in\mathscr{S}(\mathbb{R}),$ the Fourier transform
$\hat{\eta}$ of any tempered distribution $\eta\in\mathscr{S}^{'}(\mathbb{R})$
is defined by 
\[
<\hat{\eta},\varphi>_{\mathscr{S}^{'},\mathscr{S}}=<\eta,\hat{\varphi}>_{\mathscr{S}^{'},\mathscr{S}}.
\]

\paragraph{Definition \textmd{(Convolution of a Schwartz function and a tempered
distribution)}}

Given a Schwartz function $\varphi$ and a tempered distribution $\eta$,
their convolution is defined by 
\[
(\varphi\ast\eta)(x)\coloneqq<\iota\eta,T_{-x}\varphi>_{\mathscr{S}^{'},\mathscr{S}}.
\]
\\
Finally, we define the spaces of multipliers and convolutors.

\paragraph{Definition: Multipliers}

The space of \emph{multipliers}, $\mathscr{O}(\mathbb{R})$, is the
space of functions $\varphi\in\mathscr{S}(\mathbb{R})$ such that,
for every $\alpha\in\mathbb{N}$, there is a polynomial $P_{\alpha}$
such that, $\forall x\in\mathbb{R}$,
\[
|\partial_{\alpha}\varphi(x)|\le|P(x)|,
\]
i.e., whose derivatives are polynomially bounded. Pointwise multiplication
of a Schwartz function by a multiplier results in another Schwartz
function.

\paragraph{Definition: Convolutors}

The space of \emph{convolutors}, $\mathscr{O}^{'}(\mathbb{R})$ is
the space of tempered distributions $\Lambda$ for which, for any
integer $h\ge0$, there is a finite family of continuous functions,
$f_{\alpha}:\mathbb{R}\rightarrow\mathbb{C}$, with index $\alpha\in\mathbb{N}_{0}$,
such that 
\[
\Lambda=\sum_{|\alpha|\le h}\partial_{\alpha}f_{\alpha},
\]
and such that 
\[
\lim_{|x|\rightarrow\infty}(1+|x|)^{h}|f_{\alpha}(x)|=0
\]
for all $|\alpha|\le h$.

The convolutors are precisely those tempered distributions
which map Schwartz functions to Schwartz functions under convolution. The Fourier Transform is a linear bijection between the spaces $\mathscr{O}_{M}(\mathbb{R})$
and $\mathscr{O}_{C}^{'}(\mathbb{R})$.

\subsection*{The category $S'(\mathbb{R})$}

One might be tempted to try to just work with Schwartz functions.
However, restricting the set of morphisms to those which are multipliers does
not produce a category, since the delta distribution, $\delta$, which is needed to produce the identity
morphism at each object, is not in the space. Such a construct \emph{is}, however
a `category without identities' or \emph{semicategory}; see, e.g.,
\cite{tringali_plots_2016} for a general treatment of mathematical
structures, called plots, that are similar to and include those structures
found in category theory, but may lack identities.

\paragraph{Definition: \textmd{The semicategory $S(\mathbb{R})$ is the one
with objects, Schwartz functions, and morphisms, multipliers between them.}\protect \\
}  
$ $

However, if we lift to the space $S'(\mathbb{R})$, then the operation
of convolution by a delta function centered at zero - equivalently
multiplication in the spectral domain by the constant function $\boldsymbol{1}$ - serves
as the identity morphism for any signal. The rest of the category
structure then follows from the associativity of multiplication and
likewise of convolution.

\paragraph{Definition: \textmd{The category $S'(\mathbb{R})$ is the category
with objects, tempered distributions (including Schwartz functions),
and morphisms, convolutors between them.}\protect \\
\textmd{}\protect \\
}

The category $S'(\mathbb{R})$ has a filtered structure, with convolutions in
the time-domain weakly contracting spectral bandwidth.

\typeout{} 
\chapter*{Appendix B: Proof of the associativity of \ensuremath{\vartriangleleft}}

\addcontentsline{toc}{chapter}{Appendix B}

In this appendix, we show that the operation, $\triangleleft$, of
series composition of Volterra series, as defined in equations (\ref{Series composition (multivariate inputs from A to B)})
and (\ref{binary composition (kernels form)}), is associative; i.e.,
\[
(C\triangleleft B)\triangleleft A=C\triangleleft(B\triangleleft A).
\]
To do this, we make use of the following concepts from combinatorics.

\paragraph{Weak, weighted, and restricted $(n,k)-$compositions \protect \\
}

In equation (\ref{binary composition (kernels form)}), the variable
of the second sum ranges over all the possible ways of partitioning
$j$ into $k$ parts of possibly zero size. Each possibility is called
a \emph{weak $k-$composition }of $j$. The notion of weak (integer)
composition is an extension of that of integer composition, which
is a way of writing an integer, $n$, as a sum of integers $\alpha_{i},1\le i\le m$,
i.e., such that $\stackrel[i=1]{m}{\sum}\alpha_{i}=n$. We will denote
the set of $m-$compositions of $n$ as $C(n,m)$.

The $m-$compositions of $n$ are counted by the binomial coefficients
$\binom{n-1}{m-1}$, i.e., $|C(n,m)|=\binom{n-1}{m-1}$. We have as
an identity for these objects that $\stackrel[m=1]{n}{\sum}\binom{n-1}{m-1}=2^{n-1}.$
The \emph{weak }$k-$compositions of $j$ are, in contrast, counted
by the binomial coefficients $\binom{j+k-1}{j}=\binom{j+k-1}{k-1}$,
and can include parts of zero size. Thus, we can rewrite (\ref{binary composition (kernels form)})
as\\
\begin{equation}
\stackrel[k=0]{n_{B}}{\sum}\underset{p\in\binom{j+k-1}{j}}{\sum}\widehat{b}_{k}(S_{p}^{(j,k)}\boldsymbol{\Omega}_{j})\stackrel[r=1]{k}{\prod}\widehat{a}_{\alpha_{r}^{p}}(\boldsymbol{\theta}_{r}^{p})\label{binary comp (repeated)}
\end{equation}
Note that we write $\underset{p\in\binom{j+k-1}{j}}{\sum}$, rather
than, say, $\stackrel[p=1]{\binom{j+k-1}{j}}{\sum}$ , to emphasize
that, while a linear order can be given to the elements of this set,
doing so would involve breaking the symmetry exhibited by the weak
compositions.\footnote{The weak compositions are indexed by elements of the slices of the
combinatorial object, itself a kind of proto-multilinear space, known
as Pascal's simplex. That is, the set of weak $k-$compositions of
$j$ is isomorphic to the set of points in the $j^{\text{th}}$slice
of Pascal's $k-$simplex. In particular, we have that $\underset{p\in\binom{j+k-1}{j}}{\sum}\binom{j}{\alpha_{1}^{p},\alpha_{2}^{p},...,\alpha_{k}^{p}}=k^{j}$,
where $\binom{j}{\alpha_{1},\alpha_{2},...,\alpha_{k}}$ is the multinomial
coefficient, which can also be written $\frac{(\stackrel[r=1]{k}{\sum}\alpha_{r})!}{\stackrel[r=1]{k}{\prod}\alpha_{r}!}$.}\\

The basic strategy of the following proof of associativity is to exhibit
a bijection between the sets of homogeneous operators that are obtained
by substituting equation (\ref{binary composition (kernels form)})
into itself in each of the two possible ways, as depicted in Figs$.$
\ref{Ternary-Series-Composition-Type1} and \ref{Ternary-Series-Composition-Type2}.
Denote by $(v\triangleleft w)_{j}$ the $j^{\text{th}}-$order operator
of the composite series $V\triangleleft W$. Then these two ternary
compositions correspond to the following two sets of equations: \\

Type 1:
\begin{equation}
\begin{array}{c}
\widehat{(c\triangleleft(b\triangleleft a))}_{j}(\boldsymbol{\Omega}_{j})=\\
\\
\stackrel[k=0]{n_{C}}{\sum}\underset{p\in\binom{j+k-1}{j}}{\sum}\widehat{c}_{k}(S_{p}^{(j,k)}\boldsymbol{\Omega}_{j})\stackrel[r=1]{k}{\prod}(\widehat{b\triangleleft a})_{\alpha_{r}^{p}}(\boldsymbol{\theta}_{r}^{p})=\\
\\
\stackrel[k=1]{n_{C}}{\sum}\underset{p\in\binom{j+k-1}{j}}{\sum}\widehat{c}_{k}(S_{p}^{(j,k)}\boldsymbol{\Omega}_{j})\stackrel[r=1]{k}{\prod}\left\{ \stackrel[l=1]{n_{B}}{\sum}\underset{q\in\binom{\alpha_{r}^{q}+l-1}{\alpha_{r}^{q}}}{\sum}\widehat{b}_{l}(S_{q}^{(\alpha_{r}^{q},l)}\boldsymbol{\Omega}_{\alpha_{r}^{q}})\stackrel[s=1]{l}{\prod}\widehat{a}_{\alpha_{s}^{q}}(\boldsymbol{\theta}_{s}^{q})\right\} 
\end{array}\label{ternary type 1 - non normal form}
\end{equation}
where we have as the constraints, on the variables $\alpha^{p}$ and
$\alpha^{q}$, respectively, that $\stackrel[r=1]{k}{\sum}\alpha_{r}^{p}=j$
and, for each $r\in k$, $\stackrel[s=1]{l}{\sum}\alpha_{s}^{q}=\alpha_{r}^{p}$.
Thus, we have that

\begin{equation}
\stackrel[r=1]{k}{\sum}(\stackrel[s=1]{l}{\sum}\alpha_{s}^{q})=j.\label{ternary identity - type 1}
\end{equation}

Type 2:
\begin{equation}
\begin{array}{c}
\widehat{((c\triangleleft b)\triangleleft a)}_{j}(\boldsymbol{\Omega}_{j})=\\
\\
\stackrel[k=1]{n_{C\circ B}}{\sum}\underset{p\in\binom{j+k-1}{j}}{\sum}\widehat{c\triangleleft b}_{k}(S_{p}^{(j,k)}\boldsymbol{\Omega}_{j})\stackrel[r=1]{k}{\prod}\widehat{a}_{\alpha_{r}^{p}}(\boldsymbol{\theta}_{r}^{p})=\\
\\
\stackrel[k=1]{n_{C\circ B}}{\sum}\underset{p\in\binom{j+k-1}{j}}{\sum}\left\{ \stackrel[l=1]{n_{C}}{\sum}\underset{q\in\binom{k+l-1}{k}}{\sum}\widehat{c}_{l}(S_{q}^{(k,l)}\boldsymbol{\Omega}_{k})\stackrel[s=1]{l}{\prod}\widehat{b}_{\alpha_{s}^{q}}(\boldsymbol{\theta}_{s}^{q})\right\} \stackrel[r=1]{k}{\prod}\widehat{a}_{\alpha_{r}^{p}}(\boldsymbol{\theta}_{r}^{p})
\end{array}\label{ternary type 2 - normal form}
\end{equation}
where we have as the constraints, on the variables $\alpha^{p}$ and
$\alpha^{q}$, respectively, that $\stackrel[r=1]{k}{\sum}\alpha_{r}^{p}=j$
and $\stackrel[s=1]{l}{\sum}\alpha_{s}^{q}=k$; and, so, that

\begin{equation}
\sum_{r=1}^{(\sum_{s=1}^{l}\alpha_{s}^{q})}\alpha_{r}^{p}=j.\label{ternary identity - type 2}
\end{equation}
We would like to simplify these expressions; in particular, to reduce
eq. (\ref{ternary type 1 - non normal form}) to \emph{normal form},
wherein every $\prod$ is written to the right of every $\sum$. In
order to do so, we will use the following technique for manipulating
dependent sums and products:

\paragraph{Pushing $\sum$ past $\prod$}

The following algebraic equivalence between products of sums and sums
of products is useful when manipulating polynomials (\cite{poly_book},
Chp. 2) - and, so, Volterra series. Given any set $I$, $I-$indexed
set $J=\{J(i),i\in I\}$, and dependent set $X(i,j)$, we have an
equivalence:
\begin{equation}
\prod_{i\in I}\sum_{j\in J(i)}X(i,j)\cong\sum_{\bar{j}\in\underset{i\in I}{\prod}J(i)}\prod_{i\in I}X(i,\bar{j}(i))\label{pushing sigma past pi - dependent form}
\end{equation}
where $\bar{j}:(i\in I)\rightarrow J(i)$ is a \emph{dependent function}.\\

In the special case where the codomain, $J$, does not depend on $i,$
(\ref{pushing sigma past pi - dependent form}) simplifies to
\begin{equation}
\prod_{i\in I}\sum_{j\in J}X(i,j)\cong\sum_{\bar{j}:I\rightarrow J}\prod_{i\in I}X(i,\bar{j}(i))\label{eq:18}
\end{equation}
This identity will allow us to rewrite equation (\ref{ternary type 1 - non normal form})
in normal form, exhibiting it similarly to the expression (\ref{binary comp (repeated)})
for binary composition.\\

Now we begin by using eq. (\ref{eq:18}) to manipulate the equation
(\ref{ternary type 1 - non normal form}) of type 1. 
\[
\begin{array}{c}
\widehat{(c\triangleleft(b\triangleleft a))}_{j}(\boldsymbol{\Omega}_{j})=\stackrel[k=1]{n_{c}}{\sum}\underset{p\in\binom{j+k-1}{j}}{\sum}\widehat{c}_{k}(S_{p}^{(j,k)}\boldsymbol{\Omega}_{j})\stackrel[r=1]{k}{\prod}(\widehat{b\triangleleft a})_{\alpha_{r}^{p}}(\boldsymbol{\theta}_{r}^{p})\\
\\
=\stackrel[k=1]{n_{C}}{\sum}\underset{p\in\binom{j+k-1}{j}}{\sum}\widehat{c}_{k}(S_{p}^{(j,k)}\boldsymbol{\Omega}_{j})\stackrel[r=1]{k}{\prod}\left\{ \stackrel[l=1]{n_{B}}{\sum}\underset{q\in\binom{\alpha_{r}^{q}+l-1}{\alpha_{r}^{q}}}{\sum}\widehat{b}_{l}(S_{q}^{(\alpha_{r}^{p},l)}\boldsymbol{\Omega}_{\alpha_{r}^{p}})\stackrel[s=1]{l}{\prod}\widehat{a}_{\alpha_{s}^{q}}(\boldsymbol{\theta}_{s}^{q})\right\} \\
\\
=\stackrel[k=1]{n_{C}}{\sum}\underset{p\in\binom{j+k-1}{j}}{\sum}\widehat{c}_{k}(S_{p}^{(j,k)}\boldsymbol{\Omega}_{j})\underset{\bar{l}:k\rightarrow n_{B}}{\sum}\stackrel[r=1]{k}{\prod}\underset{q\in\binom{\alpha_{r}^{p}+\bar{l}(r)-1}{\alpha_{r}^{p}}}{\sum}\widehat{b}_{\bar{l}(r)}(S_{q}^{(\alpha_{r}^{p},\bar{l}(r))}\boldsymbol{\Omega}_{\alpha_{r}^{p}})\stackrel[s=1]{\bar{l}(r)}{\prod}\widehat{a}_{\alpha_{s}^{q}}(\boldsymbol{\theta}_{s}^{q})\\
\\
=\stackrel[k=1]{n_{C}}{\sum}\underset{p\in\binom{j+k-1}{j}}{\sum}\widehat{c}_{k}(S_{p}^{(j,k)}\boldsymbol{\Omega}_{j})\underset{\bar{l}:k\rightarrow n_{B}}{\sum}\underset{\bar{q}\in\stackrel[r=1]{k}{\prod}Q(r)}{\sum}\stackrel[r=1]{k}{\prod}\widehat{b}_{\bar{l}(r)}(S_{\bar{q}(r)}^{(\alpha_{r}^{p},\bar{l}(r))}\boldsymbol{\Omega}_{\alpha_{r}^{p}})\stackrel[s=1]{\bar{l}(r)}{\prod}\widehat{a}_{\alpha_{s}^{\bar{q}(r)}}(\boldsymbol{\theta}_{s}^{\bar{q}(r)})
\end{array}
\]
\\
 where $Q(r)=\binom{\alpha_{r}^{p}+\bar{l}(r)-1}{\alpha_{r}^{p}}$.
Continuing, we have\\
\\
\begin{equation}
\stackrel[k=1]{n_{C}}{\sum}\underset{\bar{l}:k\rightarrow n_{B}}{\sum}\underset{p\in\binom{j+k-1}{j}}{\sum}\underset{\bar{q}\in\stackrel[r=1]{k}{\prod}Q(r)}{\sum}\widehat{c}_{k}(S_{p}^{(j,k)}\boldsymbol{\Omega}_{j})\stackrel[r=1]{k}{\prod}\widehat{b}_{\bar{l}(r)}(S_{\bar{q}(r)}^{(\alpha_{r}^{p},\bar{l}(r))}\boldsymbol{\theta}_{r}^{p})\stackrel[s=1]{\bar{l}(r)}{\prod}\widehat{a}_{\alpha_{s}^{\bar{q}(r)}}(\boldsymbol{\theta}_{s}^{\bar{q}(r)}),\label{eq:19}
\end{equation}
\\
\\
 which can be rewritten as
\begin{equation}
\underset{(k,\text{\ensuremath{\bar{l})}}}{\sum}\underset{(p,\bar{q})}{\sum}\widehat{c}_{k}(S_{p}^{(j,k)}\boldsymbol{\Omega}_{j})\stackrel[(r,s)=(1,1)]{(k,\bar{l}(r))}{\prod}\widehat{b}_{\bar{l}(r)}(S_{\bar{q}(r)}^{(\alpha_{r}^{p},\bar{l}(r))}\boldsymbol{\theta}_{r}^{p})\widehat{a}_{\alpha_{s}^{\bar{q}(r)}}(\boldsymbol{\theta}_{s}^{\bar{q}(r)})\label{ternary type 1 - final form}
\end{equation}
\\
 where $k\in n_{C}$, $\bar{l}\in(n_{B})^{k}$, $p\in\binom{j+k-1}{j}$,
and $\bar{q}\in\stackrel[r=1]{k}{\prod}\binom{\alpha_{r}^{p}+\bar{l}(r)-1}{\alpha_{r}^{p}}$.
\\

We now ping-pong back to the other possible order of composition,
expressed in (\ref{ternary type 2 - normal form}). As mentioned,
eq. (\ref{ternary type 2 - normal form}) is already in normal form.
However, we can compress it as we did in the last stages to arrive
at eq. (\ref{ternary type 1 - final form}), in order to make its
resemblence to (\ref{binary comp (repeated)}) more apparent:\\
\[
\widehat{((c\triangleleft b)\triangleleft a)}_{j}(\boldsymbol{\Omega}_{j})=\stackrel[l=1]{n_{C\circ B}}{\sum}\underset{p\in\binom{j+l-1}{j}}{\sum}\widehat{c\triangleleft b}(S_{p}^{(j,l)}\boldsymbol{\Omega}_{j})\stackrel[s=1]{l}{\prod}\widehat{a}_{\alpha_{r}^{p}}(\boldsymbol{\theta}_{r}^{p})
\]
\[
=\stackrel[l=1]{n_{C}\times n_{B}}{\sum}\underset{q\in\binom{j+l-1}{j}}{\sum}\left\{ \stackrel[k=1]{n_{C}}{\sum}\underset{p\in\binom{l+k-1}{l}}{\sum}\widehat{c}_{k}(S_{p}^{(l,k)}(\boldsymbol{\Omega}_{l})\stackrel[r=1]{k}{\prod}\widehat{b}_{\alpha_{r}^{p}}(\boldsymbol{\theta}_{r}^{p})\right\} \stackrel[s=1]{l}{\prod}\widehat{a}_{\alpha_{s}^{q}}(\boldsymbol{\theta}_{s}^{q})
\]
\begin{equation}
=\stackrel[k=1]{n_{C}}{\sum}\stackrel[l=1]{n_{C}\times n_{B}}{\sum}\underset{q\in\binom{j+l-1}{j}}{\sum}\underset{p\in\binom{l+k-1}{l}}{\sum}\widehat{c}_{k}(S_{p,q}^{(j,k)}\boldsymbol{\Omega}_{j})\stackrel[r=1]{k}{\prod}\widehat{b}_{\alpha_{r}^{p}}(\boldsymbol{\theta}_{r}^{p})\stackrel[s=1]{l}{\prod}\widehat{a}_{\alpha_{s}^{q}}(\boldsymbol{\theta}_{s}^{q})\label{eq:21}
\end{equation}
\\
\begin{equation}
=\stackrel[(k,l)=(1,1)]{(n_{C},n_{C}\times n_{B})}{\sum}\underset{(q,p)\in(Q,P)}{\sum}\widehat{c}_{l}(S_{p,q}^{(j,k)}\boldsymbol{\Omega}_{j})\stackrel[(r,s)=(1,1)]{(k,l)}{\prod}\widehat{b}_{\alpha_{r}^{p}}(\boldsymbol{\theta}_{r}^{p})\widehat{a}_{\alpha_{s}^{q}}(\boldsymbol{\theta}_{s}^{q})\label{ternary type  2 - final form}
\end{equation}
\\
where $Q=\binom{j+l-1}{j}$ and $P=\binom{l+k-1}{l}$. Note that choosing
a weak $l-$composition of $j$, along with a weak $k-$composition
of $l$, is equivalent to choosing a weak $k-$composition of $j$;
this is similar to how, in (\ref{ternary type 1 - final form}), choosing
a weak $k-$composition of $j$, along with, for each part indexed
by $r\in k$, of the composition, a weak $\bar{l}(r)-$composition
of that part, produces a weak composition of $j$.

Equations (\ref{ternary type 1 - final form}) and (\ref{ternary type  2 - final form})
are now both in the same form as (\ref{binary comp (repeated)}).
What remains needed in order to complete the proof is to show that
the orders of and arguments to each of the Volterra kernel functions
$\widehat{c}$, $\widehat{b}$, and $\widehat{a}$ are equal. To do
this, we look first at the subindices for the kernels of the operators
of $C$, then of $B$, then $A$. To begin with, the kernels $\widehat{c}_{k}$
are manifestly equal, as are their arguments, since the variables
$j,k$ have identical ranges in each set of equations, and the matrix-vector
products $S_{p}^{(j,k)}\Omega_{j},\,\,S_{p,q}^{(j,k)}\Omega_{j}$
are each indexed by the same set - the weak $k-$compositions of $j$.

To see that the kernels $\widehat{b}_{\alpha_{r}^{p}}$ and $\widehat{b}_{\bar{l}(r)}$are
equal, note that in (\ref{ternary type 1 - final form}), the set
of functions $\bar{l}:k\rightarrow n_{B}$ has cardinality $(n_{B})^{k};$
thus, it is equal to the sum of multinomial coefficients at the $k^{\text{th}}$
level of Pascal's $n_{B}-$simplex, i.e., $(n_{B})^{k}=\underset{p\in\binom{n_{B}+k-1}{n_{B}}}{\sum}\binom{k}{\alpha_{1}^{p},\alpha_{2}^{p},...,\alpha_{n_{B}}^{p}}=\underset{p\in\binom{n_{B}+k-1}{n_{B}}}{\sum}\frac{(\sum_{b=1}^{n_{B}}\alpha_{b}^{p})!}{\prod_{b=1}^{n_{B}}\alpha_{b}^{p}!}$.
This combinatorial object measures all possible ways of partitioning
$k$ into $n_{B}$, possibly empty, parts - \emph{counting permutations};
whereas, elsewhere in our calculations we have considered only the
weak compositions, which are partitions up to a permutation of the
order of the parts.

Thus, to establish equality with equation (\ref{ternary type  2 - final form}),
we can replace $\underset{\bar{l}:k\rightarrow n_{B}}{\sum}$ in (\ref{ternary type 1 - final form})
with $\underset{\bar{l}:k\rightarrow n_{B}/\sim}{\sum},$ where $\sim$
denotes equivalence up to a permutation, of which there are exactly
$\binom{k+n_{B}-1}{k}=\binom{k+n_{B}-1}{n_{B}-1}$-many functions
in the range. We then have that 
\begin{equation}
\stackrel[k=1]{n_{C}}{\sum}\underset{p\in\binom{j+k-1}{j}}{\sum}\underset{\bar{l}:k\rightarrow n_{B}/\sim}{\sum}\underset{\bar{q}\in\stackrel[r=1]{k}{\prod}Q(r)}{\sum}\stackrel[r=1]{k}{\prod}\widehat{b}_{\bar{l}(r)}(S_{\bar{q}(r)}^{(\alpha_{r}^{p},\bar{l}(r))}\boldsymbol{\theta}_{r}^{p})=\stackrel[l=1]{n_{C}\times n_{B}}{\sum}\stackrel[k=1]{n_{C}}{\sum}\underset{p\in\binom{l+k-1}{l}}{\sum}\stackrel[r=1]{k}{\prod}\widehat{b}_{\alpha_{r}^{p}}(\boldsymbol{\theta}_{r}^{p})\label{bijection of the bs}
\end{equation}
where the left hand side instructs one to:
\begin{itemize}
\item first choose an order, $k$ (which we can picture as a tree with $k$
leaves from $C$); then choose a $k-$composition of $j$; then choose
a function assigning each element $r\in k$ an order $\bar{l}(r)$
of $B$; and finally, for all $r\in k$, form the product of the $\widehat{b}_{\bar{l}(r)}$;
\end{itemize}
and the right side instructs one to:
\begin{itemize}
\item first choose an order, $l$, of the composite $C\circ B$\footnote{$C\triangleleft B$, pictured as a tree, has a maximum of $n_{C}\times n_{B}-$many
leaves; it has one tree with $n_{B}-$many leaves for each leaf of
each tree with $n_{C}-$many leaves.}; then choose an order, $k$, of $C$; then choose a $k-$composition,
$p$, of $l$; and finally, for all $r\in k$, form the product of
the $\widehat{b}_{\alpha_{r}^{p}}.$
\end{itemize}
But the image of the function $\bar{l}$ on the left-hand side is
exactly the set of parts of the composition on the right-hand side.
Recalling that $Q(r)=\binom{\alpha_{r}^{p}+\bar{l}(r)-1}{\alpha_{r}^{p}}$,
a similar argument shows that the arguments to these functions are
also the same.

Lastly, equality of the subindices and arguments of the VFRFs of the
operators of $A$ implies that
\begin{equation}
\stackrel[k=1]{n_{C}}{\sum}\underset{p\in\binom{j+k-1}{j}}{\sum}\underset{\bar{l}:k\rightarrow n_{B}/\sim}{\sum}\underset{\bar{q}\in\stackrel[r=1]{k}{\prod}Q(r)}{\sum}\stackrel[r=1]{k}{\prod}\stackrel[s=1]{\bar{l}(r)}{\prod}\widehat{a}_{\alpha_{s}^{\bar{q}(r)}}(\boldsymbol{\theta}_{s}^{\bar{q}(r)})=\stackrel[l=1]{n_{C}\times n_{B}}{\sum}\stackrel[k=1]{n_{C}}{\sum}\underset{q\in\binom{j+l-1}{j}}{\sum}\underset{p\in\binom{l+k-1}{l}}{\sum}\stackrel[r=1]{k}{\prod}\stackrel[s=1]{l}{\prod}\widehat{a}_{\alpha_{s}^{q}}(\boldsymbol{\theta}_{s}^{q}),\label{bijection of the as}
\end{equation}
which says that the set of VFRFs of kernels from $A$, whose orders
ultimately sum to $j$, are formed equivalently by the following two
procedures: either follow the first sequence of steps above for forming
the set of VFRFs of $B$, and then, for each tree $\bar{l}(r)$ of
$B$, form the product of all the parts of previously chosen $\bar{l}(r)-$composition
of $\alpha_{r}^{p}$; or follow the second sequence of steps above,
and then form the product of all the parts in the $l-$composition
of $j$.

Recalling the identities (\ref{ternary identity - type 1}) and (\ref{ternary identity - type 2}),
and rewriting (\ref{ternary identity - type 1}) in dependent form,
we have that
\begin{equation}
\stackrel[r=1]{k}{\sum}(\stackrel[s=1]{\bar{l}(r)}{\sum}\alpha_{s}^{\bar{q}(r)})=j=\sum_{r=1}^{(\sum_{s=1}^{l}\alpha_{s}^{q})}\alpha_{r}^{p}.\label{equivalence of ternary identities}
\end{equation}
But this is exactly stating that there is a bijection between the
two different sets of ways of constructing the composite tree, as
described above. $\square$

\begin{singlespace}
\bibliography{ThesisEx}
 \bibliographystyle{ieeetr}
\end{singlespace}

\end{document}